\documentclass[aps,amsmath,amssymb,rmp,reprint,unsortedaddress]{revtex4-1}

\usepackage{graphicx} %
\usepackage{xcolor}
\usepackage{soul}  
\usepackage{tabularx} 
\usepackage{multirow} 
\usepackage{stackrel}
\begin{document}

\newcommand{\ATP}{\mathrm{ATP}}
\newcommand{\ADP}{\mathrm{ADP}}
\newcommand{\Pin}{\mathrm{P_i}}

\newcommand{\kB}{k_\mathrm{B}}
\newcommand{\kBT}{k_\mathrm{B}T}
\newcommand{\cATP}{\mathrm{[ATP]}}
\newcommand{\cADP}{\mathrm{[ADP]}}
\newcommand{\cPin}{\mathrm{[P_i]}}
\newcommand{\Keq}{\mathrm{K_{eq}}}

\newcommand{\Molar}{\mathrm{M}}
\newcommand{\mMolar}{\mathrm{mM}}
\newcommand{\uMolar}{\mathrm{\mu M}}

\newcommand{\nm}{\mathrm{nm}}
\newcommand{\pN}{\mathrm{pN}}
\newcommand{\s}{\mathrm{s}}

\newcommand{\J}{\mathcal{J}}

\newcommand{\mPstep}{\hat{\mathrm{P}}_{\mathrm{step}}}
\newcommand{\mM}{\hat{\mathrm{M}}}
\newcommand{\mS}{\hat{\mathrm{S}}}
\newcommand{\mF}{\hat{\mathrm{F}}}
\newcommand{\mB}{\hat{\mathrm{B}}}
\newcommand{\vp}{\vec{p}}
\newcommand{\vP}{\vec{P}}
\newcommand{\mI}{\hat{\mathrm{I}}}
\newcommand{\vOneT}{\vec{1}^\intercal}
\newcommand{\prob}{\mathbb{P}}

\newcommand{\fmax}{f_\mathrm{max}}
\newcommand{\fstall}{f_\mathrm{stall}}

%
\title{Theoretical Perspectives on Biological Machines}

\author{Mauro L. Mugnai}
\email{maurolm83@gmail.com}
\affiliation{Department of Chemistry, University of Texas, Austin, TX 78712}
\author{Changbong Hyeon}
\email{hyeoncb@kias.re.kr}
\affiliation{Korea Institute for Advanced Study, Seoul 02455, Republic of Korea}
\author{Michael Hinczewski}
\email{mhincz@gmail.com}
\affiliation{Department of Physics, Case Western Reserve University, OH 44106}
\author{D. Thirumalai}
\email{dave.thirumalai@gmail.com}
\affiliation{Department of Chemistry, University of Texas, Austin, TX 78712}

\begin{abstract}
Many biological functions are executed by molecular machines, which like man made motors consume energy and convert it into mechanical work. Biological machines have evolved to transport cargo, facilitate folding of proteins and RNA, remodel chromatin and replicate DNA. 
A common aspect of these machines is that their functions are driven by fuel provided by hydrolysis of ATP or GTP, thus driving them out of equilibrium. 
It is a challenge to provide a general framework for understanding the functions of biological machines, such as molecular motors (kinesin, dynein, and myosin), molecular chaperones, and helicases. Using these machines, whose structures have  little resemblance to one another, as prototypical examples, we describe a few general theoretical methods  that have provided insights into their functions.  
Although the theories rely on coarse-graining of these complex systems they have proven useful in not only accounting for many {\it in vitro} experiments but also address questions such as how the trade-off between precision, energetic costs and optimal performances  are balanced.  
However, many complexities associated with biological machines will require one to go beyond current theoretical methods. 
We point out that simple point mutations in the enzyme could drastically alter functions, making the motors  bi-directional or result in unexpected diseases or dramatically restrict the capacity of molecular chaperones to help proteins fold. 
These examples are reminders that while the search for principles of generality in biology is intellectually stimulating, one also ought to keep in mind that molecular details must be accounted for to develop a deeper understanding of processes driven by biological machines. 
Going beyond generic descriptions of {\it in vitro} behavior to making genuine understanding of {\it in vivo} functions will likely remain a major challenge for some time to come. 
In this context, the combination of careful experiments and the use of physics and physical chemistry principles will be useful in elucidating the rules governing the workings of biological machines.  
\end{abstract}

\maketitle
%

\tableofcontents

\section{Introduction}

The opening sentence of a perspective by Bustamante \cite{Bustamante11Cell} on the workings of nucleic acid translocases begins with the quote, ``The operative industry of Nature is so prolific that machines will be eventually found not only unknown to us but also unimaginable by our mind", attributed to Marcello Malpighi, who is considered the founder of microscopic anatomy, histology, and embryology. 
This statement, made over three centuries ago, is even more relevant today. It is a reminder that mechanical forces must play a fundamental role in biology. In modern times this vast subject falls under the growing field of mechanobiology. 
The pervasive role of mechanics in living systems controls motility on all length scales, from motion of a single cell on a substrate and collections of cells to dynamics at the molecular level. 
Cooperative interactions between various modules at the molecular level is thought to control functions at the mesoscale.
A number of complex dynamical processes such as transcription, translation, transport of vesicles and organelles, folding of proteins and RNA, and chromosome segregation control the sustenance and growth of cells. 
At some level all these biologically important processes involve molecular machines, whose ability to perform their functions, often but always with high efficiency, in a noisy crowded environment is truly remarkable. 
It is worth remembering that the ability to execute a variety of functions  distinguishes living and abiotic systems.  
Because of functional demands in living systems, which also includes adaptation to changing environmental conditions,  it is virtually impossible to fully describe biology without evolutionary considerations. 
Although not the focus of our perspective, the role of evolutionary constraints must also be integrated with physical models in order to discover general principles governing the functions of biological machines.  

\begin{figure}
    \centering
    \includegraphics[width=0.5\textwidth]{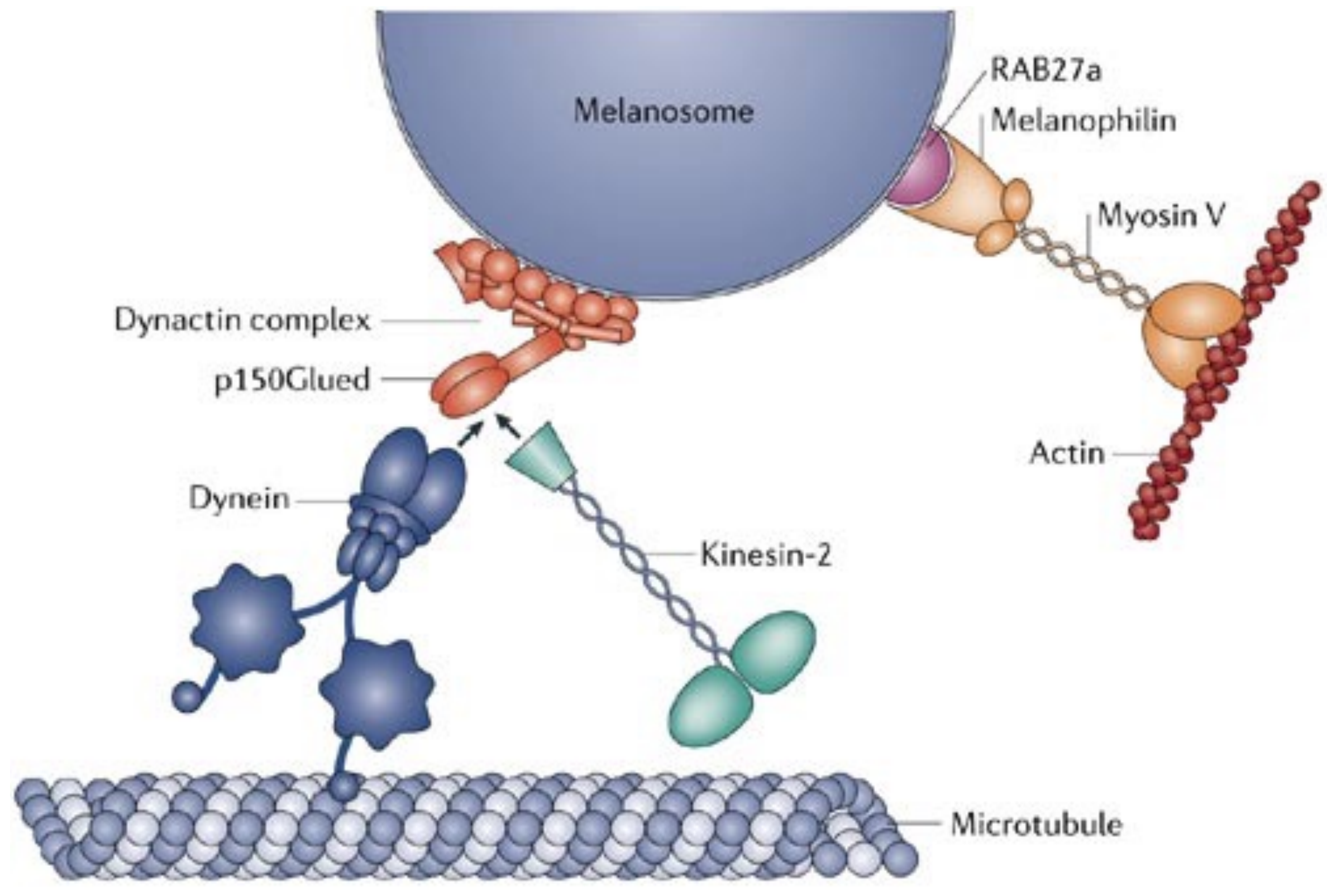}
    \caption{
    Illustration of the complexity of transportation of melanosomes, which are vesicles containing melanin. 
    Both dynein and kinesin-2 compete for the same binding site on dynactin mediated by p150Glued. 
    Depending on the function (aggregation of melanosomes or their dispersion throughout the cell) one or the other wins. 
    When melanosomes are dispersed in the cell they are transported by kinesin-2 and myosin V whereas when they aggregate dynein moves the cargo.
    Figure extracted from~\citet{Soldati06NRMCB}.
    }
    \label{Fig:KinDynMyo}
\end{figure}

What are the characteristics of biological machines? 
First, there are many varieties of machines, all of which should be thought of as enzymes, which consume energy and perform mechanical work to carry out specific tasks. 
A few of the machines that we consider here are kinesin, myosin, and dynein, which are collectively referred to as molecular motors \cite{Vale00Science,Block2007BJ,Sun11BJ,Vladislav14NatComm,Reck-Peterson18NatRevMolCellBiol}. 
These cytoskeletal motors transport cargo by walking, almost always unidirectionally, on filamentous actin and microtubules (MT). Under {\it in vivo} conditions the motors cooperate or there could be a tug-of-war in the process of transport (see Fig.~\ref{Fig:KinDynMyo})~\cite{Gross2002JCB,Levi06BJ}.
To illustrate the diversity of motor-like functions and point out certain emerging general principles,  we also provide theoretical descriptions for the functions of molecular chaperones, which assist in the folding of proteins and ribozymes (RNA enzymes) that cannot do so spontaneously, and helicases with multiple functions that includes separation of double stranded (ds) DNA strands. 
Both molecular chaperones and helicases bear no structural resemblance or sequence similarity to molecular motors. 
Nevertheless, we are convinced that by comparing the characteristics of these seemingly unrelated machines, integrating theory and experiments, unifying themes and differences between them could be elucidated.
Second, all these motors and others not covered here (for example polymerases, ribosomes, and packaging motors)  are all multidomain proteins, whose architectures are spectacularly different. 
Despite the vastly different sequences, structures, and evolutionary origins it is indeed the case that all of these machines utilize some form of chemical energy (generated by ATP or GTP hydrolysis) in order to  amplify the small local conformational changes through their structural linkages to facilitate large conformational changes for functional purposes. Such conformational amplifications are examples of remarkable allostery or action at a distance, and could also be couched in terms of information transfer between structural subunits that are spatially well separated, which in some cases (dynein for example \cite{Bhabha14Cell}) can be as large as 25nm.  

A large number of experiments have unveiled many of the details of how these machines move by converting chemical energy into mechanical work (see for example \cite{Hartman11AnnRevCellDevBiol,Spudich10NatRevMolCellBiol}~\cite{DeLaCruz2004COCB,Sweeney2010ARB,Holzbaur2010COCB}.  A remarkable number of experimental methods have been developed to address various aspects of biological machines. These include, but are not restricted to, ensemble experiments (stopped flow and fluorescent labeling) that provide the much needed data on ATP hydrolysis and ADP release rates, single molecule experiments that yield dwell time distributions in molecular motors, processivity and velocity as a function of external loads in motors and helicases. In addition, a combination of ensemble experiments and determination of structures using X-ray crystallography and cryo-EM experiments has produced a vivid picture of the molecular basis of chaperone function.  
Of particular note are optical trap experiments, which have been used to obtain mean motor velocity as a function of a resistive force and ATP concentration in kinesin, myosin, and dynein, and the dependence of velocity and processivity in a number of helicases. 
No point would be served in reviewing the detailed results from these experiments as there are many articles that the interested reader might consult.
 
Our focus here is to describe theoretical approaches that are rooted in a number of areas in physics in order to understand principally outcomes of {\it in vitro} experiments. 
The theoretical approaches in these studies were developed, and continue to be the focus of current research, in order to quantitatively explain the experimental observations, and shed light on new puzzles that seem to arise with ever improving advances in experimental techniques. 
Perhaps, the most detailed view of how biological machines operate might be obtained from molecular dynamics simulations. 
Although atomically detailed molecular dynamics simulations have been performed to get a molecular picture of  certain aspects of the functions of motors \cite{Hwang08Structure,Hwang17eLife} and other machines (see for example \cite{Ma00JMB,Stan05JMB,Elber10PNAS}, the current limitations of such approaches prevent them from making direct contact with experiments. 
Improvements in the development of accurate energy functions  (also known as force fields) and enhancement in computer capacity to enable simulations for long times will in the future bridge the gap between what is currently possible and what is needed for realistic description of molecular machines. 

\begin{figure}
    \centering
    \includegraphics[width=0.5\textwidth]{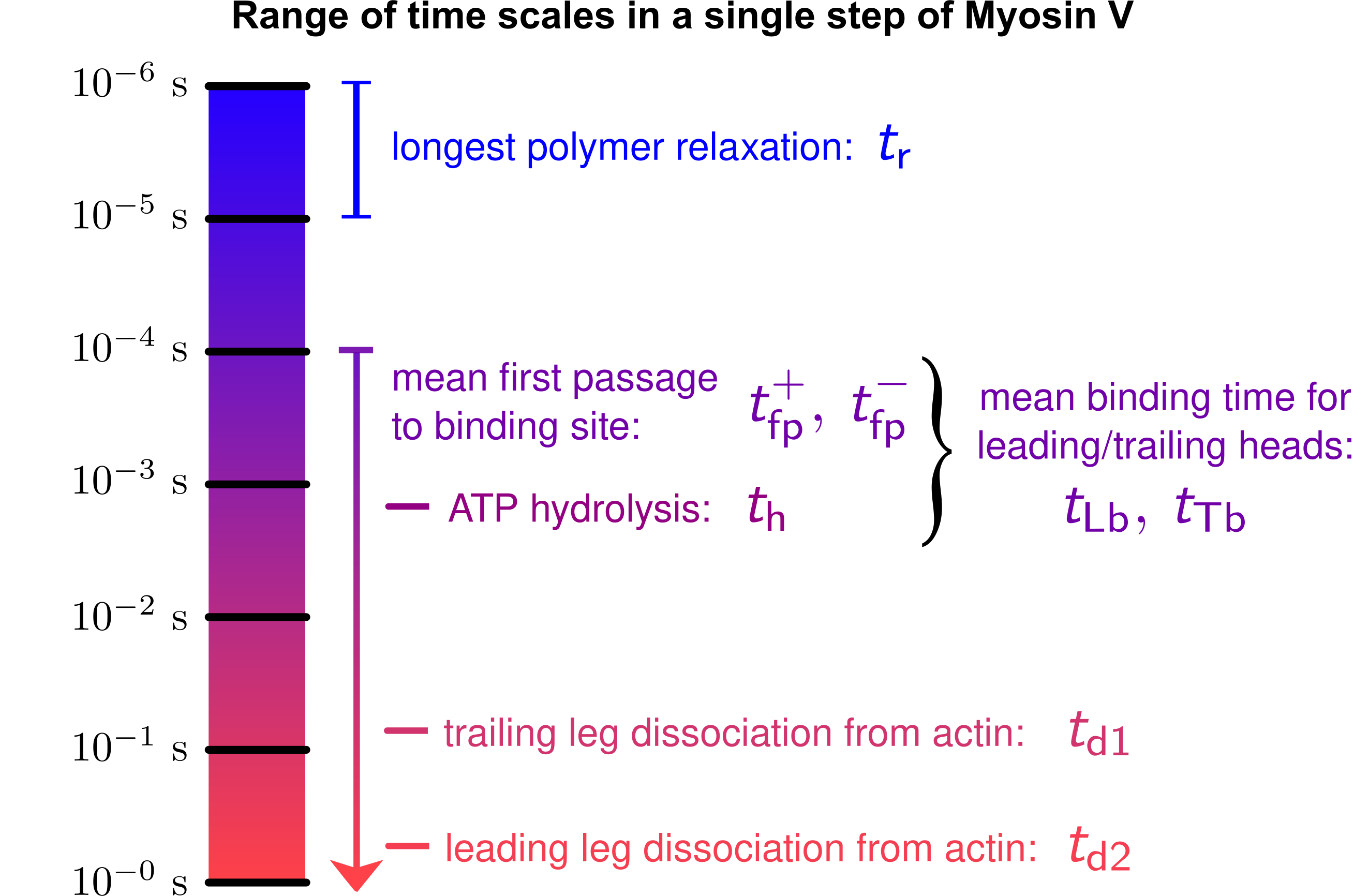}
    \caption{Significant physical timescales associated with myosin V dynamics, from the coarse-grained polymer theory model of \citet{Hinczewski13PNAS}.}
    \label{myoV_times}
\end{figure}

The actions of molecular machines, like all biological processes, are complicated, involving cooperative dynamics on multiple time scales. 
This is illustrated using the typical time scales involved in a single step of myosin V (Fig.~\ref{myoV_times}). 
In many problems in physics it is the hoped that simple models, that capture the essence of the problem  can be created, which be can be solved, in order to obtain insights into highly non-trivial systems. 
This strategy is hard to implement for biological machines because the interaction energies are highly heterogeneous (like in spin glasses), thus making it hard to construct a reasonable coarse-graining strategy. 
Nevertheless, in order to make progress one has to devise tractable coarse-grained models, which can be either simulated or solved analytically (at least approximately). 
The efficacy of such approaches can then be assessed by direct comparisons with experimental results, and their abilities to provide insights into the mechanisms of their functions.
Remarkably, inspired by advances in experiments in the last decade, several theoretical models have been proposed, which have greatly contributed to our understanding of molecular machines. 
These developments have occurred in different contexts, which belie the underlying unifying principles. 
To bring these issues to the fore, we provide our perspectives on the application of these models, which should be thought of as coarse-grained network models or their generalizations. 
Such models have been created for explaining not only the motility of motors, but also the functions of molecular chaperones, and helicases. 
We focus on the applications of these theoretical ideas in order to account for the functions of these intrinsically non-equilibrium systems.   
Collectively, these approaches show  that, by examining in detail the workings of many machines, universal principles, both at the conceptual and practical levels, might emerge.  

The literature on the functions of biological machines is vast. Therefore, we restricted ourselves to only a few topics that focus on theoretical approaches, which are sufficiently general that they can be applied to an array of problems in the field.  Rather than describe many results in detail, we walk the the reader through a selection of theoretical methods, which is necessarily biased, so that she or he can access the literature readily. Before getting to the details of this review we will be remiss if we did not point out one prescient monograph \cite{howard2001book} and two forward looking reviews \cite{Julicher97RMP,kolomeisky07arpc}. The monograph by Howard, written nearly twenty years ago, covers all aspects of cytoskeletal motors and is a landmark in this field. It provides conceptual and practical guidelines needed to understand the fundamentals of motor mechanics.  Questions of generality, such as describing movement of a generic motor either in isolation or as a collection, were addressed using principally the Brownian ratchet model by Julicher, Adjari, and Prost \cite{Julicher97RMP}, whereas Kolomeisky and Fisher \cite{kolomeisky07arpc} summarized the development and  practical applications of stochastic kinetic models. Here, our focus is on the more recent developments and applications of physical principles that have started to provide a unified perspective for not only molecular motors but a large of class of machines with vastly different functions. These new insights have become possible by expanding the scope of traditional stochastic chemical kinetics models for motors, helicases, and molecular chaperones, incorporation of polymer physics concepts to account for the architecture of motors, and use of coarse-grained models in simulating the stepping kinetics of motors and the dynamics of large scale allosteric transitions. 

\begin{figure}
    \centering
    \includegraphics[width=0.5\textwidth]{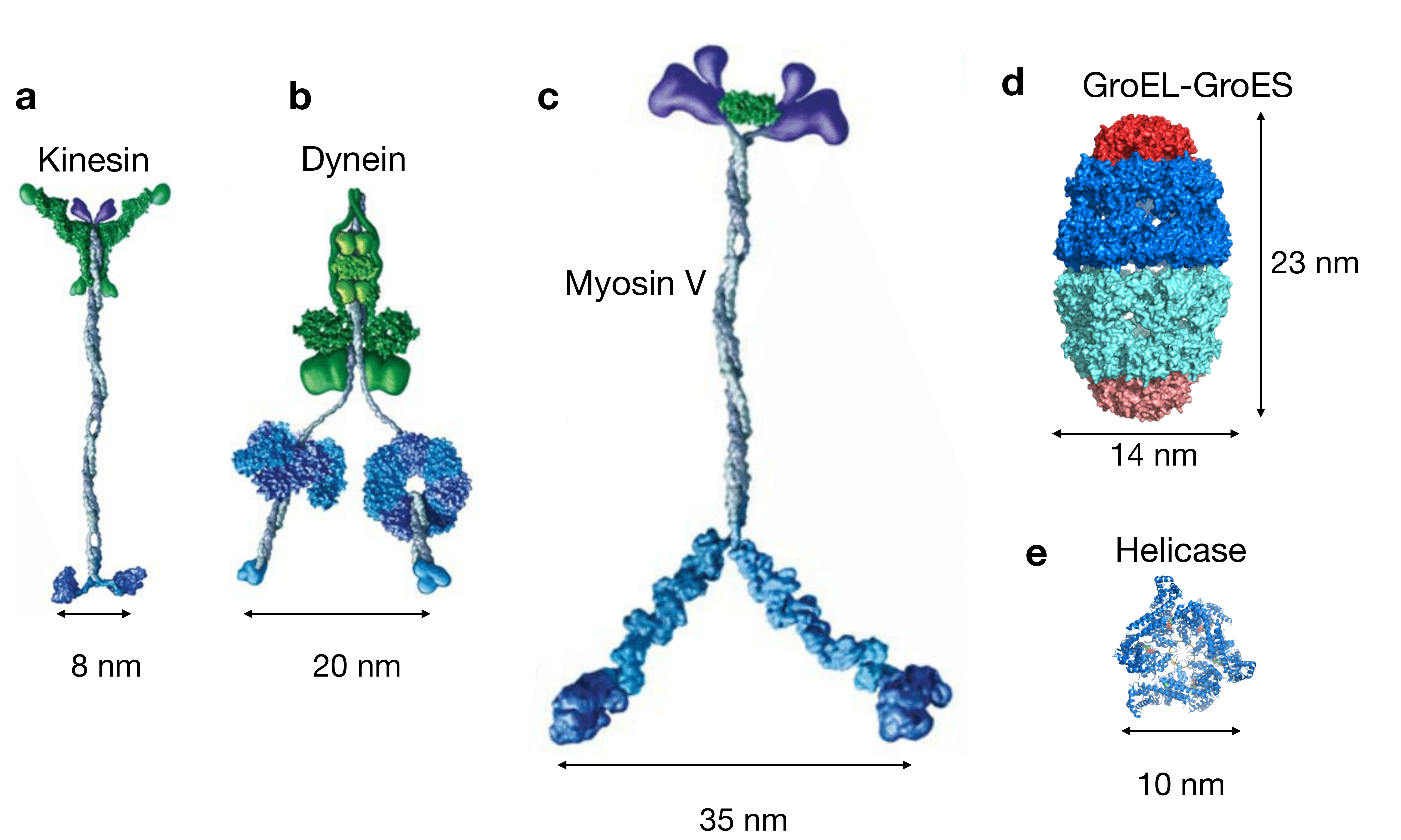}
    \caption{ Schematic representations of the structures of six biological machines. Molecular motors, like conventional kinesin (a), dynein (b), and myosin V (c), show great variations in their structures and sizes. The motor heads in kinesin and myosin V, shown in dark blue, bind directly to the microtubule  and actin, respectively  (not shown) but in dynein the microtubule binding domain (light blue) is separated from the motor domain (hexameric ring structures) by nearly twenty five nanometers. (d) Structure of the GroEL in the symmetric state (Protein Data Bank (PDB) code 4PKO~\cite{Fei14PNAS}), which is the functional state in the presence of misfolded proteins. The blue and aqua colors represent the seven fold symmetric GroEL rings. The red and the pink correspond to the co-chaperonin GroES. (e) Structure of the DnaB  helicase with six subunits assembled as a ring. The figure was created using the coordinates in the PDB code 4ESV~\cite{Itsathitphaisarn12Cell}. 
    Structures (a)-(c) are adapted from~\cite{Vale03Cell}.
    }
    \label{Structures}
\end{figure}
	
\section{Structures} 
The structures of the biological machines that we consider in our perspective (kinesin, myosin, dynein, the {\it E. Coli.} chaperonin machinery GroEL/GroES, and helicases) are shown in Fig.~\ref{Structures}. 
Inspection of the figures shows that there are considerable variations in the architectures, although kinesin-1 (or conventional kinesin) and myosin V share some structural similarities. 
 
 \subsection{Molecular motors} 
 Let us first consider the three motors whose structures are schematically shown in Fig. \ref{Structures} (a-c). 
 Although there are differences between them, the parts list is roughly the same. 
 These motors are dimeric. 
 The nucleotide binding sites are in the two motor heads, which are connected  through a mobile linker (the lever arm in myosin) to a tail domain involved in dimerization and cargo-binding.
 Changes due to nucleotide binding and hydrolysis in the motor heads result in conformational changes in the linker that propel the motor on the cytoskeletal filaments. 
 For the motors to be processive, which means they take multiple steps before disengaging from the track, one head has to be bound till the detached head rebinds to a site on the track. This involves communication between the heads and is referred to as gating, the origin of which is still not fully understood at the molecular level.  
 
 {\it Kinesin:} There are at least forty five members belonging to the kinesin superfamily in  mouse and human genomes \cite{Hirokawa09NatRevMolCellBiol}. They are all microtubule (MT) bound motors, which walk unidirectionally towards the plus end of the MT (for example kinesin-1) or to the minus end (Ncd motor). Kinesin-1 takes precisely 8.2 nm steps, which corresponds to the distance between two adjacent $\alpha/\beta$ tubulin dimers, which are the building blocks of the MT. The size of kinesin-1 is a few nanometers whereas the length of the stalk is in the range of 30-40 nm.   
 The kinesin-1 velocity  depends on the concentration of ATP, saturating at high values. 
 The motor moves towards the plus end of the MT  with maximal velocity of approximately 800 nm/s~\cite{Visscher99Nature}, and is capable of resisting forces on the order of  7 pN ~\cite{Visscher99Nature,Cross05Nature} ($k_B T$ = 4.1 pN$\cdot$nm where $k_B$ is the Boltzmann constant and $T$ is the temperature).  
  
 {\it Myosin:}  The number of genes encoding for myosin motors in roughly thirty five \cite{Sellers00BBA}. The superfamily of the actin bound myosins are divided into fifteen classes \cite{Hartman12JCS,Sellers00BBA}. With the exception of  myosin VI all other members of this family walk towards the plus end of actin. 
 The structure of myosin V (Fig. \ref{Structures}) shows that the motor heads are connected to the lever arms, which are made up of six IQ motifs. 
 The lever arm is  $\approx 36$ nm-long stiff unit (the persistence length exceeds 100 nm), whose size is commensurate with  half the helical repeat length of F-actin.  The  maximal velocity of myosin V, whose motor head is larger than kinesin-1, is roughly 500 nm/s~\cite{baker2004myosin} with a stall force between 2-3 pN~\cite{Mehta99Nature,Veigel2002NCB,Uemura04NSMB,Kad2008JBC}.
 
 {\it Dynein:} Cytoplasmic dynein, discovered over fifty years ago \cite{Gibbons65Science} and pictured in Fig. \ref{Structures}b,  walks somewhat erratically with a broad step-size distribution \cite{Reck06Cell,Dewitt2012Science} on the MT towards the minus end.  
 The structural features of dynein are different when compared with other cytoskeletal motors (see Fig. \ref{Structures}).  First,  the motor head belongs to the class of AAA+ family, which means dynein must have evolved from a different lineage compared with myosin and kinesin. Second, other AAA+ enzymes, such as bacterial chaperonin GroEL (Fig.\ref{Structures}d) and protein degradation machines, are oligomeric assemblies. In contrast, the hexameric ring that constitutes the motor domain assembles from a single  polypeptide chain! 
 Third, the size of the dynein motor head is significantly larger than those of kinesin and myosin. The length of the motor head of dynein along its longest axis is about 25 nm, which is in contrast to the diameter of kinesin, which is only about 5 nm. Finally, although there are six nucleotide binding sites in dynein, hydrolysis in only two (perhaps three) are relevant for its motility.   
 The velocity of dynein  at 1 mM ATP is roughly $\sim$100 nm/s~\cite{Reck2006Cell} with a stall force that is  $\approx$ 7pN~\cite{Toba2006PNAS,Gennerich2007Cell}.
 
 {\it Chaperonins:}
 The beautiful and unusual structure with seven fold symmetry of the {\it E. Coli.} chaperonin machinery, consisting of a complex between GroEL and GroES, resembles an American football (Fig. \ref{Structures}d). The GroEL/GroES machine helps in the folding of recalcitrant proteins that do not fold spontaneously. The structure in Fig. \ref{Structures}d (reported in~\cite{Fei14PNAS}) is the functional state of this stochastic machine that is one of the populated states during the catalytic cycle of GroEL in the presence of misfolded substrate proteins (SPs). There are two chambers in which the SPs could be sequestered. The volume of each of the chambers in the structure in Fig. \ref{Structures}e is about 185,000 \AA$^3$, which is about twice the volume in the relaxed structure in the absence of nucleotides and GroES, the co-chaperonin. The spectacular change in the  chamber volume that occurs during the catalytic cycle, with accompanying alterations in the chemical nature of the cavity interior, changing from hydrophobic to polar, is the mechanism of annealing action of this machine \cite{Todd96PNAS}.
 
 {\it Helicases:}
 Helicases, which are involved in all aspects of nucleic acid metabolism~\cite{ lohman1992,lohman_mechanisms_1996}, function by coupling nucleoside triphosphate (NTP)  hydrolysis, to either translocate on single strand nucleic acids (ssNAs) or unwind double-stranded (ds) DNA.  Depending on their sequences they are classified into six super families (SFs). The structural diversity of helicases can be appreciated by noticing that sequences classified under the SF1 and SF2 families are non-ring forming whereas those in SF3-SF6 form ring structures. The hexametric structure \cite{Itsathitphaisarn12Cell} of the replicative helices (DnaB) from bacteria, which does belong to the AAA+ family, is shown in Fig. \ref{Structures}d. However, unlike GroEL (see Fig. \ref{Structures}d) DnaB has the expected crystallographically allowed six fold symmetry. How the NTP chemistry is coupled to translocation and unwinding remains an outstanding unsolved theoretical problem.

\subsection{Catalytic Cycle} 
\begin{figure}
    \centering
    \includegraphics[width=0.5\textwidth]{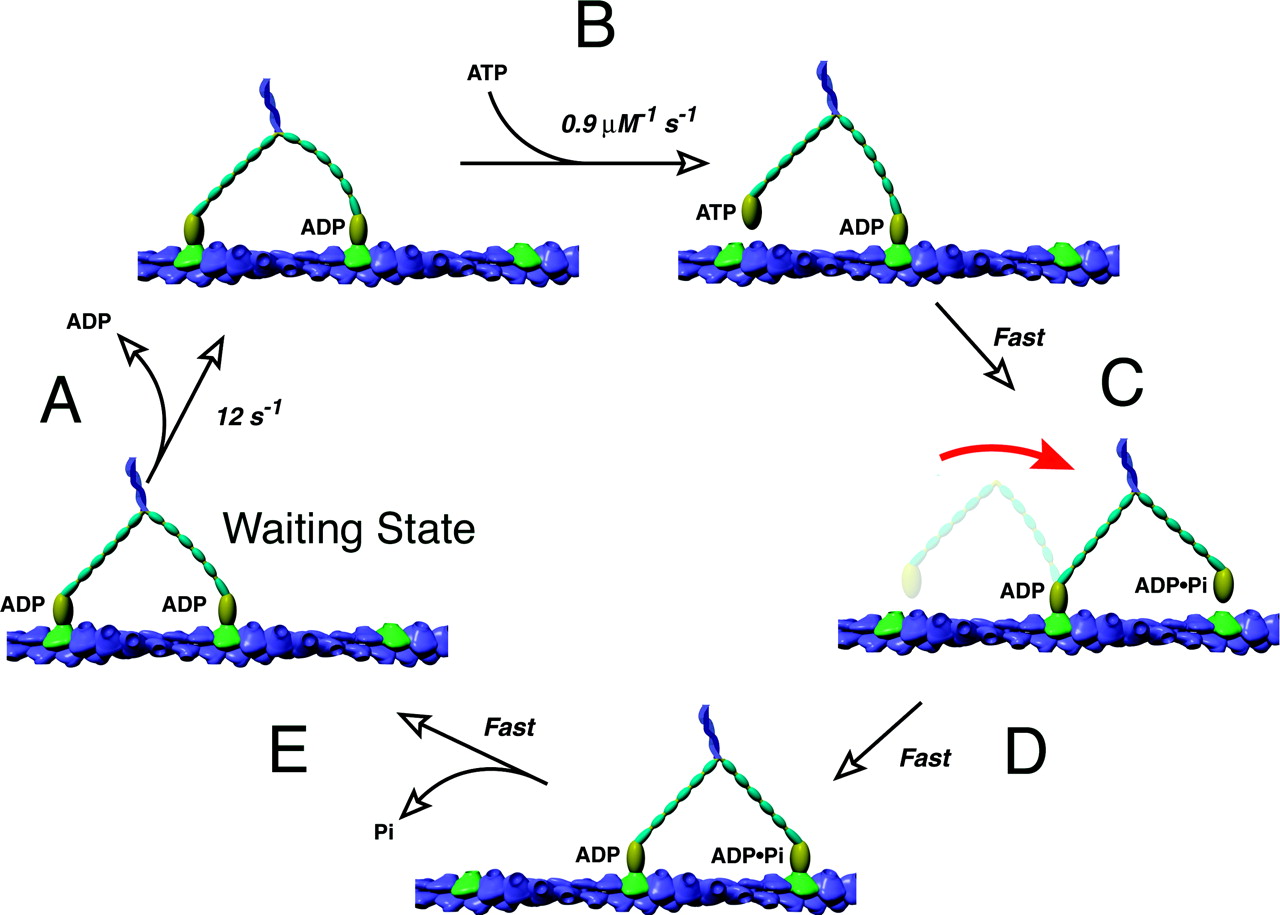}
    \caption{A simple representation of the catalytic cycle of myosin V describing the stepping of the trailing head towards the plus end of actin, which is the dominant pathway in the absence of resisting force. 
    The main text describes the details. 
    As described later in this perspective there are four other pathways that have to be accounted for in order to produce a quantitative theory of the stepping kinetics of this motor. 
    The figure is reproduced from~\cite{Vale2003JCB}}
    \label{Fig:MyoVCycle}
\end{figure}

All machines undergo a catalytic cycle, not unlike man-made motors, in which fuel, usually in the form of ATP, bound to a nucleotide binding site (or sites) is hydrolyzed. 
These events trigger conformational changes that produce motion, which in  motors and helicases results in stepping on the polar tracks or translocation on single stranded nucleic acids. 
In chaperones the nucleotide chemistry is linked to conformational changes, which in turn perform work on the protein or ribozyme to be folded. 
The link between ATPase cycle and function is somewhat different in GroEL, which we describe below. The catalytic cycle for myosin V in the simplest form, which suffices for our purposes, is reproduced in Fig.~\ref{Fig:MyoVCycle}. 
There are four crucial steps (see Fig. \ref{Fig:MyoVCycle}). 
In the initial  state,  ATP  binds to the trailing head (TH), with ADP in the (LH). 
The premature release of ADP from the LH is slowed by rearward tension, which is an example of gating. Upon ATP binding to the TH the interaction with F-actin is weakened, resulting in its detachment from F-actin. 
During the diffusive search by the TH for the forward binding site, ATP is hydrolyzed, producing ADP and the inorganic phosphate, P$_i$. 
In this state, the TH binds to F-actin after which P$_i$ is released from the new LH followed by ADP release from the TH, and the cycle continues till the processive run ends.
Of course, this simple description is incomplete because the rates for nucleotide binding vary depending on the nucleotide concentration. 
Nevertheless, this simple reaction cycle, whose main features hold for all myosin motors, is sufficient to nearly quantitatively characterize many experimental observables (see below). 
The catalytic cycle for kinesin is similar except that the interaction between the MT and kinesin is weakest if the motor head contains ADP. 
All other nucleotide states (the no nucleotide {\it apo}, ATP bound state, the state with ADP and P$_i$) bind strongly to the MT.

\section{Biological machines are active systems. }
Biological machines are non-equilibrium systems that are driven by non-conservative forces  require constant supply of energy. Therefore, it is not surprising that  detailed balance (DB) relation and the fluctuation-dissipation theorem (FDT) are violated~\cite{Battle2016Science,Gladrow2016PRL}.
 These are the key features characterizing the out-of-equilibrium nature of biological machines driven by non-conservative forces. 
Consequently, molecular machines can be thought of as active systems, and hence a note on diffusive motion is warranted. 
Chemical free energy released upon hydrolysis of ATP or GTP is the driving force for directed motion of molecular motors.  
However, if the energy source is not explicitly modeled  in the description of the associated dynamics, molecular motors could be regarded as self-propelled active particles.  To our knowledge such an approach has not been pursued to calculate experimentally measurable quantities, such as the force-velocity relation or the distribution of run lengths of motors.

The dynamics of active particles is fundamentally different from passive particles at equilibrium or moving under the influence of a conservative external field.  
For example, the  mean drift velocity of a spherical colloidal particle with a drag coefficient $\gamma$, and charge $q$ subject to a constant electric field $E$ is $V_{\text{D}}=qE/\gamma$. This relation is readily obtained  by balancing  the Stokes force ($\gamma V_{\text{D}}$) and the force exerted by the  field ($qE$).  The diffusion constant is $D=k_BT/\gamma$, which is the Stokes-Einstein relation. 
Hence, the distribution of the particle position is given by the probability density,  
\begin{equation}
P(x,t)\sim \exp{\left[-\frac{(x-x_0- V_{\text{D}}t)^2}{4Dt}\right]}. 
\end{equation}
Starting from an initial position $x_0$, the particle moves on average at velocity $V_{\text{D}}$ and the  distribution of positions spreads with time as $\langle(\delta x)^2\rangle= 2(k_BT/\gamma) t$. 
Of particular note is that $D$ is defined independently of the particle velocity $V_{\text{D}}$, so that the magnitude of $D$ is not altered by increasing the field strength. Conversely, the velocity of the particle does not depend on the ambient temperature $T$. 
Thus, $V_{\text{D}}$ and $D$ are mutually independent of one another. 

In stark contrast, for an active particle, whose motion is powered by the \emph{internal} fuel, the diffusivity is no longer independent from the driving velocity. 
The \emph{effective diffusion constant}, $D_{\text{eff}}$, defined as $\lim_{t\rightarrow\infty}\langle(\delta x)^2\rangle/2t$, depends on a set of non-thermal parameters  and violates the FDT \cite{Tailleur_2008_PRL,Liu2011Science}. 
 For transport motors exhibiting one-dimensional movement along the cytoskeletal filament, both $V=\lim_{t\rightarrow\infty}d\langle x(t)\rangle /dt$ and $D=(1/2)\lim_{t\rightarrow\infty}d\langle \delta x(t)^2\rangle /dt$ can be measured directly using a given time trace of a motor, which can be measured using single molecule optical tweezer experiments. 
Indeed, such measurements on kinesin-1 \cite{Visscher99Nature} were used to obtain the velocity ($V$) and diffusion coefficient ($D$) from the global analysis of the stepping trajectories. 
These values were used \cite{Visscher99Nature} to estimate the randomness parameter $r=2D/d_0V$ where $d_0(\approx 8.2$ nm) is the step size of kinesin-1. When the diffusion constant $D$ is calculated from the randomness parameter with the knowledge of $V$ and $d_0$, and $D$ is compared with $V$ measured under the same conditions, it can be shown that $D$ increases monotonically with $V$~\cite{Hwang2017JPCL} in contrast to passive diffusion. 
This result is a consequence of the enzyme catalytic turnover~\cite{Hwang2017JPCL}.
As already discussed the effective diffusivity $D$ for kinesin-1, which is an active particle whose dynamics is fueled by ATP hydrolysis free energy, does not obey the FDT.  This illustration shows that these energy consuming enzymes operate out of equilibrium. 

\section{Stochastic Kinetic Models}
In this section we discuss the development of stochastic kinetic models (SKMs) as a means of describing the function of different types of molecular machines.
As stated above and shown explicitly in Fig. \ref{Fig:MyoVCycle} for myosin V, all molecular machines go through cycles during which they hydrolyze ATP (or GTP) and perform a some function.
Bulk-kinetics, single-molecule experiments, and structural studies have shown that the intermediates explored by the molecular machines during the cycle have distinguishable structural or kinetic features and have provided the rate for going from one state to the next.
The temporal resolution in these studies is on the order of milliseconds [although new techniques enable the exploration of time-scales as fast as $\approx 55 \:\mathrm{\mu s}$~\cite{Isojima2016NCB}].
It follows that the dynamics of the molecular machine can be described as a discrete-state (the distinguishable intermediates should be experimentally observed), continuous-time Markov model -- a framework that falls under the rubric of the ``master equation.''
In this model, the cycle of a molecular machine is reduced to a network of states connected by edges that identify the available transitions between the intermediates.
Thus, by solving the dynamics associated with an appropriate network it is possible, in principle, to account for experimental observables in terms of the underlying network parameters.

SKMs have a number of appealing features: (i) they have a direct connection with the biochemistry associated with the molecular machines; (ii) the rates that constitute the model are measurable; (iii) it is sometimes possible to find an analytical solution, and (iv) more complex networks may be solved numerically; (v) finally, SKMs can be directly related with thermodynamics, which makes the model instructive and appeals to the interests, and intuition of a vast scientific community.

Nevertheless, SKMs also have some shortcomings: first and foremost, they often have a large number of parameters.
For instance, a simple model for a molecular machine performing mechanical work against a fixed load and described by a single-cycle network with $N$ intermediate states has $4N-2$ independent parameters, which have to be determined by fitting to appropriate experiments.
However, typically the number of observables are very few, making matching the predictions of the SKM to experiments is difficult.
Furthermore, it is often difficult to interpret the physical meaning of the extracted parameters in terms of the underlying motor architecture and the underlying biochemical cycle. As a consequence, it is necessary to strike a balance between the ``minimality'' of a model, which reduces the risk of over-fitting and increases the generality of SKMs and the predictive power, and the ``comprehensiveness'' of the network considered.
This comes at some risk; for instance, one may be tempted to neglect certain transitions that are ``fast'' compared to others, and therefore are not expected to contribute to the overall phenomenology of the machine.
In addition, it may seem reasonable to ignore off-pathway, slow and rare transitions.
On the other hand, in order to understand the function of molecular machines, experimentalists often probe their response to changes in the environment (for instance modifying the ATP concentration), or by applying mechanical perturbations and studying the response of the motor.
As the environment or perturbations change, the nature of the ``rate-limiting'' step as well as the likelihood of alternative stepping pathways might change. Therefore, a simplified description that ignores certain intermediate states might only work for a limited set of experimental conditions, thereby reducing the number of measurements that can be used to train the parameters or to falsify the predictions. Finally, we note {\it en passant} that the determination of an appropriate catalytic cycle is predicated on a few experimental observables only, and it may not be complete enough for all the states that a machine might sample during its function.

Another issue with SKMs is that structural information is incorporated in the model only in a very approximate way, normally by introducing a parameter that allows the rate to depend on the load applied according to the Bell model~\cite{Bell78Science}.
This reduces the capability of the model to incorporate the wealth of experimental structural information, and decreases the predictive power of the proposed paradigm. However, in some cases a tractable analytical theory that incorporates lever arm structural information into a kinetic model is possible~\cite{Hinczewski13PNAS}, as discussed in more detail below.
Despite these limitations, the framework underlying SKMs not only provides a convenient way to analyze experiments, but also has been used to address conceptual questions related to the efficiency of biological machines.
For these reasons, the theory underlying SKMs is an integral part of describing the function of molecular motors.


\subsection{The Master Equation}
We assume that the system is described by $N$ distinguishable intermediate states.
The probability of being in state $i$ at time $t$ is given by $p_i(t)$, and the time evolution of this probability is governed by the master equation,
\begin{equation}
\frac{dp_i(t)}{dt}=\sum_j p_j(t) w_{ji} -\sum_j p_i(t) w_{ij},
\label{Eq:Master}
\end{equation} 
where $w_{ij} \ge 0$ is the rate for the transition $i\rightarrow j$, subject to the constraint $\sum_i p_i(t)=1$.
Because of microscopic reversibility, if $w_{ij} \ne 0$ then $w_{ji} \ne 0$.
As $t\rightarrow\infty$, the probabilities $p_i(t)$ become independent of time.
This stationary solution of the master equation describes an equilibrium system ($p_i^{eq}$) if detailed balance holds, that is if across all the edges of the network the net flux is zero: $p_i^{eq} w_{ij} - p_j^{eq} w_{ji} = \Delta J_{ij} = 0$.
However, $dp_i/dt = 0$ is also satisfied by the less stringent relation $\sum_j (p^{ss}_iw_{ij} - p^{ss}_jw_{ji}) = \sum_j \Delta J_{ij} = 0$.
Under these conditions the system could be in a non-equilibrium steady state (NESS).
An isolated system is expected to reach an equilibrium state, whereas coupling with an external energy source enables the creation and persistence of a NESS.

\begin{figure}
    \centering
    \includegraphics[width=0.5\textwidth]{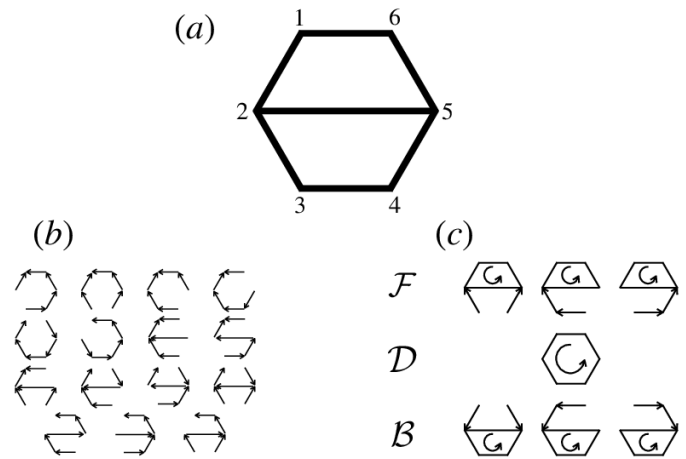}
    \caption{Example of kinetic network.
    The network has 6 states and 7 edges [see panel (a)].
    Panel (b): the partial graphs oriented in order to extract the stationary probability of state 1.
    Panel (c): There are three cycle fluxes, $\mathcal{F}$ (upper three figures), $\mathcal{B}$ (bottom three), and $\mathcal{D}$ (central figure).
    From the figure we extract the values of $\Sigma_\nu$ in Eq.~\ref{Eq:Js} for the three cycles. 
    These are given by the product of the rates flowing towards the cycle [see T. L. Hill~\cite{hill2005free} for details]: $\Sigma_\mathcal{F} = w_{32}w_{45} + w_{43}w_{32} + w_{34}w_{45}$; $\Sigma_\mathcal{B} = w_{12}w_{65} + w_{61}w_{12} + w_{16}w_{65}$; $\Sigma_\mathcal{D} = 1$. }
    \label{Fig:6StateModel}
\end{figure}

From a mathematical standpoint, given a kinetic network described by the master equation, the stationary probabilities $p_i^{ss}$ (or $p_i^{eq}$) can be obtained using graph-theoretical arguments.
The details may be found in the works of Hill~\cite{hill2005free,Hill1966JTB,Hill1975PNAS}.
Briefly, one may construct the set of partial diagrams such that all the states are visited but no cycles are formed (see Fig.~\ref{Fig:6StateModel}b).
The stationary probability of being in state $i$ is proportional to the sum of all the partial graphs oriented in such a way that the fluxes converge towards state $i$, or $\sigma_i$.
The normalization factor is $\Sigma = \sum_i \sigma_i$, so that the stationary probability is,
\begin{equation}
p^{ss}_i = \frac{\sigma_i}{\Sigma}.
\label{Eq:Ps}
\end{equation}

The stationary flux along one edge may be written as a sum of contributions from all the cycle fluxes of the system.
Each cycle can be completed in two directions: one, counterclockwise, labeled as ``+''; the other, clockwise, termed ``-''.
The cycle fluxes in the ``+'' and ``-'' directions are given by the following relationships~\cite{hill2005free},
\begin{equation}
J_{\nu^\pm} = \frac{\Sigma_{\nu} \Pi_{\nu^\pm}}{\Sigma},
\label{Eq:Js}
\end{equation}
where $\Sigma_{\nu}$ is a combination of rates that are specific for cycle $\nu$, $\Pi_{\nu^\pm}$ is the products of the rates of the cycle performed in the ``+'' ($\Pi_{\nu^+}$) or ``-'' ($\Pi_{\nu^-}$) direction, and the denominator $\Sigma$ is defined above (see Fig.~\ref{Fig:6StateModel}c).
It follows that,
\begin{equation}
\frac{J_{\nu^+}}{J_{\nu^-}} = \frac{\Pi_{\nu^+}}{\Pi_{\nu^-}} = e^{\beta A_{\nu}},
\label{Eq:DefAffinity}
\end{equation}
where $A_{\nu}$ (or $\beta A_\nu$) has been termed the action functional~\cite{Lebowitz:1999} or affinity~\cite{Schnakenberg1976RMP} of the cycle $\nu$.
Note that the net direction of completion of cycle $\nu$ is given by the sign of $\Delta J_{\nu} = J_{\nu^+} - J_{\nu^-}$.
It is easy to show that~\cite{Hill1976PNAS},
\begin{equation}
\Delta J_{\nu} A_{\nu} \ge 0.
\label{Eq:SignFluxAffinity}
\end{equation}
The above equality holds only when $J_{\nu,+} = J_{\nu,-}$.
In other words, $\Delta J_\nu$ and $A_\nu$ have the same sign, which means that the value of the affinity dictates the direction of the cycle~\cite{Hill1976PNAS}. 

In order to identify the physical interpretation of these mathematical identities we need to connect the rates with thermodynamic quantities, such as energy, entropy, and the chemical potential.
Following three different approaches we show that the affinity (Eq.~\ref{Eq:DefAffinity}) is related to the entropy produced during a cycle.
The first two strategies involve the Shannon entropy~\cite{Schnakenberg1976RMP,Liepelt2007EPL,Lipowsky2008JSP} and fluctuation theorems~\cite{Crooks1998JSP,Seifert2005EPL,Seifert2005PRL,Seifert2012RPP}, and they will be discussed without specifically referring to the experimental hallmarks for motor velocity described in the previous sections.
The last method, which is based on establishing a connection between pseudo-first-order rate constants and the chemical potential~\cite{hill2005free}, will be developed in the context of molecular motors.\\

\subsection{Thermodynamics}
Molecular machines are enzymes ($E$) that catalyze the chemical transformation of substrate molecules ($S\rightarrow P$), which in general can be described by the Michaelis-Menten kinetics ($ E+S\rightleftharpoons ES\rightarrow E+P $).
Of course, the process is reversible, and the enzyme also catalyzes the reverse reaction -- the transformation of $P$ into $S$.
In the case of molecular machines, the substrate is ATP, and the products of ATP hydrolysis are ADP and orthophosphate ($\Pin$).
(Some molecular machines catalyze the hydrolysis of GTP; mutatis mutandis our considerations do not change.)

We imagine the following experimental setup: a molecular machine operates in a solution containing ATP, ADP, and $\Pin$. 
The chemical potentials of these three species are given by~\cite{hill2004free},
\begin{equation}
\begin{aligned}
\mu_\mathrm{ATP} &= \mu^0_\mathrm{ATP} + \kBT \ln \cATP  \\
\mu_\mathrm{ADP} &= \mu^0_\mathrm{ADP} + \kBT \ln \cADP \\
\mu_\mathrm{P_i} &= \mu^0_\mathrm{\Pin} + \kBT \ln \cPin \\
X &= -\Delta \mu_\mathrm{hyd} = \mu_\mathrm{ATP} - \mu_\mathrm{ADP} - \mu_\mathrm{P_i} =\\
&= \kBT \ln\frac{\Keq \cATP}{\cADP\cPin},
\end{aligned}
\label{Eq:ChemPot}
\end{equation}
where $\mu_\mathrm{c}$ ($\mu^0_\mathrm{c}$) is the (standard) chemical potential of species $c$, $[c]$ is the  concentration in solution, and $\Keq\approx 4.9\cdot10^5 \:\Molar$~\cite{howard2001book} is the equilibrium constant for ATP hydrolysis.
(Note that we should use activities $a_c$ in Eq.~\ref{Eq:ChemPot} instead of concentrations $[c]$; throughout the review we will assume that $a_c \approx [c]$.)
At equilibrium, $(\cADP \cPin/\cATP)_\text{Eq} = \Keq$, and $\Delta \mu_\text{hyd} = 0$.
On the other hand, if $\cATP \mathrm{K_{eq}} \gg \cADP\cPin$, then $\Delta \mu_\text{hyd} < 0$, and the hydrolysis of ATP is a spontaneous reaction.
We exclude un-catalyzed hydrolysis/synthesis of ATP, as it occurs over time-scales beyond our interest~\cite{Hulett1970Nature}; as a consequence, in the absence of molecular machines the concentrations $\cATP$, $\cADP$, and $\cPin$ are constant.
The presence of the molecular machine does not alter the equilibrium features of ATP hydrolysis (namely, $\mathrm{K_{eq}}$), however it accelerates the rate of ATP synthesis/hydrolysis. 
Let the initial concentrations of ATP, ADP, and $\Pin$ make ATP hydrolysis a spontaneous reaction.
After each catalytic cycle the enzyme attains the same conformation that it had at the start. However, the solution conditions have changed as a substrate (ATP) has been consumed and products (ADP and $\Pin$) have been created.
Thus, in the presence of the molecular machine the solution approaches the equilibrium ratio of concentrations of ATP, ADP, and $\Pin$, and monitoring the dynamics under these conditions corresponds to studying the time-dependent relaxation towards equilibrium.
At equilibrium, the rate of hydrolyzing and synthesizing ATP is the same.

However, this is not what happens in a cell, where the concentration of ATP is maintained under homeostatic control~\cite{Wang2017Elife} far away from equilibrium; typical concentrations are $\cATP\approx 1\mMolar$, $\cADP\approx 10\uMolar$, $\cPin \approx 1\mMolar$~\cite{howard2001book}, resulting in a chemical potential $X=-\Delta \mu_\text{hyd}$ of $\approx 25$ ~\cite{howard2001book}, which makes ATP hydrolysis a spontaneous reaction ($\Delta \mu_\mathrm{hyd} < 0$).
This means that the enzymatic cycle of the molecular machine will be driven in the direction that consumes ATP, and ATP hydrolysis provides the driving force that enables the molecular machine to perform work.
Without accounting explicitly for the whole cellular apparatus involved in dictating and maintaining the set-point ATP level, in the simplest theoretical model the system is assumed to be in contact with some devices referred to as ``chemostats'' capable of maintaining the initial concentrations of ATP, ADP, and $\Pin$~\cite{Qian2005_BiophyChem,Lipowsky2008JSP,Lipowsky2009JSP,Seifert2011EPJE}.
After each cycle the chemostat removes the products from solution and replenishes the substrates, thereby ensuring that after each enzymatic cycle the initial condition is reset,  which enables the creation of a NESS for $t\rightarrow \infty$.

The system is also in contact with a thermal reservoir with which it exchanges heat~\cite{Lipowsky2008JSP,Lipowsky2009JSP,Seifert2011EPJE}.
We assume that the thermal reservoir operates ``quasi-statically'' on the time-scale of the heat exchange, and the combination of the system and the thermal reservoir is energetically isolated.
Because after a cycle the motor and the solution have not changed, the entropy produced is equal to the ratio between the heat absorbed by the thermal reservoir, $Q$, and the constant temperature $T$.
We exclude $pV$ work from our formulation, but we include the possibility that the molecular machine performs work against a fixed load, so that for a forward (backward) step $W_\text{mech} = f d_0$ ($W_\text{mech} = - f d_0$), where the choice of the sign implies that a positive force opposes forward movement.
Note that the load $f$ is assumed to be clamped, so at every step, regardless of the position of the motor, the cycle is repeated under identical conditions.

The energy $X=-\Delta \mu_\text{hyd} > 0$ consumed during a cycle is partitioned into work performed ($-W_\text{mech}$) and heat released ($-Q$), that is,
\begin{equation}
\Delta \mu_\text{hyd} + Q + W_\text{mech} = 0, 
\end{equation}
implying that  the total change of entropy of the system plus environment over one cycle is given by,
\begin{equation}
\Delta S = \frac{Q}{T} = \frac{-\Delta \mu_\text{hyd} - f d_0}{T} \ge 0.
\label{Eq:DeltaS}
\end{equation}
(For backward steps, the sign in front of $f d_0$ is the opposite).  The inequality is due to the second law of thermodynamics, which states that we should to find an increase in entropy during the cycle because the combination of system and environment is isolated. 
Eq.~\eqref{Eq:DeltaS} leads to the observation that the maximum amount of force that the motor can resist is given by,
\begin{equation}
    \fmax = \frac{-\Delta \mu_\text{hyd}}{d_0}.
    \label{Eq:fmax}
\end{equation}
 A number of authors have discussed variations of this thermodynamic framework.
In particular, Seifert elucidated the contribution of the thermodynamic contribution of chemostats~\cite{Seifert2011EPJE}; Qian and Beard discussed a network of reactions in which some metabolites are clamped, while other are introduced at a constant rate~\cite{Qian2005_BiophyChem}.
\\


\subsection{Rate of Entropy Production}
Let the entropy be,
\begin{equation}
S = - \kB \sum_i p_i \ln p_i,
\end{equation}
where the sum extends over all the available conformations of the system.
Let us take a time derivative of $S$, and after imposing the conservation of probability ($\sum_i dp_i/dt = 0$) condition, and plugging in the master equation we get,
\begin{equation}
\frac{dS}{dt} = \kB \sum_i \sum_j (p_i w_{ij} - p_j w_{ji}) \ln p_i.
\end{equation}
Note that in a NESS, $\sum_j (p_i w_{ij} - p_j w_{ji}) = \sum_j J_{ij} = 0$, and as a consequence $dS/dt = 0$, as expected.
We first symmetrize the result with respect to the indexes $i$ and $j$, and then add and subtract $1/2\kB\sum_i\sum_j (p_i w_{ij} - p_jw_{ji}) \ln(w_{ij}/w_{ji})$, which leads to~\cite{Schnakenberg1976RMP,Lipowsky2008JSP},
\begin{equation}
\begin{aligned}
\frac{dS}{dt} = \kB \frac{1}{2} \sum_i \sum_j J_{ij} \ln \frac{p_i w_{ij}}{p_j w_{ji}} - \kB \frac{1}{2} \sum_i\sum_j J_{ij} \ln \frac{w_{ij}}{w_{ji}}.
\end{aligned}
\label{Eq:dSdt}
\end{equation}
The first of the two terms on the r.h.s. of Eq.~\ref{Eq:dSdt} is always non-negative, and it has been identified as the rate of entropy production, $d_i S/dt$~\cite{Schnakenberg1976RMP,Lipowsky2008JSP}.
We note parenthetically that some authors~\cite{Hill1976PNAS} do not include the factor $1/2$ as they interpret the summation to be carried out over the edges of the network, and not over the states.
Because in a NESS $dS/dt = 0$, it follows that the second term of the r.h.s of Eq.~\ref{Eq:dSdt} is the rate of entropy out-flux, and,
\begin{equation}
\begin{aligned}
\frac{d_i S}{dt} &= \kB \frac{1}{2} \sum_i \sum_j J_{ij} \ln \frac{p_i w_{ij}}{p_j w_{ji}}  =\\
&=\kB \frac{1}{2} \sum_i\sum_j J_{ij} \ln \frac{w_{ij}}{w_{ji}} \ge 0.
\end{aligned}
\end{equation}

Consider now a cyclic network of $N$ states; the only transitions allowed from state $i$ are $i \rightleftharpoons i+1$ and $i \rightleftharpoons i-1$.
(The cyclic nature of the network implies periodic boundary conditions, and so state $i=0$ is the same as state $i=N$).
In a NESS, the condition $\sum_j J_{ij} = 0$ becomes $\Delta J_{i,i+1} = \Delta J_{i-1,i}$, so that the net flux across all the edges is the same and equal to the net cycle flux, $\Delta J = J_+ - J_-$.
It follows that,
\begin{equation}
\frac{d_i S}{dt} = k_B \Delta J \ln \prod_{i=0}^{N-1}\frac{w_{i,i+1}}{w_{i+1,i}} = \frac{\Delta J A}{T}
\label{Eq:EntropyProd1Cycle}
\end{equation}
where $w_{i,i+1}$ and $w_{i+1,i}$ are the forward and backward rates, respectively.
The term $A = \kBT \ln \prod_{i=0}^{N-1}\frac{w_{i,i+1}}{w_{i+1,i}}$ is the affinity (see Eq.~\ref{Eq:DefAffinity}), which  Hill refers to as a thermodynamic force driving the system out of equilibrium and imposing a NESS~\cite{hill2005free}.
Liepelt and Lipowky identified the thermodynamic force with the entropy production~\cite{Liepelt2007EPL}; if the rate of completing a cycle is $\Delta J >0$ (which implies that the cycle is preferentially completed in the forward, or ``+'', direction), one can identify the entropy  $\Delta S$ produced over one cycle as,
\begin{equation}
T \Delta S = \frac{T}{\Delta J} \frac{d_i S}{dt} = k_BT  \ln \prod_{i=0}^{N-1}\frac{w_{i,i+1}}{w_{i+1,i}} = A \ge 0.
\end{equation}
The argument can be extended to the case in which the kinetic network is characterized by multiple cycles~\cite{Liepelt2007EPL}; if $\Delta J_{\nu}^{-1}$ is the average time for completing a cycle $\nu$, and $\Delta S_{\nu}$ is the entropy produced with that cycle, we can write,
\begin{equation}
\frac{d_i S}{dt} = \sum_{\nu} J_{\nu} \Delta S_{\nu} \ge 0.
\end{equation}
Equation~\ref{Eq:SignFluxAffinity} indicates that each term in the summation is non-negative~\cite{Hill1976PNAS}. 

To summarize, in this section we showed how the action functional (or affinity) can be identified with the entropy produced.
We now present a different argument based upon fluctuation theorems leading to the same conclusions.\\

\subsection{Fluctuation Theorems}
\begin{figure}
    \includegraphics[width=0.5\textwidth]{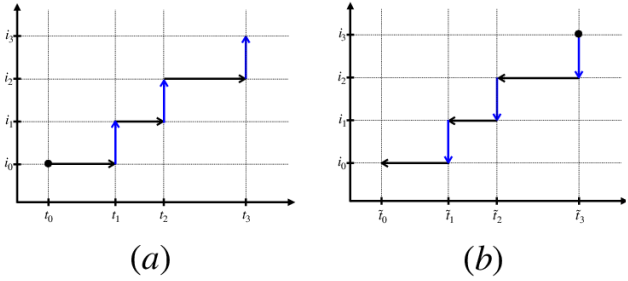}
    \caption{ Path in state space executed forward (a) and backward (b) in time.
    The starting point is the dot, waiting times at a fixed state are in black, transitions are in blue.}
    \label{Fig:PathFwdBwd}
\end{figure}

Let the probability density of taking a forward path of $n$ steps be
$\mathcal{P}_F(i_0,t_0;i_1,t_1;\cdots;i_n,\tau)$,
in which the system is in state $i_0$ at $t_0$, transitions to state $i_1$ at $t_1$ and so on until it reaches state $i_n$ at time $\tau$ (see Fig.~\ref{Fig:PathFwdBwd}a).
We use the subscript ``$F$'' to indicate that the transitions are completed forward in time.
Using the Markov property, we can rewrite this joint probability as~\cite{Crooks1998JSP},
\begin{equation}
\begin{aligned}
    &\mathcal{P}_F(i_0,t_0;i_1,t_1;\cdots;i_n,\tau) = \\
    &= p_{i_0}(t_0)P(i_1,t_1|i_0,t_0)\cdots P(i_n,\tau|i_{n-1},t_{n-1}).
\end{aligned}
\end{equation}
Assuming that the system is in a steady state at $t=t_0$, we replace $p_{i_0} =  p^{ss}_{i_0}$. 
The conditional probabilities are given by $P(j,t_j|i,t_i) = w_{ij}e^{-W_i(t_j-t_i)}$,
where $W_i = \sum_j w_{ij}$.
The conditional probability $P(j,t_j|i,t_i)$ is the product of two probabilities.  The first is $w_{ij} / W_i$, the probability of making a transition to state $j$ among all the other possibilities when starting in state $i$.  The second is $W_i e^{-W_i(t_j-t_i)}$, the probability that this transition happens after time $(t_j-t_i)$.  The reason for the $W_i$ in the exponent is that the mean waiting time to transition from state $i$ to any other state is $W_i^{-1}$, independent of state $j$.  This independence is a well-known property of escape times in Markov networks, which may be computed  through well-known methods~\cite{vankampen}.  Therefore, the joint probability density becomes~\cite{Seifert2012RPP},
\begin{equation}
\begin{aligned}
    &\mathcal{P}_F(i_0,t_0;i_1,t_1;\cdots;i_n,\tau) = \\
    &=\mathcal{N}^{-1} p_{i_0}^{ss}
    \prod_{\alpha=0}^{n-1} w_{i_{\alpha},i_{\alpha+1}}e^{-W_{i_{\alpha}} (t_{\alpha+1}-t_{\alpha})},
\end{aligned}
\end{equation}
where $t_{n} = \tau$ and $\mathcal{N}(t_0,\tau)$ is a normalization factor to ensure that $ \sum_{i_0,i_1,\cdot,i_n}\int_{t_0}^{\tau}dt_1\int_{t_1}^{\tau}dt_2\cdots\int_{t_{n-1}}^\tau dt_{n} \mathcal{P}_F =1$.
The probability of completing the time-reversed path (Fig.~\ref{Fig:PathFwdBwd}b) is,
\begin{equation}
    \begin{aligned}
    &\mathcal{P}_B(i_n,\tilde{\tau},i_{n-1},\tilde{t}_{n-1};\cdots;i_0,\tilde{t}_0) = \\
    &= \tilde{\mathcal{N}}^{-1}p_{i_n}^{ss}
    \prod_{\alpha=0}^{N-1}w_{i_{\alpha+1},i_{\alpha}}e^{-W_{i_\alpha}(\tilde{t}_\alpha-\tilde{t}_{\alpha+1})},
    \end{aligned}
\end{equation}
in which we started in state $i_n$ at time $\tilde{t}_n = \tau - t_n$ and retrace the forward path until we reach state $i_0$ at time $\tilde{t}_0 = \tau - t_0$.
Note that the initial probability is now $p^{ss}_{i_n}$ and because each path could be performed forward or backward in time, the normalization factor $\mathcal{N} = \tilde{\mathcal{N}}$.
Following Seifert~\cite{Seifert2005EPL}, we define $R$ to be the logarithm of the ratio between the probability for forward and backward paths, and for the sake of simplicity we consider only cyclical pathways, in which $i_0 = i_n$.
Therefore,
\begin{equation}
    R = \ln\frac{\mathcal{P}_F}{\mathcal{P}_B} = 
    \ln\prod_{\alpha=0}^{n-1} \frac{w_{i_\alpha,i_{\alpha+1}}}{w_{i_{\alpha+1},i_{\alpha}}}.
\end{equation}
Note that the time-dependent terms in the joint probabilities cancel out, so we neglect them in the following.
(In alternative, we may consider the time at which these jumps occur to be fixed.)
Here, $R$ has a structure analogous to that of the affinities introduced in the previous sections.
More precisely, given a set of states visited we can rewrite,
\begin{equation}
R = \sum_{\nu} n_{\nu}\beta A_{\nu},
\label{Eq:RtoAffinity}
\end{equation}
where $n_{\nu}$ is the net number of times the cycle $\nu$ has been completed during the $n$-step path associated with $R$.
In order to justify this expression, note that every time that during the path a cycle is not completed, the path re-traces it self and those branch-like excursions do not contribute to $R$.
If we average over all paths that complete a cycle of length $n$, we find the following sequence of identities~\cite{Seifert2005EPL},
\begin{equation}
    \langle e^{- R} \rangle = \sum_\text{paths} P_F e^{-R} = \sum_\text{paths} P_B = 1.
\end{equation}
Using Jensen's inequality, $\langle e^{- R} \rangle \ge e^{-\langle  R \rangle}$, leads to,
\begin{equation}
    \langle R \rangle \ge 0.
    \label{Eq:R_ineq}
\end{equation}
From Eq.~\ref{Eq:RtoAffinity} we obtain,
\begin{equation}
\langle R \rangle = \sum_{\nu} \langle n_{\nu} \rangle \beta A_\nu \ge 0.
\label{Eq:Rave}
\end{equation}

We now introduce an Arrhenius-type relationship between the rates, $w_{ij}/w_{ji} = e^{-\beta \Delta F_{ij}}$, where $\Delta F_{ij} = F_{j} - F_{i}$ is the free energy difference between state $i$ and state $j$.
The free energy difference accounts for three contributions: (i) the intrinsic free energy of a state may change; (ii) the motor may bind/release ATP, ADP, and $\Pin$; (iii) the motor may perform work against an external load.
After a cycle, the motor returns to the initial state, and as a consequence the intrinsic free energy does not carry any contribution to the cycle.
The hydrolysis of ATP contributed with the release of energy equal to $-\Delta \mu_\text{hyd}$, and the work performed per each displacement of size $d_0$ against a load $f$ is equal to $-fd_0$.
It follows that~\cite{Lipowsky2008JSP},
\begin{equation}
    \prod_{|i,j\rangle \in \nu} \frac{k_{ij}}{k_{ji}} = e^{-n_{\nu,\mathrm{ATP}}\beta \Delta \mu_h - \beta f l_{\nu}} = e^{\frac{\Delta S_\nu}{\kB}},
\end{equation}
where, following the notation of Lipowsky and Liepelt~\cite{Lipowsky2008JSP}, $|i,j\rangle$ corresponds to a directed edge from state $i$ to $j$, and the sum is extended over all the directed edges belonging to cycle $\nu$.
Here, $n_{\nu, \mathrm{ATP}}$ is the net number of ATP hydrolyzed during cycle $\nu$, and $l_{\nu}$ is the net displacement along the track (a multiple of $d_0$).
The last identity follows from Eq.~\ref{Eq:DeltaS}, and from Eq.~\ref{Eq:Rave} gives,
\begin{equation}
\kB \langle R \rangle = \sum_\nu \langle n_\nu \rangle \Delta S_\nu \ge 0.
\end{equation}

\section{Molecular Motors -- Models without Detachment}
In this section we develop models of molecular motors with increasing complexity, from one-state models, to uni-cyclic models, ending with multi-cycle kinetic networks.
We show that experimental evidence and thermodynamic insights suggest that the introduction of multiple cycles creates models that are physically more sensible.
We discuss more in detail the efficiency of the motors, and we conclude the section by describing models that account for the detachment of the molecular motor from the track.\\

\subsection{Molecular Motors are Processive Enzymes}
The catalytic cycle of kinesin, dynein, and myosin proceeds via multiple steps, in which ATP binding, hydrolysis, and release of ADP and P$_i$ lead to substantial changes in the motor conformation. 
The presence of the cytoskeletal filament, CF (F-actin in the case of myosin, microtubules for dynein and kinesin), increases the rate of ATP hydrolysis~\cite{Hackney1988PNAS,DeLaCruz1999PNAS,DeLaCruz2001JBC,Homma2005JBC,Ori2010NCB}.
When the structural changes are properly rectified under the specially designed interactions with the appropriate CFs, the conformational fluctuations in M are transduced to a predominantly uni-directional motion along the track, and generates mechanical forces against an external load.
The free energy source necessary to perform this movement (or work) is provided by the hydrolysis of ATP.

Enzymes that work in conjunction with substrates (as is the case for Ms and CFs) are said to be ``processive'' if they perform multiple catalytic cycles without fully disengaging from the CFs~\cite{Schnitzer1995ColdSpringHarbor}.
Processive motors move long distances along the CFs by hydrolyzing one ATP molecule per step without  dissociating from the CF.
In general (although with some fascinating exceptions~\cite{Inoue2002NCB,Post2002JBC}) processive movement requires the cooperation of multiple enzymes: some members of the ensemble proceed forward while the others hold tight onto the CF. 
From here on, we focus our discussion exclusively on dimeric processive motors made of two identical enzymes of the same family, which we refer to as ``heads.'' 
With this we may identify the leading head (LH) as the one that is in front of the dimeric complex, while the other is the trailing head (TH). We reserve the word ``motor'' for the motile construct, which is a dimer in the case of myosin V~\cite{Mehta99Nature}, VI~\cite{Rock01PNAS}, and X~\cite{Sun2010NSMB}, conventional kinesin (kinesin-1)~\cite{howard1989Nature}, and cytoplasmic dynein~\cite{Reck2006Cell}.

In the well-accepted hand-over-hand stepping mechanism~\cite{Yildiz2003Science,Yildiz2004Science,Yildiz2004JBC,Okten2004NSMB,Toba2006PNAS,Sun2010NSMB}, the TH of the motor detaches from the filament, overcomes the bound head, and advances the motor by reaching a forward target binding site (TBS).
After the step is completed, the role of the two heads is reversed, and the process is repeated identically, until the motor detaches from the CF.
A few key ingredients are necessary for processive, hand-over-hand movement to occur.
First, in order to prevent premature end of the processive run, the head that remains bound to the track during a step should not dissociate from the CF at least until the detached head reaches the TBS.
Second, the two heads should go through their catalytic cycles  out-of-phase~\cite{Block2007BJ}, so that the TH is more likely to detach from CF than the LH. 
This requirement necessarily implies that the two heads ought to communicate with one another through action at a distance, which may be viewed as a complex form of allostery~\cite{Thirumalai19ChemRev}.
Third, during a step a combination of conformational transitions in the two heads biases the diffusive motion of the free head towards the TBS unless a large resistive force is applied.

Meeting the first two conditions requires specific features of the hydrolytic cycles that the two heads must undergo.
For instance, if the rate-limiting step occurs in a CF-bound conformation, the chances of prematurely ending the processive run are diminished.
In addition, the interaction between the two heads is believed to be key in ensuring that the TH steps first, and that the LH is unlikely to detach during a step.
This is possible if the interplay between the two heads slows down (or ``gates'') some of the steps of the LH catalytic cycle, or accelerates them in the trailing head (also referred to as ``gating'')~\cite{Block2007BJ,Sweeney2010ARB,Hancock2016BJ}.
The structural bias towards the TBS is provided by a conformational transition of the CF-bound motor that projects forward the stepping head. 
For the three classes of motors mentioned before, the lever arm swing in myosin, the neck-linker docking in kinesin, and the bent$\rightarrow$straight transition of the linker domain in dynein provide the requisite forward bias. It is remarkable that these structural transitions, induced by ATP binding and hydrolysis, which occur on relatively small length scales are amplified by responses in other structural elements.
The terms ``power stroke''~\cite{Dominguez1998Cell} and ``Brownian ratchet''~\cite{Rice1999Nature} have been used to schematize this forward bias, depending on whether the emphasis is placed on the mechanistic (order$\rightarrow$order) or stochastic (disorder$\rightarrow$order) nature of the conformational transitions.

In the simplest cases, the step-size of a motor is commensurate with the filament repeat -- for motors moving along the MT, $d_0 \approx 8.2\,\nm$, whereas $d_0 \approx 36\,\nm$ for actin-based motors (myosins).
However, not all motors walk precisely by following this rule: myosin VI and dynein, for instance, display a broad step-size distribution, in which hand-over-hand steps are intertwined with inchworm-like movements and frequent backward steps~\cite{Reck06Cell,nishikawa2010switch}. Of course, the variability in the step sizes is also a consequence of the architecture of the motor. Thus, both the structural design of the motor and the coupling to catalytic cycles determine not only the stepping kinetics but also the precision of the step sizes.\\

\subsection{Processive Motor Velocity}
A variety of single-molecule techniques are able to inform on the motile properties of molecular motors. 
The motor or the track may be labeled with a fluorescent~\cite{Yildiz2003Science,nishikawa2010switch} or refractive~\cite{mickolajczyk2015kinetics,Isojima2016NCB} probe.  
Monitoring the time-dependent changes in the location of the such probes enables the determination of the velocity of the motor as a function of nucleotide concentration. Alternatively, optical trapping techniques may be used to follow the displacement of the motor or the filament, and to investigate the motor response to the external load~\cite{howard1989Nature,Block90Nature,Finer1994Nature}. 
Much of our current knowledge about kinesin comes from single molecule experiments \cite{Block90Nature,howard1989Nature}, which were developed approximately at the time of the discovery of kinesin \cite{Brady85Nature,Vale85Cell}.  
It has been shown that each step of kinesin motors along MTs is tightly coupled with hydrolysis of one ATP molecule, that is, a single event of ATP hydrolysis by kinesin leads to $\sim$8 nm step along MTs \cite{Schnitzer97Nature}. 
Thus, within the Michaelis Menten (MM) scheme, the velocity of kinesin is associated with enzyme turnover rate is, 
    \begin{equation}
    V=d_0\frac{k_{\text{cat}}[\text{ATP}]}{K_M+[\text{ATP}]},
    \label{eqn:MM}
    \end{equation}
    where $d_0=8$ nm is the step size of the kinesin, $k_{\text{cat}}$ and $K_M$ are the rate of catalysis and the Michaelis constant, respectively. 
    For kinesin-1, the typical experimental value for the Michaelis
constant is $K_M\gtrsim  50$ $\mu$M, and for $[\text{ATP}] \gtrsim 1$ mM the velocity of kinesin is saturated to its maximum value $V_{\text{max}} = d_0k_{\text{cat}}\approx 8$ nm/10 ms $=0.8$ $\mu$m/s \cite{Visscher99Nature,kolomeisky07arpc}. 
Similar single-molecule studies have been conducted on other motors, showing that the Michealis-Menten scheme provides a good paradigm to rationalize the ATP concentration dependence of velocity.
However, the parameters of the fit depend on the motor: for myosin V it was found that $V_\mathrm{max} \approx 0.55 \, \mu$m/s, with $K_M\approx 163\,\mu$ M~\cite{baker2004myosin}, whereas the slower myosin VI has $V_\mathrm{max} \approx 0.16 \, \mu$m/s and $K_M\approx 274 \, \mu$M [fit of Eq.~\ref{eqn:MM} to data from~\citet{elting2011detailed}].
 
Experiments employing optical tweezers have shown that external loads affect the
ATP chemistry in the catalytic site as well as the motility of the motor \cite{Visscher99Nature,block03pnas,Veigel2005NCB,oguchi2008load}. 
Thus, the resulting force-velocity-ATP relationship has been a landmark measurement, which is used as a constraint to decipher the mechanism underlying kinesin motility, and to construct the appropriate SKMs.
Incorporation of the effect of load into MM model was suggested by making the parameters $k_{\text{cat}}$ and $K_M$ force-dependent \cite{Schnitzer00NCB},
\begin{equation}
V(F,[\text{ATP}])=\frac{d_0k_{\text{cat}}(F)[\text{ATP}]}{k_{\text{cat}}(F)/k_b(F)+[\text{ATP}]}
\label{eqn:Block_fit}
\end{equation} 
where 
$k_{\text{cat}}(F)=k_{\text{cat}}^o/(p_{\text{cat}}+(1-p_{\text{cat}})e^{F\delta_{\text{cat}}/k_BT})$ and $k_b(F)=k_b^o/(p_b+(1-p_b)e^{F\delta_b/k_BT})$. 
The fit of the $F$-dependent velocity data of motor using Eq.\ref{eqn:Block_fit} is reasonable as long as the magnitude of load ($F$) is small \cite{Schnitzer00NCB}. 
However, the conventional MM model 
has an intrinsic drawback when the external load approaches the stall force value ($\approx 7$ pN for kinesin) and becomes greater than the stall force. 
An increase of load ought to induce backward stepping ($V <0$); yet no modification of Eq.\ref{eqn:MM} produces a negative velocity!
One possible fix for the failure of the MM model at large load is to permit in the model a reverse reaction current, which is realized by rendering every elementary reaction step within the kinesin cycle ``reversible." 
We discuss later how such models have been constructed in the context of Markov jump processes. \\

\subsection{Periodic Lattice Model}
In a molecular motor (i) binding, release, and chemical transformations of ATP, ADP and P$_i$ from the motor head domain, and (ii) the advancement along the track occur over much slower times (ms - s) compared to the fluctuations of molecular conformations (ns - $\mu$s). 
The time-scale separation enables the construction of a Markov jump processes in which the state of the system is defined by two coordinates: (i) one of the $N$ intermediates defining the chemical state of the substrate, and (ii) the location $l$ along the track. 
The probability density of the $i$-th state ($i=1,2,\ldots, N$) at filament site $l$ obeys the master equation, 
\begin{equation}
\frac{dP_i(l,t)}{dt}=\sum_{l'=l-1}^{l+1}\sum_j [w_{j,l\rightarrow i,l'}P_j(l',t)- w_{i,l\rightarrow j,l'}P_i(l,t)]
\end{equation} 
where $\sum_l\sum_iP_i(l,t)=1$, and $w_{j,l\rightarrow i,l'}$ is the rate for going from state $j$ in location $l$ to state $i$ on F site $l'=l,l\pm1$.
Under the assumption that the track is infinite and periodic, that is that the rates do not depend on $l$, one may sum the r.h.s and the l.h.s of the master equation and eliminate $l$.
As a result, we obtain the master equation in Eq.~\ref{Eq:Master}, and we can use the theoretical framework presented before.

Periodic one-dimensional models, which were originally suggested by Derrida \cite{Derrida83JSP} in order to study the mean velocity and diffusion constant of random systems, have been widely adopted in describing the dynamics of molecular motors.      
In particular, Fisher and Kolomeisky \cite{fisher99pnas,fisher01pnas,kolomeisky07arpc} have pioneered this approach and specifically used the reversible kinetic model to describe the motility of kinesin-1 and myosin V.
Numerous other studies have used variants of these models to discuss the thermodynamic features of stochastic motors.
We review the main successes and shortcomings of these models in the context of molecular motors as their complexity is increased. \\

\begin{figure}
\includegraphics[width=0.5\textwidth]{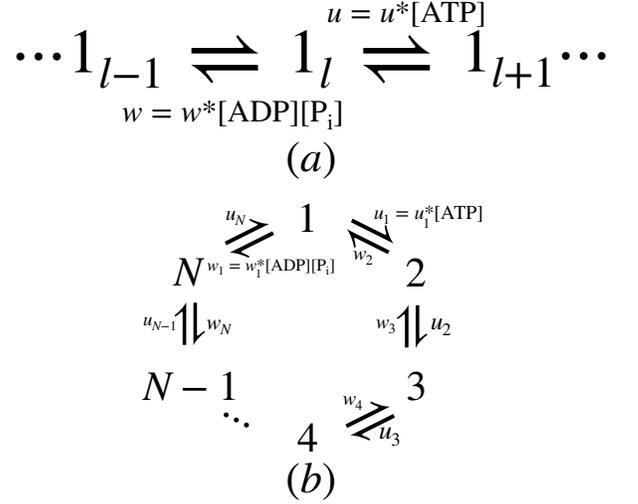}
\caption{Example of simple kinetic networks.
Panel (a): the network has only one statte, $N=1$. 
The forward rate is given by $u=u^*\cATP$, the backward rate is $w=w^*\cADP\cPin$.
Both forward and backward rates are pseudo-first-order, with units $\mMolar^{-1} \s^{-1}$ and $\mMolar^{-2} \s^{-1}$, respectively.
Panel (b): multi-state model.
Here, the rates $u_1$ and $w_1$ are arbitrarily chosen to depend on $\cATP$ and $\cADP\cPin$, respectively.
Panel (c): a cycle corresponding to the multi-state model.
}
\label{Fig:1State-NState1Cycle}
\end{figure}

\subsubsection{One-state Models}
In the simplest possible model, the motor is defined by a single ($N=1$) chemical state at each site along the track (Fig.~\ref{Fig:1State-NState1Cycle}a).
The rate of forward stepping, $u$, depends on the concentration of ATP.
Microscopic reversibility demands that the backward transition is possible as well, which is then nominally associated with ATP synthesis.
The rates of forward and backward stepping are modeled as pseudo-first order processes, 
\begin{equation}
u = u^{*} \cATP ,\quad w = w^{*} \cADP \cPin,
\label{Eq:PseudoFirst}
\end{equation}
leading to $u/w = u^{*}/w^{*} \cdot \cATP/(\cADP\cPin)$.
In equilibrium, the probability of going forward is identical to the probability of backward stepping, which implies that $(u/w)_{\mathrm{Eq}} = 1$.
From Eq.~\ref{Eq:ChemPot}, at equilibrium $\Delta\mu_\mathrm{hyd} = 0$ and $(\cADP\cPin/\cATP)_{\mathrm{Eq}} = \Keq$.
As a consequence, using Eq.~\ref{Eq:ChemPot} we obtain that the ratio between forward and backward rates is given by,
\begin{equation}
\begin{aligned}
&(a) \quad \frac{u}{w} = \frac{u^{*}}{w^{*}} \frac{\cATP}{\cADP \cPin} = e^{\beta X}\\
&(b) \quad \frac{u^{*}}{w^{*}} = \Keq, 
\end{aligned}
\label{Eq:u/w}
\end{equation}
where we recall that $X = - \Delta\mu_\text{hyd}$.
The average, stationary velocity is given by the step-size times the difference between the forward and backward rates,
\begin{equation}
v = d_0 (u -w) = d_0 w \Big(e^{\beta X} - 1\Big).
\label{Eq:v1state}
\end{equation}
The crucial insight from this expression pertains to the interplay between the non-equilibrium steady-state (NESS) homeostatically maintained by the cell and molecular motor function: in the cytosol $X > 0$, thus $v>0$; at equilibrium, $X = \Delta \mu_\mathrm{hyd} = 0$, implying that $v=0$.

In the absence of applied load, the motor moves without performing any work.
In this case, the entire free energy extracted from the hydrolysis of ATP is dissipated into heat, $Q = -\Delta\mu_\mathrm{hyd}$.
From Eq.~\ref{Eq:EntropyProd1Cycle}, the rate of heat dissipation per step is given by,
\begin{equation}
T \frac{d_i S}{dt} = \dot{Q} = \kBT(u-v)\ln\frac{u}{v} = (u-v) (-\Delta \mu_\text{hyd}) \ge 0.
\end{equation}
Clearly, $\dot{Q}\ge 0$, and the equality holds in equilibrium, where $u=v$. 
If instead the motor is subjected to a force, $f$, opposing its movement, at each step the motor performs mechanical work $W = f d_0$.
In this case, the free energy released by ATP hydrolysis is partitioned between work and heat, $Q = -\Delta\mu_\mathrm{hyd} - W$.
In order to account for the presence of applied load, the rates may be modified using the Bell model~\cite{fisher1999PhysA}, 
\begin{equation}
\begin{split}
&(a) \quad u(F) = u^* \cATP e^{-\beta \theta f d_0}, \\
&(b) \quad w(F)= w^* \cADP\cPin e^{\beta (1-\theta) f d_0}. \\
&(c) \quad \frac{u}{w} = e^{\beta (X-fd_0)}
\end{split}
\label{Eq:u/w_F}
\end{equation}
Here, the coefficient $\theta$ indicates how the load is partitioned between the forward and backward steps, and it identifies the position of the transition state along the direction of motion~\cite{fisher01pnas}.
For $\theta = 0$, the forward step is unaffected by force, whereas backward steps are insensitive to load if $\theta=1$.
These two extreme cases have been referred to as ``power stroke" and ``ratchet", respectively~\cite{howard2006protein}.
In a ``power stroke", the chemical step (crossing the transition state) occurs before the movement forward of the motor.
In contrast, in the case of a ``ratchet" the motor fluctuates between the sites $l$ and $l+1$ until it is captured in the forward site by  ATP hydrolysis.
In the presence of backward load the velocity becomes,
\begin{equation}
v = d_0 w(0) e^{\beta(1-\theta)fd_0} \Big(e^{\beta X - \beta fd_0} - 1\Big).
\label{Eq:v1state_F}
\end{equation}
It is clear that the motor stalls under a resistive load of,  
\begin{equation}
\fstall = \frac{X}{d_0} = - \frac{\Delta \mu_\mathrm{hyd}}{d_0},
\label{Eq:Fs}
\end{equation}
at which the mean motor velocity becomes $v=0$.
It is clear that $\fstall = \fmax$ for this model.
Using Eq.~\ref{Eq:Fs} together with the value of $\Delta\mu_\mathrm{hyd}$ in the cell and the average step-size of a motor, one can predict the value of $\fstall$: for kinesin $\fstall \approx 12.5 \pN$, for myosin V $\fstall \approx 2.8 \nm$.
In the case of myosin V, $\fstall$ approximates the value of the stall force measured experimentally; for kinesin, the $\fmax$ is about twice as large as the stall force.
Note also that according to Eq.~\ref{Eq:Fs}, $F_s$ depends on $\cATP$; a few experiments have suggested that this is not the case  for a variety of motors~\cite{Nishiyama02NCB,Uemura04NSMB,Cross05Nature,Toba2006PNAS,Gennerich2007Cell}, although other studies have observed such adependence~\cite{Visscher99Nature}. 
The independence of $F_s$ on $\cATP$ might be reasonable because each head has only one binding pocket for ATP, and hence $\cATP$ should not determine $F_s$.
The relationship $v(f,\Delta\mu)$ allows one also to compute the power output of the motor, defined as $P=vF$, and the motor efficiency, 
\begin{equation}
\eta(f,X) = \frac{fd_0}{X},
\end{equation}
that is the ratio of the work produced over the energy released by the hydrolysis of ATP.
Note that the power is zero when $f=0$ and at the stall force (where $v=0$), thus for some force $0<f^*<F_s$ it reaches a maximum $P^*$. 
The force-velocity curve, power output, efficiency, and efficiency at maximum power ($\eta^*$) have been used to compare ratchet-like motors ($\theta = 1$) with power-stroke-driven ($\theta=0$).
Given a value of the driving force $X$, power strokes result in larger velocity at fixed $F$, and display higher efficiencies and exert more power at a given velocity~\cite{Wagoner16JPCB}.
Furthermore, both $\eta^{*}$ and $P^*$ at fixed $X$ increase as $\theta\rightarrow0$~\cite{Schmiedl08EPL}. 
Finally, from Eq.~\ref{Eq:EntropyProd1Cycle} we write,
\begin{equation}
T\frac{d_i S}{dt} = \dot{Q} = (u-w) \ln \frac{u}{w} = (u-w)(-\Delta\mu_\text{hyd}-fd_0),
\end{equation}
where we used Eq.~\ref{Eq:u/w_F}c and Eq.~\ref{Eq:v1state_F}.
Note that, again, $\dot{Q} \ge 0$, and the equality holds at equilibrium ($u=w$).

One-state models are appealing because of their simplicity; for instance, they have only two fitting parameters with straightforward physical interpretations: $w^*$ sets the time-scale (see Eq.~\ref{Eq:v1state}), and $\theta$ establishes the location of the transition state during a step.
However, these models fail in reproducing the dependence of the velocity on ATP concentration observed in experiments (Eq.~\ref{eqn:MM}).
According to Eq.~\ref{Eq:v1state} $v$ grows without saturating as $\cATP$ increases, which is not physical.
In order to solve this problem, one must use kinetic models with $N>1$. \\

\subsubsection{Multi-state, Uni-cycle Models}
Figure~\ref{Fig:1State-NState1Cycle}b shows a kinetic model for molecular motors with $N>1$ intermediates.
Assuming that the filament is periodic, we can drop the label $l$ and construct an equivalent uni-cycle in Fig.~\ref{Fig:1State-NState1Cycle}c, where the rate $u_N$ is associated with a forward step, and $w_1$ with a backward displacement.
From Eq.~\ref{Eq:Js} we obtain the following relationships,
\begin{equation}
\begin{aligned}
&(a) \quad J^+ = \frac{\prod_{i=1}^N u_i}{\Sigma(\{u,w\})}, \quad J^- = \frac{\prod_{i=1}^N w_i}{\Sigma(\{u,w\})}, \\
&(b) \quad \frac{\prod_{i=1}^N u_i}{\prod_{j=1}^Nw_j} = \frac{J^+}{J^-} = e^{\beta  X}, \\
&(c) \quad \frac{u_1^* \prod_{i>1}^N u_i}{w_1^* \prod_{j>1}^N w_j} = \Keq,\\
\label{Eq:ThermoLinkN}
\end{aligned}
\end{equation}
where $J^+$ and $J^-$ are the fluxes to complete a cycle in the clockwise and counterclockwise direction, respectively, and $\Sigma(\{u,w\})$ is a function of all the rates (see Eq.~\ref{Eq:Ps}).
The connection with thermodynamics is provided by Eq.~\ref{Eq:ThermoLinkN}b, which is equivalent to Eq.~\ref{Eq:u/w}a; Eq.~\ref{Eq:ThermoLinkN}c holds because in equilibrium $X=0$ (see Eq.~\ref{Eq:u/w}b for the $N=1$ case).
For the case of uni-cyclic network models, the net steady state flux ($\Delta J = J^+ - J^-$) along the cycle is obtained by calculating 
$\Delta J=w_{i,i+1}p_i^{ss}-w_{i+1,i}p_{i+1}^{ss}$ at any edge between the two neighboring chemical states along the cycle. 
Therefore, the velocity, which is obtained as the flux across edge $N\stackrel[w_1]{u_N}{\rightleftarrows} 1$, is given by,
\begin{equation}
v = d_0 \Delta J = d_0 J^- (e^{\beta X} - 1) = d_0 \frac{\prod_{i=1}^N w_i}{\Sigma(\{u,w\})} (e^{\beta X} - 1),
\label{Eq:vNstates}
\end{equation}
The similarity between Eq.~\ref{Eq:v1state} and Eq.~\ref{Eq:vNstates} is clear: both of the expressions relate the velocity to the exponential of the free energy released upon the hydrolysis of ATP, and in both cases at equilibrium the motor does not move.
However, the expression in Eq.~\ref{Eq:vNstates} saturates at large $\cATP$, in agreement with experiments (see Eq.~\ref{eqn:MM}).

Compared to the case of a single chemical state, new load distribution factors are necessary for each transition rate, i.e.
\begin{equation}
u_i(f) = u_i(0) e^{-\beta \theta_i^{+} f d}, \quad w_i(f)= w_i(0) e^{\beta \theta_i^{-} f d}.
\end{equation}
The distribution factors obey the relationship $\sum_{i=1}^N (\theta_i^+ +\theta_i^-) = 1$~\cite{fisher1999PhysA}.
It follows that,
\begin{equation}
\frac{\prod_{i=0}^{N-1} u_i(f)}{\prod_{i=0}^{N-1} w_i(f)} = \frac{J^+(f)}{J^-(f)} = e^{\beta X - \beta fd_0},
\label{Eq:u/wN_F}
\end{equation}
and given $\Delta J(f) = J^+(f) - J^-(f)$, the force-velocity relation is,
\begin{equation}
v(f) = d_0 \Delta J(f) = d_0\frac{\prod_{i=1}^N w_i(f)}{\Sigma(f)}\Big(e^{\beta X - \beta fd_0} - 1\Big).
\label{Eq:VvsFN}
\end{equation}
Although $v(F)$ is a complicated function of all the parameters~\cite{fisher1999PhysA,Wagoner16JPCB}, the expression for a theoretical estimation of the stall force is the same as the one found for $N=1$, given by Eq.~\ref{Eq:Fs}, and it is again equal to $\fmax$ in Eq.~\ref{Eq:fmax}.

In principle, one could design kinetic networks with an arbitrary number of intermediates.
However, because the number of free parameters becomes $4N-2$, it is desirable to establish the minimal $N$ that enables an accurate description of the experimental data as accurately as possible.
Let $D=\lim_{t\rightarrow\infty} \frac{d}{dt}(\langle x(t)^2\rangle - \langle x(t) \rangle^2)$ be the dispersion, and $v = \lim_{t\rightarrow\infty} \frac{d}{dt}\langle x \rangle$ is the velocity; the randomness parameter for a quantity $x$ is defined as,
\begin{equation}
r = \frac{2D}{d_0v}.
\end{equation}
It can be shown that $N\ge1/r$~\cite{koza2002general,kolomeisky07arpc}, and therefore $r^{-1}$ sets a lower bound to $N$ that is needed to account for measurements. Note that $r$ is a function of ATP concentration and external load.

Following Eq.~\ref{Eq:EntropyProd1Cycle}, the rate of heat released during a cycle~\cite{hill2005free,Qian2005_BiophyChem,Qian07ARPC,Toyabe2010PRL,Zimmermann_NJP_2012} is, 
\begin{equation}
\dot{Q} = \Delta J \kBT \ln\frac{J^+}{J^-} = \Delta J(-\Delta\mu_\mathrm{hydr} - fd_0) = \dot{E} + \dot{W} \ge 0,
\label{eq:Qtot}
\end{equation}
where $\dot{E}$ is the rate of free energy expended per cycle. 
Note that in uni-cyclic network models, $\dot{W}=-\Delta J(f)Fd_0=0$ either at $f=0$ or if the motor is subject to an opposing stall force ($f=F_s$), where $\Delta J(F_s)=0$; thus the work production ($\dot{W}$) is a non-monotonic function of $f$, whereas $\dot{E}$ and $\dot{Q}$ decrease monotonically with $F$.

Although the conventional N-state unicyclic models~\cite{fisher99pnas,fisher01pnas,kolomeisky2003simple} appear successful in describing the motility and thermodynamics of molecular motors, unicyclic models encounter two serious problems, especially when the molecular motor is stalled or starts taking backward steps at large hindering loads \cite{astumian1996bj,liepelt07prl,hyeon09pccp}. 
First, the backward step in the unicyclic network, by construction, is produced by a reversal of the forward cycle, which  implies that the backward step is always realized via the synthesis of ATP from ADP and P$_i$.  
This is the case for rotary ATP-synthases, which function as transducers of electrochemical potential into the synthesis of ATP in the mitochondria~\cite{AlbertsBook}, and depending on the sign of $\Delta \mu_\mathrm{hyd}$ and on environmental conditions (e.g. the presence and strength of a proton gradient and applied load), these spectacular motors can reverse their function~\cite{Turina2003EMBOJ,Itoh2004Nature}.
Although ATP hydrolysis is reversible for linear molecular motors~\cite{bagshaw1973reversibility,Hackney05PNAS},  back-stepping has not been associated with ATP synthesis. Rather they are associated with ATP-independent ``slippage'' or coupled to fuel consumption just as forward stepping~\cite{Cross05Nature,Gebhardt2006PNAS,clancy2011nsmb,ikezaki2013spontaneous}.
This requires an update of the kinetic network.

In addition, the unicyclic network models lead to $\dot{Q}=0$ under stall conditions (Eq.~\ref{eq:Qtot}), which contradicts physical reality. 
For example, an idling car still burns fuel and dissipates heat ($\dot{Q}\neq 0$)! 
There certainly exist fundamental differences between molecular world and macroscopic counterpart in that the former is subject to a large degree of fluctuations, which permit the reverse process of ATP hydrolysis (i.e., synthesis) or negative heat dissipation for a sub-ensemble of the entire realizations. 
Yet, the mean of entropy production $\dot{Q}/T$ is still bound to be positive as demanded by thermodynamics, and the probability of a local violation of this principle becomes vanishingly small as the system size grows. 
To ameliorate the aforementioned physically problematic interpretation, associated with the backward step mechanism, 
it has been proposed to extend the unicyclic network into a network with multiple cycles \cite{liepelt07prl,yildiz08cell,hyeon09pccp,clancy2011nsmb}, so that the kinetic pathway of ATP-induced (fuel-burning) backward step or ATP-consuming stall can naturally be considered in the  model.\\

\subsubsection{Multi-cycle Models}
\begin{figure}
	\centering
	\includegraphics[width=0.4\textwidth]{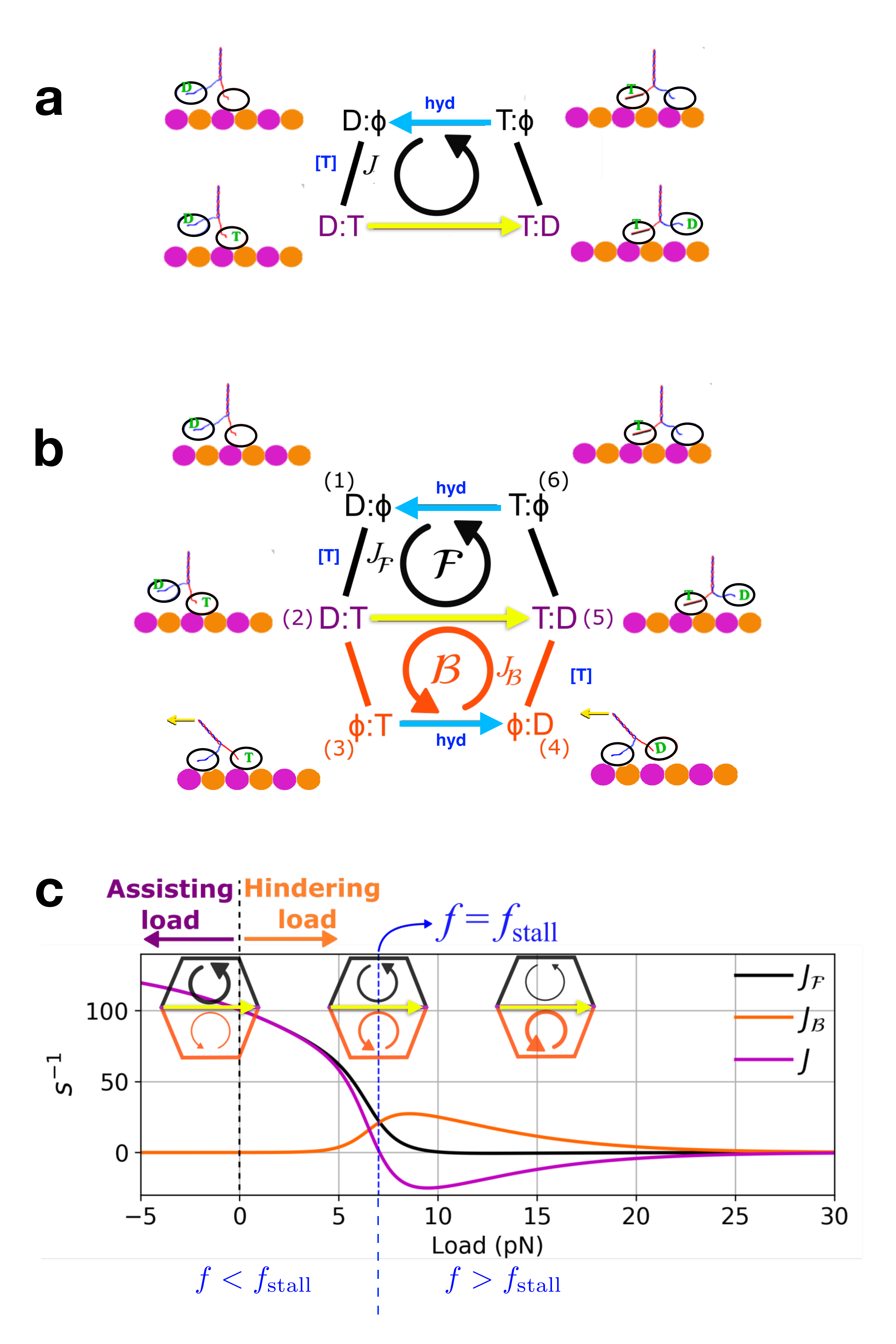}
	\caption{
	{\bf b.} Schematics of the network model representing the dynamics of double-headed kinesin-1. $T$, $D$, and $\phi$ denote ATP-, ADP-bound, and apo state, respectively.
		Through ATP binding [(1)$\rightarrow$(2)], mechanical step [(2)$\rightarrow$(5)], release of ADP [(5)$\rightarrow$(6)], and hydrolysis of ATP [(6)$\rightarrow$(1)], kinesin moves forward in the $\mathcal{F}$-cycle [$(1) \rightarrow (2) \rightarrow (5) \rightarrow (6) \rightarrow (1)$], backward in the $\mathcal{B}$-cycle [$(4) \rightarrow (5) \rightarrow (2) \rightarrow (3) \rightarrow (4)$]. 
		The arrows in the figure depict the direction of major counterclockwise reaction current in each cycle.
		In both cycles, each chemical step is reversible and $k_{ij}$ defines the transition rate from the $i$-th to $j$-th state.
	{\bf c.} Reaction current $J_{\mathcal{F}}$, $J_{\mathcal{B}}$, and $J$ as a function of load. The three cartoons illustrate the amount of current flowing along the $\mathcal{F}$ and $\mathcal{B}$ cycles as a function of $f$. 
		$f<0$ and $f>0$ are the assisting and hindering load, respectively. 
		}
\label{double_cycle}
\end{figure}

The $(N=6)$-state double cycle network in Fig.~\ref{double_cycle}b is discussed here as a minimal kinetic model to account for subtleties in the physics of kinesin under external load.   
In the model, the rate constants at the edges between the adjacent chemical states are given as 
\begin{align}
&k_{25}(f)=k_{25}^oe^{-\theta fd_0/k_BT}\nonumber\\
&k_{52}(f)=k_{52}^oe^{(1-\theta) fd_0/k_BT}\nonumber\\
&k_{ij}(f)=2k_{ij}^o(1+e^{\chi_{ij}fd_0/k_BT})^{-1}\text{ for $ij\neq 25$ or $52$}
\label{eqn:kij_f}
\end{align}
For the sake of simplicity, it is assumed that $(2)\rightleftharpoons(5)$, the step associated with the switching of the trailing and leading head positions (the yellow arrow in Fig.\ref{double_cycle}b), obeys the Bell-like force dependence. When $f>0$, the force exerted on the motor hinders the forward step and enhances the backward step. 
Other steps ($ij\neq 25$ or $52$) are abolished ($k_{ij}\rightarrow 0$) at large $f(>0)$, 
which corresponds to the physical situation where the large external force impedes the chemical steps such as ATP binding, hydrolysis, and ADP release by deforming the conformation of molecular motor.  
The model consists of two  sub-cycles  
$\mathcal{F}$ ($(1)\rightleftharpoons(2)\rightleftharpoons(5)\rightleftharpoons(6)\rightleftharpoons (1)$) and 
$\mathcal{B}$ ($(2)\rightleftharpoons(3)\rightleftharpoons(4)\rightleftharpoons(5)\rightleftharpoons (2)$).  
At small ($f\approx 0$) or assisting force ($f<0$), which renders $k_{25}\gg k_{52}$, 
the reaction current is mainly formed along the counterclockwise direction of $\mathcal{F}$ cycle ($\mathcal{F}^+$); in contrast, at large hindering force ($f>0$) above the stall condition ($f=f_s$), the current flow along the counterclockwise direction of $\mathcal{B}$ cycle ($\mathcal{B}^+$), which corresponds to the ATP hydrolysis-induced backstep.  

Owing to the microscopic reversibility, each cycle can in principle be performed in both counterclockwise ($+$ sign, $\mathcal{F}^+$,$\mathcal{B}^+$) and clockwise ($-$ sign, $\mathcal{F}^-$,$\mathcal{B}^-$) directions.
Forward (backward) steps occur via the $(2)\rightarrow (5)$ ($(5)\rightarrow (2)$) transition; 
the edges $(1)\rightarrow (2)$ ($(2)\rightarrow (1)$) and $(4)\rightarrow (5)$ ($(5)\rightarrow (4)$) are associated with ATP binding (dissociation); 
the edges $(6)\rightarrow (1)$ ($(1)\rightarrow (6)$) and $(3) \rightarrow (4)$ ($(4) \rightarrow (3)$) are associated with ATP hydrolysis (synthesis) (The cyan arrows in $\mathcal{F}$ and $\mathcal{B}$ cycles highlight ATP hydrolysis).
The model in principle accommodates 4 different pathways:
(i) ATP-hydrolysis-induced forward step ($\mathcal{F}^+$);
(ii) ATP-hydrolysis-induced backward step ($\mathcal{B}^+$);
(iii) forward step that synthesizes ATP ($\mathcal{B}^-$);
(iv) backward step that synthesizes ATP ($\mathcal{F}^-$). 
The reaction currents $J_\mathcal{F}$ flowing through $(6)\rightleftharpoons(1)$ and $J_\mathcal{B}$ through $(3)\rightleftarrows(4)$ can be calculated by the generating function technique by Koza \cite{Koza1999JPA,Hwang2017JPCL}, or utilizing the approach based on the large deviation theory \cite{Lebowitz:1999}.
The expressions of $J_{\mathcal{F}}$ and $J_{\mathcal{B}}$ in terms of $\{k_{ij}\}$ for the double-cycle network are generally lengthy and too complicated to be shown here (see Eq. S25 in \cite{Hwang2017JPCL}).

In the double-cyclic network model 
the total heat generated from the kinetic cycle depicted in Fig. \ref{double_cycle}b is decomposed into two contributions from the subcycles, $\dot{\mathcal{Q}}_{\mathcal{F}}$ and $\dot{\mathcal{Q}}_{\mathcal{B}}$, each of which is the product of reaction current and affinity \cite{liepelt07prl,barato2015PRL,Seifert2012RPP,Qian2005_BiophyChem,Qian_PRE_2004,Qian2010PRE}
\begin{equation}
\dot{Q} = J_{\mathcal{F}} \mathcal{A_F} + J_{\mathcal{B}} \mathcal{A_B},
\label{eq:dotQ}
\end{equation}
where the two driving forces (affinities) $\mathcal{A}$ are given by,
\begin{equation}
\begin{aligned}
&(a) \quad \mathcal{A}_\mathcal{F} = k_BT\log{\left(\frac{k_{12}k_{25}k_{56}k_{61}}{k_{21}k_{52}k_{65}k_{16}}\right)}=-\Delta\mu_{\text{hyd}} - f d_0\\
&(b) \quad \mathcal{A}_\mathcal{B} = k_BT\log{\left(\frac{k_{23}k_{34}k_{45}k_{52}}{k_{32}k_{43}k_{54}k_{25}}\right)}=-\Delta\mu_{\text{hyd}} + f d_0\\
\end{aligned}
\label{Eq:A}
\end{equation}
Note that at $f=0$, the chemical driving forces for $\mathcal{F}$ and $\mathcal{B}$ cycles are identical to be $-\Delta \mu_{\text{hyd}}$. 
The above decomposition of affinity associated with each cycle into the chemical driving force and the work done by the motor follows naturally from the $f$-dependent expression of $\{k_{ij}\}$ given in Eq.\ref{eqn:kij_f} \cite{liepelt07prl,fisher99pnas,fisher01pnas}.   
The entropy production from the double cycle model is expressed as the sum of entropy produced from the two sub-cycles $\mathcal{F}$ and $\mathcal{B}$.
\begin{figure}
\begin{align}
\beta\dot{Q}&=\beta\left(J_{\mathcal{F}}\mathcal{A}_{\mathcal{F}}+J_{\mathcal{B}}\mathcal{A}_{\mathcal{B}}\right)\nonumber\\
&=J_{\mathcal{F}}\log{\left(\frac{k_{12}k_{25}k_{56}k_{61}}{k_{21}k_{52}k_{65}k_{16}}\right)}+J_{\mathcal{B}}\log{\left(\frac{k_{23}k_{34}k_{45}k_{52}}{k_{32}k_{43}k_{54}k_{25}}\right)}\nonumber\\
	&=\underbrace{J_{\mathcal{F}}\log{\left(\frac{k_{12}k_{56}k_{61}}{k_{21}k_{65}k_{16}}\right)}
	J_{\mathcal{B}}\log{\left(\frac{k_{23}k_{34}k_{45}}{k_{32}k_{43}k_{54}}\right)}}_{\text{chemical}}+\\
	&+\underbrace{(J_{\mathcal{F}}-J_{\mathcal{B}})\log{\left(\frac{k_{25}}{k_{52}}\right)}}_{\text{mechanical}}
	\label{eqn:betaQ}
	\end{align}
\end{figure}
Or simply from Eq.~\ref{eq:dotQ} and Eq.~\ref{Eq:A}, $\dot{Q}$ can be recast into the differnece between the total free energy input [$\dot{E} = (J_\mathcal{F} + J_\mathcal{B}) (-\Delta \mu_{\text{hyd}})$] and work production [$\dot{W} = (J_\mathcal{F} -  J_\mathcal{B} ) f d_0$]:   
\begin{align}
\dot{Q} &=J_{\mathcal{F}} \mathcal{A_F} + J_{\mathcal{B}} \mathcal{A_B} \nonumber\\
&= (J_\mathcal{F} + J_\mathcal{B}) (-\Delta \mu_{\text{hyd}}) - (J_\mathcal{F} -  J_\mathcal{B} ) f d_0 \nonumber\\
&= \dot{E} - \dot{W}. 
\end{align}
	
	Notice that in order to reiterate our earlier remark that the mechanical equilibrium at the stall condition does not correspond to the thermodynamic equilibrium, the last line of $\beta\dot{Q}$ (Eq.\ref{eqn:betaQ}) is decomposed into the contributions by chemical and mechanical processes. 
	In the proposed double-cycle kinetic model, 
	the average velocity of the motor is given by $V = d_0 (J_\mathcal{F} - J_\mathcal{B})$. 
	Thus, if the numbers of forward and backward steps taken over time are balanced to each other, satisfying $J_{\mathcal{F}}=J_{\mathcal{B}}$ \cite{Cross05Nature}, we expect no net directional movement of the motor ($\langle x(t)\rangle=0$), which is equivalent to say $V=0$.  
	It is important to recognize that displacement (or travel distance) $x(t)$ is the most direct observable to an external observer if one were to use optical tweezers or fluorescence dyes.  
In the stall condition ($J_{\mathcal{F}}=J_{\mathcal{B}}\equiv J_s\neq 0$), however, because the chemical processes associated with ``two" ATP hydrolysis events, one along the forward and the other along the backward step, are still at work,   
the heat production is finite ($\dot{Q}>0$) 
\begin{align}
\beta\dot{Q}
	&=J_s\log{\left(\frac{k_{12}k_{23}k_{34}k_{45}k_{56}k_{61}}{k_{21}k_{32}k_{43}k_{54}k_{65}k_{16}}\right)}=2J_s(-\Delta\mu_{\text{hyd}}). 
\end{align}
This is a point of great importance, which cannot be capture by the uni-cyclic network model (Fig.\ref{double_cycle}a). 	

	Figure \ref{double_cycle}c depicts the currents flowing through the two sub-cycles $J_{\mathcal{F}}$ and $J_{\mathcal{B}}$ calculated from the set of rate constants ($\{k_{ij}(f)\}$)  determined against the motility data of kinesin-1. 
Under small hindering ($f\gtrsim 0$) or assisting load ($f<0$), kinesin-1 predominantly moves forward through the $\mathcal{F}$-cycle, whereas it starts taking more number of backsteps through the pathway $\mathcal{B}^+$ as the load increases further. 
At stall conditions, 
the net current associated with the mechanical stepping defined between the states (2) and (5) vanishes ($J=J_\mathcal{F}-J_\mathcal{B}=0$); however, non-vanishing current along the futile cycle hydrolyzing ATP still persists along the reaction path of $(1)\rightleftharpoons(2)\rightleftharpoons(3)\rightleftharpoons(4)\rightleftharpoons(5)\rightleftharpoons(6)\rightleftharpoons(1)$ (see Figure \ref{double_cycle}b). 
A further increase of $f$ beyond $f_{\text{s}}$ leads to $J_{\mathcal{F}}< J_{\mathcal{B}}$, increasing the chance of backsteps via ATP hydrolysis. 
Although a backward step satisfying $J_{\mathcal{F}}<0$ could be realized through an ATP synthesis, a theoretical analysis \cite{hyeon09pccp} of the experimental data \cite{Nishiyama02NCB,Cross05Nature} suggest that such event ($J_{\mathcal{F}}<0$, ATP synthesis induced backstep) is practically negligible compared with the one associated with ATP hydrolysis-induced backstep ($J_{\mathcal{B}}>0$), so that $|J_{\mathcal{B}}|\gg |J_{\mathcal{F}}|$.
This is in agreement with energetic analysis suggesting that backward and forward steps are not the reverse of each other~\cite{Hackney05PNAS}.

\subsection{Additional remark on non-equilibrium nature of motors}
It is worth mentioning a couple of recent studies that further highlight the non-equilibrium aspect of molecular motors. 
First,  the Harada-Sasa equality \cite{harada2006PRE} quantifies the heat generated from Langevin processes out of equilibrium as follows, 
\begin{align}
\dot{Q}=\sum_{i=1}^N\gamma_i \left\{\overline{v}_i^2+\gamma\int_{-\infty}^{\infty}[\tilde{C}_{ii}(\omega)-2k_BT\tilde{R}'_{ii}(\omega)]\frac{d\omega}{2\pi}\right\}.
\end{align}
It equates the total heat dissipation rate from the system $\dot{Q}$ with the expression in terms of friction coefficient $\gamma_i$ of an $i$-th variable $x_i$, the Fourier component of autocorrelation of the velocity fluctuation $C_{ii}(t)\equiv\langle (\dot{x}_i(t)-\overline{v}_i)(\dot{x}(0)-\overline{v}_i)\rangle_0$, and the real part of the response function $\tilde{R}'_{ii}(\omega)$.
At equilibrium, when the detailed balance condition is satisfied, $\overline{v}_i=0$, which leads to the standard  fluctuation dissipation theorem, $\tilde{C}_{ii}(\omega)=2k_BT\tilde{R}'_{ii}(\omega)$. Thus, the heat dissipation is zero.  
Therefore, the equality relates the extent of violation of the fluctuation-response relation in NESS with the rate of energy dissipation from the system into the bath. 
At least two studies have so far used this equality to experimentally assess the heat dissipation from molecular motors, one for kinesin and the other for F1-ATPase.  
Ariga~\cite{Ariga2018PRL} have adapted this equality in the form,  
\begin{align}
\dot{Q}_x=\gamma_x \left\{\overline{v}_x^2+\int_{-\infty}^{\infty}[\tilde{C}_{vv}(\omega)-2k_BT\tilde{R}'_{vv}(\omega)]\frac{d\omega}{2\pi}\right\}
\label{Ariga_formula}
\end{align}
 where the displacement of kinesin motor $x(t)$ was experimentally monitored, and the Fourier component of auto-correlation function and response function of velocity fluctuation were directly calculated using the time traces obtained from single molecule measurements.  
They showed that the total heat dissipation assessed using Eq.\ref{Ariga_formula} and the work done by the kinesin do not add up to the total chemical free energy input to the motor. 
The authors suggested that there are other elements of heat dissipation that do not involve the dynamics of the motor along the direction of motion $x(t)$ that can be monitored by their experiment.
Although the proposed double cycle model in Fig.\ref{double_cycle}b has already accommodated other possibilities, such as ATP-hydrolysis involved futile cycle without stepping, other scenarios such as slippage induced by mechanical force without chemical process \cite{yildiz08cell} are not included. 
Further elements of heat dissipation could be included in kinetic models in order to describe the function of the motors. 

The finding by Ariga \emph{et al.} should be contrasted with another experimental study using the Harada-Sasa equality on a single F$_1$-ATPase~\cite{Toyabe2010PRL}. These authors showed that the total free energy input to the enzyme was partitioned to the heat and work production with little loss. 
This indicates that the cycle of ATP synthesis corresponds to the reversed cycle of the hydrolysis-driven motor rotation. 
However, a recent careful theoretical analysis by Sumi and Klumpp on F$_1$-ATPase \cite{sumi2019NanoLett} suggested that the reversibility (100\% efficiency) of the rotary motor is only attained under certain conditions. 
Mechanical slip can occur in the presence of high external torque  without chemomechanical coupling, the effect of which is amplified at low ATP and ADP concentrations. This reduces the previously estimated 100 \% free-energy transduction efficiency. It was argued that in addition to the viscous dissipation of the probe, heat dissipation could occur from the rotary motor itself as the torque applied to the biological nanomachine induced mechanical slip. This occurs as a result of  deformation of  molecular conformation,  which is best optimized to function in the absence of torque. 
\\

\subsection{Applications}
\subsubsection{Myosin V and Kinesin-1}
SKMs have been successfully used to model the function of molecular motors.
Sometime ago  Leibler and Huse~\cite{leibler1993porters} devised a model to investigate the differences between motors that act as ``porters'' and those that function as ``rowers.''
The former, like the processive molecular motors described so far, work in small groups in order to transport cargoes; myosin II and axonemal dynein are examples of the latter, and work in large teams in order to slide or bend filaments.

\begin{figure}
    \centering
    \includegraphics[width=0.5\textwidth]{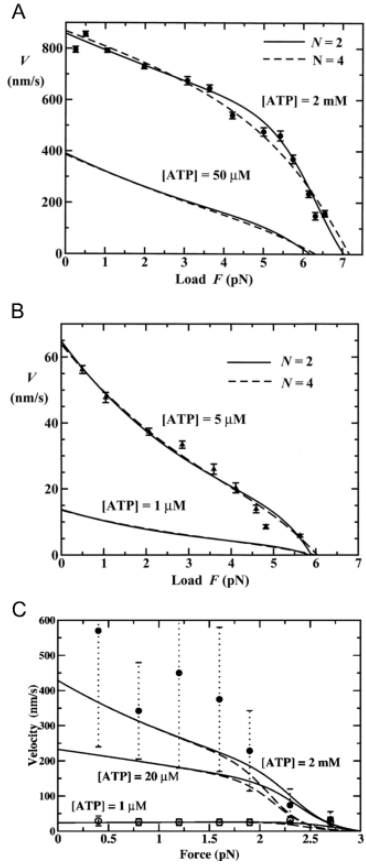}
    \caption{Unicycle models for kinesin-1 and myosin V.
    Panel (a-b): comparison for kinesin-1 of the results of a uni-cycle model  with experiments from~\citet{Visscher99Nature}.
    Panel (c): myosin V, comparison between a $N=2$ model for myosin V and the velocity from experiments, obtained from the mean dwell times $\tau$~\cite{Mehta99Nature} via the relationship $v=36\nm/\tau$.
    Panels (a-b) and panel (c) are adapted from~\citet{fisher01pnas} and from~\citet{kolomeisky2003simple}, respectively.
    }
    \label{Fig:MEF}
\end{figure}

In two landmark papers Fisher and Kolomeisky  analyzed the data for kinesin-1~\cite{fisher01pnas} and myosin V~\cite{kolomeisky2003simple} in order to train the parameters in their SKMs.
A 2-state and a 4-state model for kinesin-1 described accurately the motor velocity as a function of ATP and resistive load (see Fig.~\ref{Fig:MEF}a-b); although the 2-state model was also used to describe the run-length of kinesin (see the section~\ref{Section:ModelsWithDetachment} for details about the run length), the analysis of the randomness parameter as a function of $\mathrm{[ATP]}$ and $f$ highlighted the need for a $N=4$ model. 
In the case of myosin V, Kolomeisky and Fisher considered the data for ATP and load-dependent dwell time before a forward step~\cite{Mehta99Nature}; they fit the parameters of their 2-state model (Fig.~\ref{Fig:MEF}c) and predicted the value of the randomness parameter as a function of $\mathrm{[ATP]}$ and $f$.
The analysis of the parameters trained with the experimental data revealed in both circumstances the existence of a substep, and that most of the load dependence on the rates was carried by the reverse processes, indicating that the transition states are closer to the initial state than to the final state.

\begin{figure}
    \centering
    \includegraphics[width=0.5\textwidth]{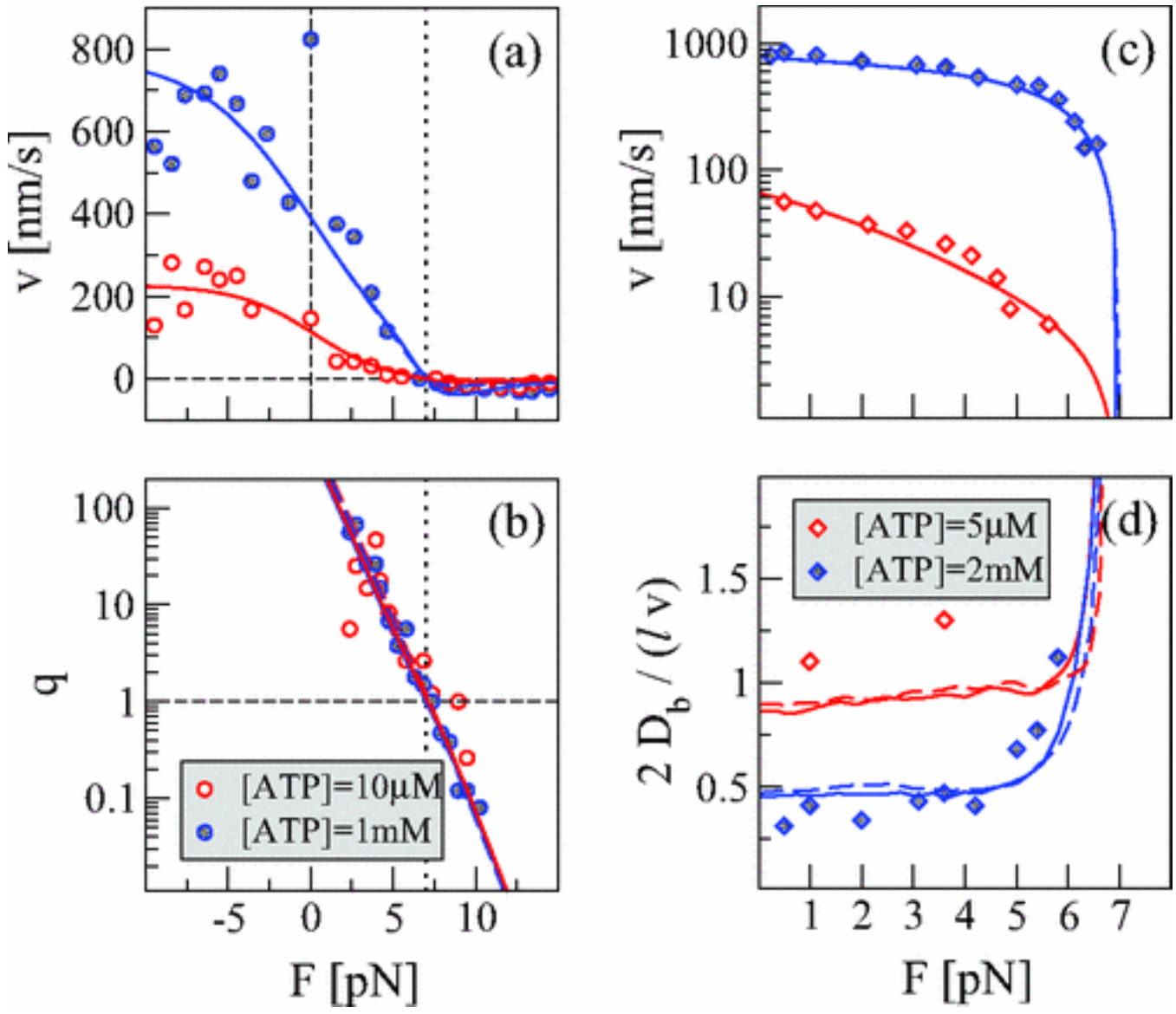}
    \caption{Fit of experimental data for kinesin-1 with a di-cycle, $N=6$ model.
    The data in panels (a) and (b) is from~\citet{Cross05Nature}, whereas the experiments considered in panels (c) and (d) are from~\citet{Visscher99Nature}.
    The figure is reproduced from~\citet{liepelt07prl}.
    }
    \label{fig:LL_Kin}
\end{figure}

Liepelt and Lipowsky~\cite{liepelt07prl} devised a multi-cycle model (see Fig.~\ref{double_cycle}b), which allowed them to incorporate backward steps fueled by ATP hydrolysis. 
The experimental results of~\citet{Cross05Nature} and~\citet{Visscher99Nature} were recovered with a di-cyclic, $N=6$ model (see Fig.~\ref{fig:LL_Kin}), although Liepelt and Lipowsky showed that another state (and a new cycle) was necessary in order to account for the velocity as a function of ADP concentration~\cite{Schief04PNAS}. 

A further development was put forth by~\citet{hyeon09pccp}, where a more sophisticated network was proposed in order to consider an alternative pathway for backward stepping based upon known structural features of the kinesin-1 dimer.
\citet{clancy2011nsmb} proposed yet another cycle for kinesin, in this case to fit the data for a mutant with an extended neck linker; the model successfully recovered the velocity, randomness parameter, and ratio between number of forward and backward steps as a function of ATP and resistive load, even in the super-stall regime.

\subsubsection{Dynein -- an erratic motor}
In part due to its complex architecture and the paucity of details of the nucleotide chemistry much less is known about the kinematics of dynein stepping kinetics. 
We should note that both from the perspective of structure determination \cite{Schmidt2015Nature} and single molecule experiments \cite{Reck06Cell,Dewitt2012Science,Belyy14NatComm} there have been spectacular advances in recent years, which could be used to create new theories. 
A theoretical model different from those described in this perspective, which by necessity treated the structural and transition kinetics between pre- and post-power stroke states approximately,  was introduced \cite{Tsygankov11BJ} to calculate the distribution of step size. 
Despite several untested approximation, experiments~\cite{Reck06Cell} and kinetic Monte Carlo simulations of the model were in fair agreement (see Fig. 3A in \cite{Tsygankov11BJ}). 
A much more elaborate model that couples structural aspects of dynein with a model for the catalytic cycle, which is similar in spirit to the theory for myosin V \cite{Hinczewski13PNAS}, was proposed more recently \cite{Sarlah14BJ}. 
The predictions of the model, which has a large number of parameters, were successful in obtaining  the step-size distribution as well as an estimate of the stall force. 
However, simple theories that can account for the unusually broad step size distributions reflecting the erratic nature of this motor, force-velocity curves as a function of ATP concentration, as has been done for kinesin and myosin V, are lacking.

\section{Molecular Motors -- Models with Detachment}
\label{Section:ModelsWithDetachment}
All molecular motors take only a finite number of steps along their tracks before detaching. 
Therefore, prominent features of motor motility are the run-length, $L$, and the run-time, $T$, which correspond to the distance covered during a processive run, and the amount of time spent bound to the filament before detaching, respectively.
In addition, the ratio between $L$ and $T$ constitutes an alternative and intuitive definition of velocity, $v = L/T$, which is simply the ratio between the spatial displacement of a motor and the amount of time taken to complete the movement.
Processive motors take many steps along the filament before dissociating from the track. 
The number of steps depends on the motor, and on a number of parameters such as the concentration of nucleotides, and the value of the applied load.
This underscores the importance of accounting for the end of the processive run in a way that is consistent with the enzymatic cycle of a specific motor and the experimentally determined run-length, run-time etc.

The models described in the previous sections assume that the motor is permanently attached to the track, and ignore the end of the processive run.
We discuss here a number of strategies that have been used in order to account for the detachment from the filament.
(i) The most direct way is to increase the number of states by $N_d$, the number of detached states, and to introduce the rates of releasing and attaching to the filament.
This method, which was used to model the function of kinesin-1 with $N_d=1$~\cite{liepelt07prl}, enables a steady-state treatment of the corresponding Markov jump process, but requires an increase in the number of states, the number of edges, and the number of cycles.
(ii) Alternatively, it is possible to enforce a new stationary state in the Markov jump process by diverting the flux into the detached state towards the filament-bound conformations.
Although this approach does not require any other state ($N_d=0$), it is necessary to increase the number of edges, which results in the creation of new cycles.
A theoretical description of this steady-state approach can be found in the remarkable work by Hill~\cite{hill2005free}.
It is unlikely that the methods described by Hill could be used to calculate velocity and run-length distribution. 
However, one could calculate their averages as a function of nucleotide concentration or external load.
(iii) It is also possible to account for the detached trajectories by re-normalizing quantities such as the average velocity over the decreasing number of bound motors.
This strategy, put forth by Kolomeisky and Fisher, is described in~\cite{kolomeisky2000periodic}, and has been used to test an approximate equation for kinesin run length~\cite{fisher01pnas}.
(iv) One could define the average run length as the ratio between the average velocity $v$ and the rate of detachment $\gamma$ ($L\approx v/\gamma$).
In order to do so, $\gamma$ could be estimated as the product between the stationary probabilities of occupying the ``vulnerable'' states and newly introduced detachment rates ($\gamma=\sum_{i\in\text{vulnerable}} p^S_i k_{i,\text{det}}$)~\cite{fisher01pnas,maes2003JSP}.
(v) Finally, the detached state could be treated as an absorbing state.
A stationary solution is not possible: after a sufficient time ($t>>\gamma^{-1}$) all the motors will be absorbed to the detached state.
However, one may account for all of the possible trajectories leading to absorption, and perform averages over the ensemble of these paths.
In the remainder of this section we focus on this last method.\\

\subsection{Velocity Distribution}
\begin{figure}
\centering
\includegraphics[width=0.5\textwidth]{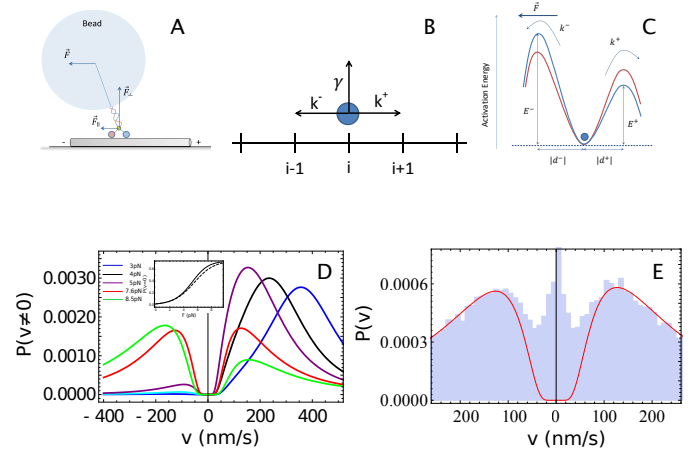}
\caption{One-state model for kinesin motility with detachment.
(A) Kinesin walks along the MT, subject to an external load $F$.
(B) The MT is represented as a linear discrete set of sites separated by $d_0=8.2$nm; the motor advances towards the + end with rate $k^+$, backsteps with rate $k^-$, and at every site $\gamma$ establishes the detachment rate.
(C) Load affects the rates according to Bell model, which in turn corresponds to a modification of the energy profile: $k^+$ diminishes as load increases, whereas the resistive force favors back-stepping (increases $k^-$). The transition state location with respect to the starting state, $\theta = \frac{|d^-|}{(|d^-| + |d^+|)}$.
(D) The velocity distribution displays a markedly bimodal shape at a variety of resistive forces, with the negative peak that increases as the load approaches stall.
(E) Simulations performed assuming an experimentally motivated error on the determination of $d_0$ maintain the bimodal structure of $P(v)$.
Figure adapted from~\citet{vu2016discrete}.}
\label{Fig:1State_Detach}
\end{figure}
The simplest model of molecular motor with detachment is shown in Fig.~\ref{Fig:1State_Detach}, in which forward (backward) steps occur with rate $k^+$ ($k^-$), and the detachment rate is $\gamma$.
For this model it is possible to determine analytically the probability distribution $p(\bar{v})$, where the velocity $\bar{v}$ is the ratio between the net number of steps taken ($n=m-l$, where $m$ is the number of forward steps and $l$ is the number of backward steps) divided by the run-time ($\bar{v} = n/T=v/d_0$).
Vu {\it et al.} showed that~\cite{vu2016discrete},
\begin{equation}
\begin{aligned}
&p(\bar{v}\gtrless 0) =\\
&=\frac{\gamma}{|\bar{v}|}\sum_{n=0}^{\infty} \Big(\frac{n}{|\bar{v}|}\Big)^{n+1} \frac{1}{n!} (k^\pm e^{-k_\mathrm{T}/|\bar{v}|})^n 
                 { }_{0}F_1(;n+1;\frac{n^2k^+k^-}{|\bar{v}|^2}),
\end{aligned}
\label{Eq:PofV_OneState}
\end{equation}
where ${ }_{0}F_1(;n+1;\frac{n^2}{|\bar{v}|^2}k^+k^-)$ is a hypergeometric function, and $k_\mathrm{T} = k^+ + k^- + \gamma$.
For an alternative derivation, see a more recent study~\cite{Zhang18JPCB}.
This model was adopted to describe the stepping mechanism of kinesin-1; the parameters $k^+$ and $k^-$ were established from experiments and the run length and velocity distribution~\cite{walter2012tubulin} were fitted using only one parameter, $\gamma$.
The result in Eq.~\ref{Eq:PofV_OneState} hold also in the presence of force, which increases the probability of back-stepping and detaching from the microtubule. 
Using this model a number of noteworthy results were obtained~\cite{vu2016discrete}: (i) even at zero load, the velocity distribution is not Gaussian, although it approaches a normal distribution if the run length is large (that is, $k^+/\gamma >> 1$ and $k^+/k^- >> 1$); (ii) the probability distribution of the instantaneous velocity (the size of the step divided by the dwell time) differs from Eq.~\ref{Eq:PofV_OneState}, and does not match the experimental distribution~\cite{walter2012tubulin}, suggesting that it is not the correct way to compute velocity; (iii) most surprisingly, the velocity distribution is bimodal, with distinct peaks for $\bar{v}>0$ and $\bar{v}<0$.
The prediction that $p(v)$ is bimodal, becoming most prominent at stall force, was unexpected.
At $f=\fstall$, the na\"{i}ve expectation would be that $p(v)$ would have a peak at $v=0$, and the area under $p(v)$ with $v>0$ and $v<0$ would be the same.
The bimodal distribution is a consequence of the discrete nature of kinesin step size, and it is exaggerated at large $f$, when the probability of taking backward steps and the probability of detaching are larger.

\begin{figure}
\centering
\includegraphics[width=0.5\textwidth]{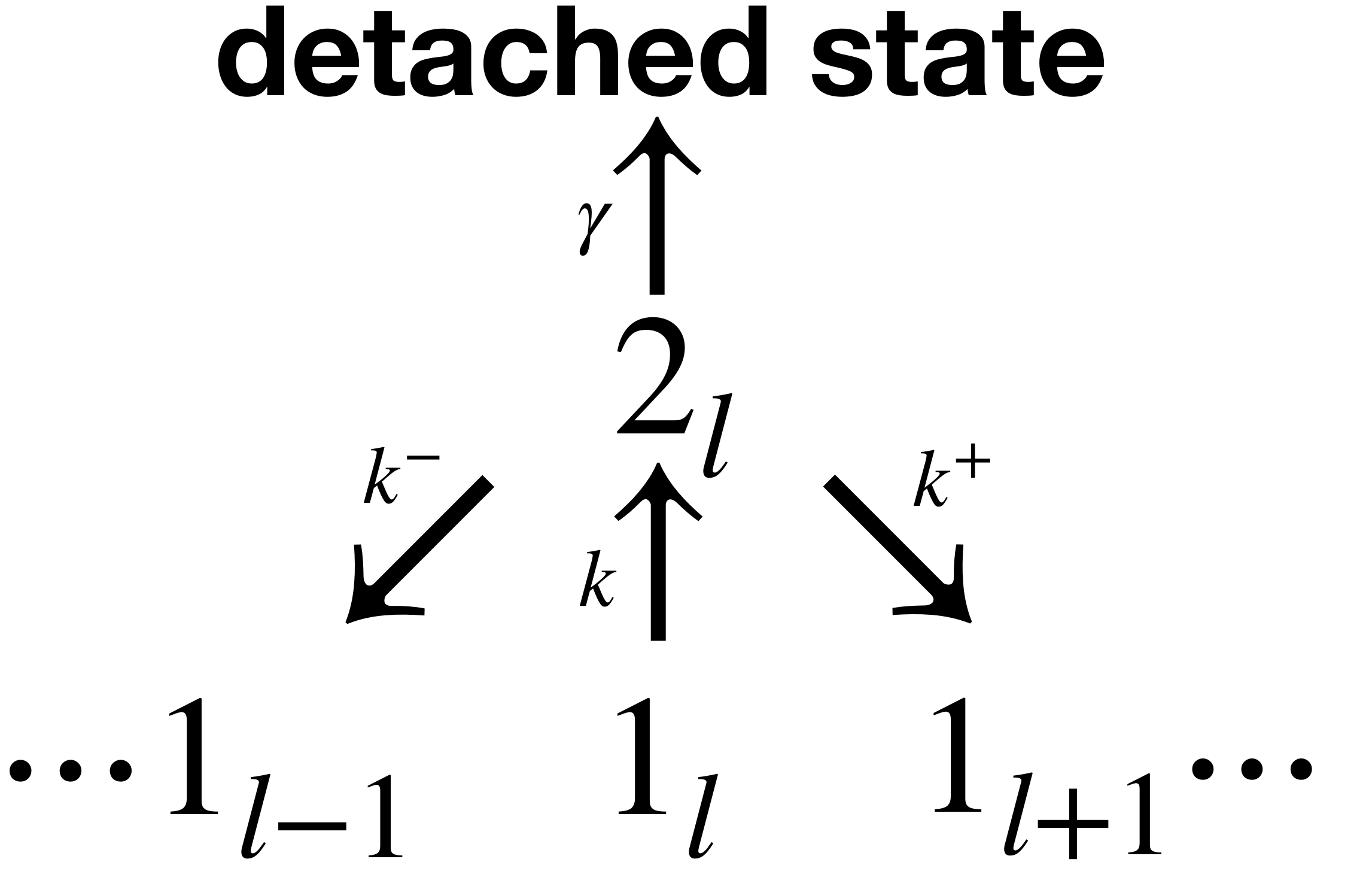}
\caption{Two-state model with detachment.
The forward and backward stepping rates are $k_+$ and $k_-$, respectively, and they connect state $2$ to state $1$.
The $1\rightarrow2$ transition occurs with rate $k$.
Detachment occurs only from state $2$, with rate $\gamma$.}
\label{Fig:2State_Detach}
\end{figure}

The model in Fig.~\ref{Fig:1State_Detach} was recently extended in order to include an intermediate step (Fig.~\ref{Fig:2State_Detach}), which was used to describe the kinetics of a molecular motor as a combination of an ATP-dependent and an ATP-independent transition~\cite{Takaki19preprint}.
The resulting velocity distribution is,
\begin{equation}
\begin{aligned}
&p(\bar{v} \gtrless 0) = \\
&\frac{\gamma}{\bar{v}} \sum_{\stackrel[m\gtrless l]{m,l}{}} \frac{m-l}{\bar{v}} \frac{\sqrt{\pi}}{m!l!} \frac{k^{n+1} (k^+)^m(k^-)l}{|k-(k^++k^-+\gamma)|^{n+1/2}}\Big(\frac{m-l}{\bar{v}}\Big)^{n+1/2} \\
&e^{-\frac{k+k^++k^-+\gamma}{2}\frac{m-l}{\bar{v}}}I_{n+1/2}\Big(\frac{|k-(k^++k^-+\gamma)|}{2}\frac{m-l}{\bar{v}}\Big),
\end{aligned}
\end{equation}
where $I$ is a modified Bessel function of the first kind.
Interestingly, the bimodal structure of the velocity distribution is robust to changes of the nucleotide concentration, and it is independent of which step ($1\rightarrow2$ or $2\rightarrow1$) is ATP-dependent.
Therefore, an experiment aimed at testing the predicted multi-modality of $p(v)$ may be conducted under arbitrary ATP concentration and resisting load.

The model in Fig.~\ref{Fig:2State_Detach} was used to tackle a vexing conundrum concerning the stepping mechanism of kinesin~\cite{Takaki19preprint}.
Recently, two groups have performed similar experiments in which they monitored the movement of one kinesin head conjugated with a gold-nanoparticle (AuNP) by tracking the position of AuNP~\cite{mickolajczyk2015kinetics,Isojima2016NCB}.
From the analysis of the trajectories, one group proposed that ATP binds to kinesin when the trailing head is still bound to the MT (or at least in the vicinity of the pre-stepping site)~\cite{mickolajczyk2015kinetics}, the other group suggested that the dissociation from the MT and forward movement of the trailing head detachment precede ATP binding~\cite{Isojima2016NCB}.
Using the model in Fig.~\ref{Fig:2State_Detach}, Takaki et al.~\cite{Takaki19preprint} predicted that the ATP-dependent profile of the randomness parameter may be used in order to determine whether ATP binds to the two-head-bound or one-head-bound conformation of the kinesin-1 dimer.\\

\subsection{Alternate Models with Detachment}
\begin{figure}
    \centering
    \includegraphics[width=0.5\textwidth]{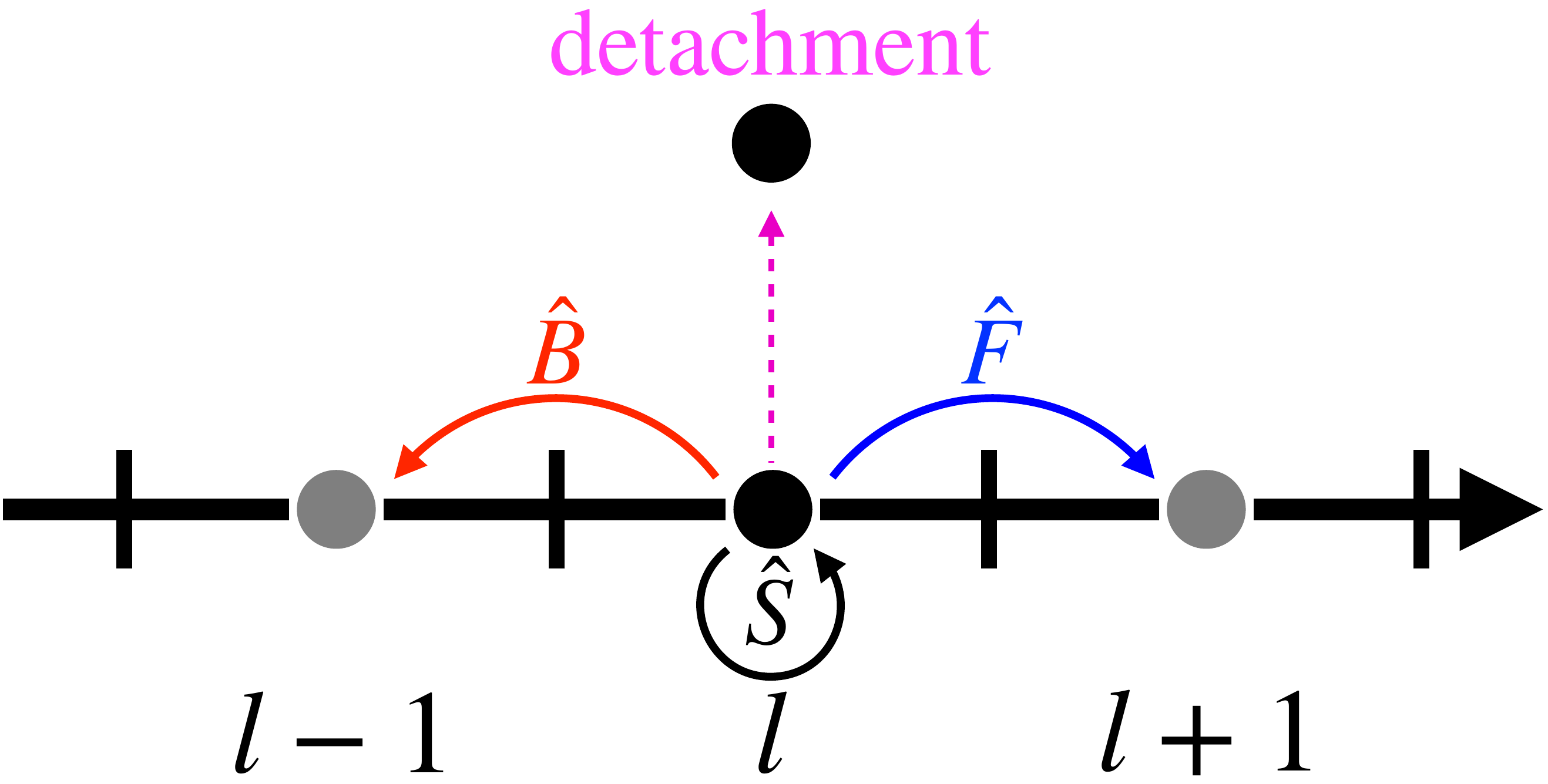}
    \caption{Model for processive motor moving on a periodic track.
    From each site $l$ of a filament the motor can change its conformation (e.g. the chemical state) within the same location (transitions included in the matrix $\hat{S}$, in black) move forward (the matrix $\hat{F}$, in red, accounts for these steps) or backward (the matrix $\hat{B}$, in blue, refers to these displacements) along the track, or detach (magenta).}
    \label{Fig:Processivity}
\end{figure}

More complicated kinetic schemes that include motor detachment have been used in order to compute the processivity and velocity of molecular motors.
Elting et al.~\cite{elting2011detailed} found an analytical solution for a 8-state model of myosin VI which they proposed in order to investigate the nature of the gating mechanism.
This model did not account for the possibility of backward stepping, which was instead included by Caporizzo et al.~\cite{caporizzo2018antiparallel} in  a kinetic model that was devised in order to investigate the motility of myosin X as a function of the structure of the filament (individual versus bundled actin) and the geometry of the tail of the dimer (parallel versus anti-parallel).
More recently, a new theoretical framework capable of extracting motility characteristics such as average velocity, average and distribution of number of steps, and probability of backward stepping for a kinetic network of arbitrary geometry has been proposed~\cite{mugnai2019processivity}.
The dynamics of the motors is described in terms of the following Markov chain,
\begin{equation}
\vP_{x+1} = (\mS + \mF + \mB)\vP_x = \mM\vP_x,
\label{Eq:MarkovChain}
\end{equation}
where the $N$-dimensional vector $\vP_x$ contains the probability of occupying any of the $N$ filament-bound states of the motor after $x$ transitions have occurred, and $\mS$, $\mF$, and $\mB$ are the $N\times N$ transition probability matrices for undergoing a transition at a fixed location, stepping forward, or backward, respectively (see Fig.~\ref{Fig:Processivity}).
Eq.~\ref{Eq:MarkovChain} holds under the assumption that the track is periodic, which makes $\mS$, $\mF$, and $\mB$ independent on the location of the motor along the filament.
By accounting for all of the possible stepping pathways, it was found that the probability of taking $n$ steps (forward or backward) before detaching is,
\begin{equation}
\prob(n) = \vOneT (\mI - \mPstep)(\mPstep)^n \vP_0,
\label{Eq:Pofn}
\end{equation}
where $\vOneT$ is the transpose of an $N$-dimensional vector of ones, $\vP_0$ is the initial probability (appropriately normalized, $\vOneT \cdot \vP_0 = 1$), and $\mPstep = (\mF+\mB)(\mI-\mS)^{-1}$.
This expression is a generalization of a well-known result for $N=1$, in which $\prob(n) = (1-\pi)\pi^n$, where $\pi<1$ is the probability of stepping and $1-\pi$ is the probability of detaching.
The average number of steps $\langle n \rangle$ may be computed using simple linear algebra, as well as the distribution and average number of forward and backward steps, $\langle m \rangle$ and $\langle l \rangle$, respectively, with $\langle m \rangle + \langle l \rangle = \langle n \rangle$.
The average run length is then $\langle L \rangle = d_0 (\langle m \rangle - \langle l \rangle)$, and by using a classical result for the mean-first-passage-time to absorption it is possible to compute $\langle \tau \rangle$~\cite{oppenheim1977stochastic}, and define an average velocity $ v = \langle L \rangle/\langle \tau \rangle$.
With Kinetic Monte Carlo simulations it is possible to show that $\langle L \rangle/\langle \tau \rangle$ differs from $\langle L / \tau \rangle$, which was adopted in~\cite{vu2016discrete,Takaki19preprint}.
On the other hand, it was shown empirically that the differences become smaller as the average number of steps taken by the motor increases~\cite{mugnai2019processivity}.
The advantages of this theoretical framework are its flexibility (it works with any network) and ease of implementation (it only requires matrix algebra). Therefore, it can be used to fit complicated models against experimental data.
The model was used to solve a complicated model for myosin VI motility inspired by  model of Yanagida and coworkers~\cite{nishikawa2010switch,ikezaki2012simultaneous,ikezaki2013spontaneous},  which accounted for a motor taking both hand-over-hand and inchworm-like steps~\cite{mugnai2019processivity}, and predicted the gating mechanism [in agreement with~\cite{dunn2010contribution,elting2011detailed}], size of the backward steps [matching experimental observations~\cite{altman2004mechanism,nishikawa2010switch} within discrepancies likely related to fluctuations and the site of probe attachment], pathway for backward stepping [as suggested by~\cite{ikezaki2013spontaneous}], and importance of foot-stomping in breaking the tight-coupling of myosin VI stepping.

\section{Polymer Physics-based Approaches Incorporating Structural Features Into Kinetic Theories}

Even with coarse-grained numerical models like  the one proposed by~\citet{Craig09PNAS},which is described below, collecting statistics from multiple simulations covering entire motor trajectories (i.e. for myosin V each run averaging tens of steps until detachment) can be computationally expensive.  To more comprehensively explore how motor architecture affects stepping dynamics, particularly in light of experiments that perturb structural features like lever arm length~\cite{Sakamoto2005Biochem,Oke2010PNAS}, alternative approaches are needed.  We focus on one particular example, an analytical theory for myosin V dynamics~\cite{Hinczewski13PNAS}, that also highlights several aspects discussed earlier:  the starting point of the theory is a six-state stochastic kinetic network model consisting of multiple cycles that explicitly includes detachment of the myosin V dimer from the actin filament.  Incorporating detachment allows us to model finite run lengths of the motor on actin, while the multi-cycle network topology captures the dominant pathways of myosin V dynamics under load forces below or near stall ($\lesssim 2-3$ pN):  these include both forward and backward steps as well as so-called ``stomps'', where either the trailing or leading head detaches and then reattaches near its original binding location.  Stomps are challenging to detect with single-molecule fluorescence techniques, but have been observed experimentally using high-speed atomic force microscopy~\cite{Kodera2010Nature}.

\begin{figure*}
    \centering
    \includegraphics[width=1.0\textwidth]{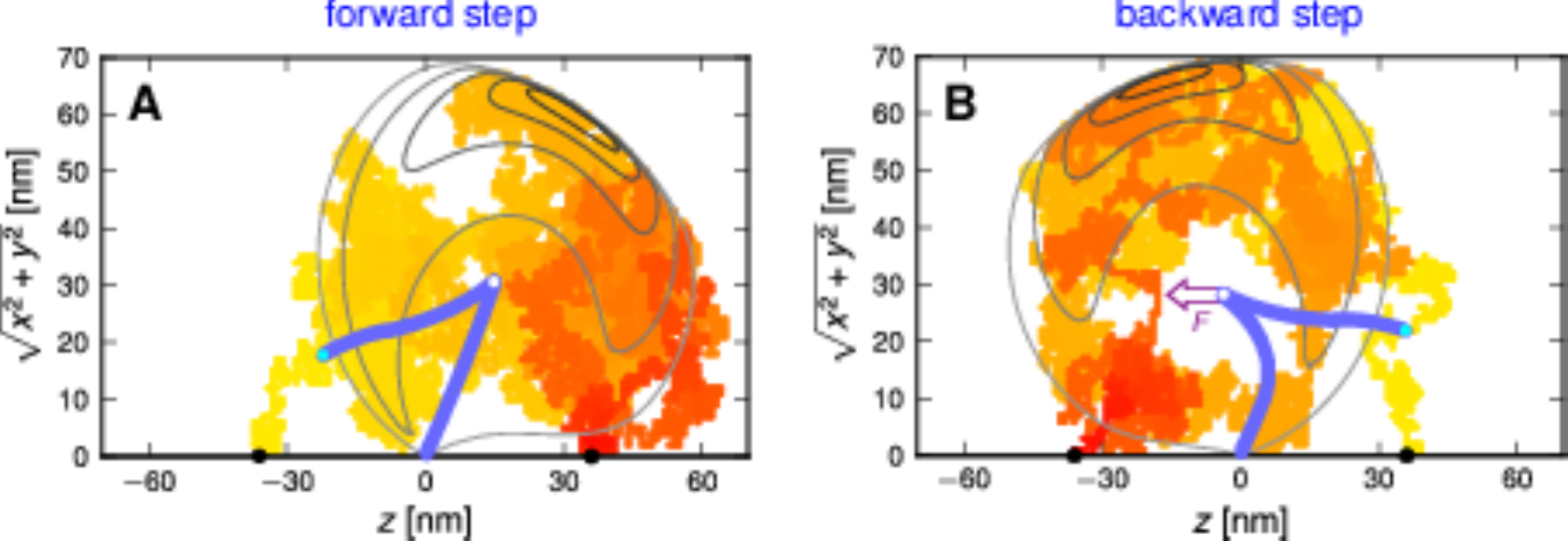}
    \caption{Examples of the diffusive search trajectory of the detached myosin V head for two different values of backwards load force: {\bf A.} $F=0$; {\bf B} $F=2$ pN.  The actin filament lies along the $z$ axis, and the trajectories are projected onto the radial distance $\sqrt{x^2+y^2}$ away from the actin axis.  The colors represent time, with lighter colors (yellow) occurring earlier than darker colors (red).  In panel A the trajectory ends in a forward step (the detached head going from the backward binding site at $z=-36$ nm to the forward binding site at $z=36$ nm.  In contrast, panel B shows a backwards step, which becomes increasingly favored as a kinetic pathway under larger load forces.  Superimposed on the trajectories are cartoon snapshots of the two polymer legs (each leg representing a lever arm plus head) of the coarse-grained model for myosin V, at a time shortly after detachment.  The white dot is the junction of the two legs, and the cyan dot is the location of the detached head at that time step.  The contours correspond to the equilibrium probability distribution ${\cal P}(\mathbf{r})$ of the detached head, calculated from the analytical polymer theory (darker contours correspond to higher probabilities).  As load force is increased, the attached leg and junction are pulled backwards, biasing the entire distribution away from the forward binding site, and thus increasing the chance of backwards versus forward steps (see also the corresponding experimental data of Fig.~\ref{myoV_exp}A). Adapted from~\citet{Hinczewski13PNAS}. 
    }
    \label{myoV_traj}
\end{figure*}

However unlike the kinetic models treated so far, the force-sensitive transition rates will not be described through a phenomenological Bell-like exponential dependence as in Eq.~\eqref{Eq:u/w_F}.  Instead, the goal of the approach by~\citet{Hinczewski13PNAS} is to model the structural mechanics underlying this force dependence through a coarse-grained polymer theory for the myosin V dimer.  This replaces the discrete semiflexible polymer description of Craig and Linke~\cite{Craig09PNAS} (interacting monomers representing the motor head and lever arm IQ domains) with an even simpler construct: a single continuous semiflexible polymer that represents the combined motor head and lever arm domains.  Thus, the dimer becomes two polymer ``legs'' attached at a flexible junction.  The load force transmitted through the cargo domain is modeled by an effective force $F$ applied at the junction.  Because the load force changes the ensemble of conformations for the two polymer legs, and hence the three-dimensional distribution of positions explored by the unbound motor head (see Fig.~\ref{myoV_traj}), it affects the rates at which the unbound head reaches potential binding sites.

The most direct advantage of this simpler coarse-grained description is that the effects of force on stepping dynamics become (to an excellent approximation) analytically tractable.  It is also readily generalized to more complex contexts.  For example, while the model we focus on here considers only a flexible junction, backwards load forces parallel to the actin axis, and restricts steps to actin binding sites separated at actin helical half lengths (the most probable step separation for myosin V), all of these assumptions can be relaxed.  A recent extension of the polymer theory approach~\cite{Hathcock19} explores the effects of a possible structural constraint at the junction (inspired by experimental evidence~\cite{Andrecka15eLife}), considers off-axis forces, and incorporates the full distribution of steps at all possible actin binding sites.  This allows us to understand previously observed step distributions of mutant myosins with various lever arm lengths~\cite{Sakamoto2005Biochem,Oke2010PNAS}, as well as the robustness of myosin V dynamics in experiments where various off-axis load forces are applied to the motor via optically trapped cargo~\cite{Oguchi2010NCB}.  Moreover the polymer approach is not limited to myosin V: an analogous treatment of dynein, with the two motor domains approximated as rigid rods (the large stiffness limit of semiflexible polymers), can successfully reproduce the complex details of dynein step distributions on microtubules~\cite{Goldtzvik2019}.  For both myosin V and dynein each half of the dimer is modeled as a single polymer / rod, and this description is likely to be applicable to other dimeric processive motors with fairly stiff lever arms, like myosin XI~\cite{Tominaga2012}.  However there are cases, like myosin VI or X, where the lever arm structure is more heterogeneous, mixing stiff and flexible domains~\cite{Sun2011BJ}.  Here any future attempt to apply coarse-grained polymer modeling would need to represent each lever arm as a series of polymer chains with different bending rigidities.

\begin{figure*}
    \centering
    \includegraphics[width=1.0\textwidth]{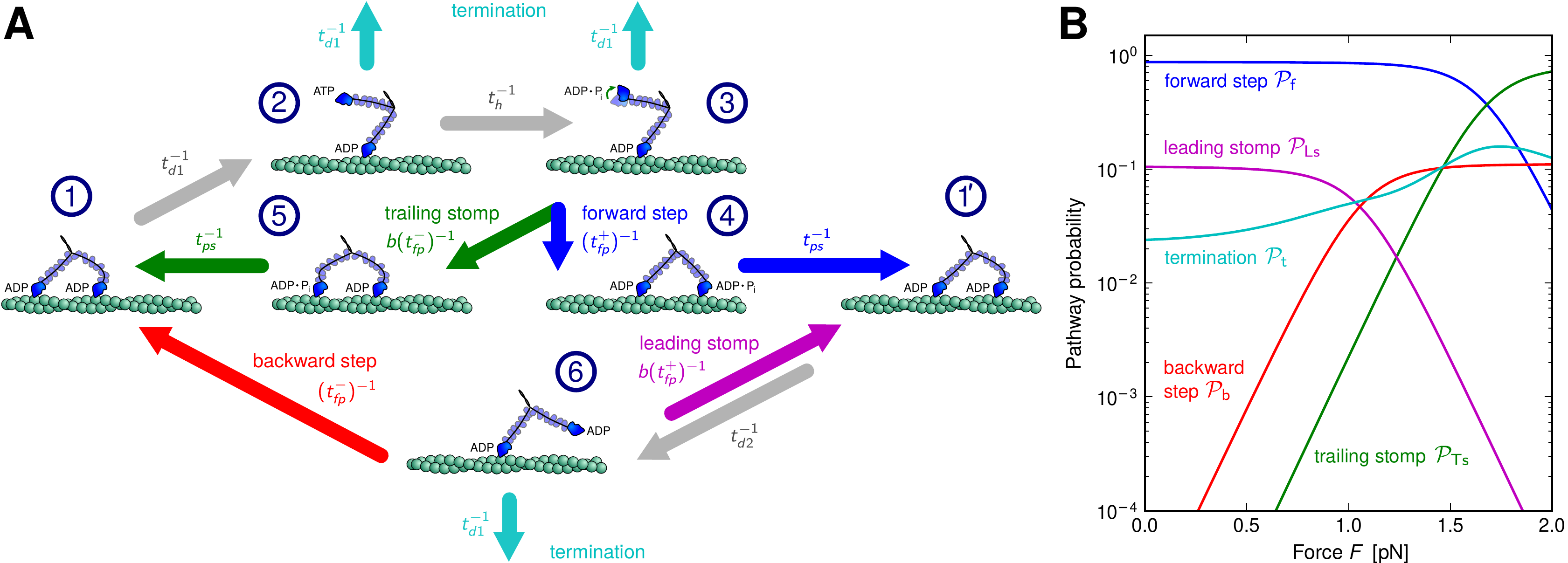}
    \caption{{\bf A.} Six-state kinetic model for myosin V that underlies the coarse-grained polymer analytical theory of~\citet{Hinczewski13PNAS}.  States 1 and $1^\prime$ are identical waiting states, with both motor heads bound to ADP and actin, except that $1^\prime$ is displaced 36 nm (a half helical length of actin) toward the plus end of the filament.  Arrows are marked by inverse timescales that denote the transition rates, the details of which are described in the text.  Colored arrows are transitions that belong to specific kinetic pathways: forward step (dark blue); trailing stomp (green); leading stomp (purple); and backward step (red); termination (light blue).  Gray arrows are transitions that are shared between multiple pathways.  {\bf B.} Predicted pathway probabilities from the model, using the kinetic network together with mean first passage-times for the diffusive search step $t_{fp}^\pm$ from polymer theory.  Adapted from~\citet{Hinczewski13PNAS}.}
    \label{myoV_net}
\end{figure*}

To understand the basic details of the polymer theory for myosin V in its simplest form, we first summarize the underlying six-state kinetic model, illustrated in Fig.~\ref{myoV_net}A.  
State 1 is the waiting state, where both motor heads have bound ADP and are strongly attached to actin.  The lever arms for both heads are in the post-powerstroke conformation:  in other words in the absence of other constraints the lever arms would orient in the forward direction (toward the plus end of actin), but because the junction pulls the lever arm of the leading head backwards the entire structure is in a strained state known as the telemark or reverse arrowhead stance~\cite{Walker2000Nature,Kodera2010Nature}.  State $1^\prime$ is also a waiting state, but with the entire motor displaced forward by one half helical length of actin ($\Delta = 36$ nm).  The possible kinetic pathways are as follows:
\begin{enumerate}
    \item  {\it Forward step} ($1 \to 2 \to 3 \to 4 \to 1^\prime$):  ADP release from the trailing head is followed by ATP binding, which leads to detachment of the head from actin ($1 \to 2$).  We assume saturating ATP conditions, so ATP binding is fast compared to ADP release, and hence the entire process is subsumed into a single transition described by a rate $t_{d1}^{-1}$.  Here $t_{d1}$ is the mean detachment time scale, dominated by the waiting time for ADP release, $t_{d1} \sim 12$ s$^{-1}$~\cite{DeLaCruz1999PNAS}.  The next transition ($2 \to 3$) involves ATP hydrolysis into ADP$+$P$_\text{i}$, along with a recovery stroke where the head / lever-arm orientation changes from post- to pre-powerstroke.  This transition occurs at a rate $t_\text{h}^{-1} = 750$ s$^{-1}$~\cite{DeLaCruz1999PNAS}.  While there are scenarios where the reverse hydrolysis rate is significant (for example in motors with modified light chain composition~\cite{Dunn2007NSMB}), for this discussion we make the typical assumption that the forward hydrolysis rate dominates.  In general, while every transition arrow in Fig.~\ref{myoV_net} has an associated reverse transition in principle, here we make the simplifying assumption that the reverse rates are negligible relative to the forward rates.  This assumption captures the dominant kinetic pathways of the motor (our focus here), but we would need to explicitly consider reverse rates to look at thermodynamic features of the system (like entropy production).  Once ATP is hydrolyzed, the motor head has the ability to associate with actin again.  The three-dimensional diffusive search can result in either binding to the forward site (36 nm ahead of the bound leg), or the original binding site.  The rate at which it binds to the forward site is $(t_{fp}^+)^{-1}$, which depends on the mean first passage time $t_{fp}^+$ to the forward site.  The dependence of $t_{fp}^+$ on the load force is related to how the force biases the diffusive search, and the underlying physics is discussed in more detail below.  A binding to the forward site ($3 \to 4$) is quickly followed by P$_\text{i}$ release and a powerstroke ($4 \to 1^\prime$), which occurs with a rate $t_{ps}^{-1}$.  We assume this step is rapid relative to the others in the network, such that $t_{ps} \ll t_{d1}$, $t_{fp}^+$, $t_h$, and hence $t_{ps}$ will not enter explicitly into our estimates for motor properties below.
    
    \item {\it Trailing stomp} ($1 \to 2 \to 3 \to 5 \to 1$): This pathway starts in the same way as the forward step, with unbinding of the trailing head ($1 \to 2$) followed by hydrolysis and the recovery stroke ($2 \to 3$), but the diffusive search now ends at the original binding site from where the head detached ($3 \to 5$).  The mean first passage time to this site is $t_{fp}^-$ (which also depends on force as described below).  However the head no longer binds in the same configuration as before, because the lever arm has undergone a recovery stroke and is now in the pre-powerstroke orientation.  Thus binding requires not only first passage to the site but overcoming the energetic barrier of an unfavorable geometric orientation.  We effectively model this through a factor $0 < b < 1$ that reduces the rate, so that the overall rate from $3 \to 5$ is given by $b(t_{fp}^-)^{-1}$.  From fitting to experimental data, described below, the value of $b \approx 0.065$.  As in the forward step, binding is followed by P$_i$ release and a power stroke ($5 \to 1$) occurring at a fast rate $t_{ps}^{-1}$.
    
    \item {\it Leading stomp} ($1^\prime \to 6 \to 1^\prime$): This pathway is initiated when the leading head detaches with ADP still bound ($1^\prime \to 6$).  The reason for this alternative head detachment mechanism is an asymmetry that arises from the strain in the myosin V dimer in the waiting state.  The backward tension on the leading lever arm in the telemark stance (estimated to be on around 2.7 pN~\cite{Hinczewski13PNAS}) inhibits ADP release by a factor of 50-70~\cite{Rosenfeld2004JBC,Kodera2010Nature}.  Rather than releasing ADP and binding ATP to detach from actin, the dominant pathway for a head under backward load is direct detachment retaining the ADP.  This has been observed experimentally in single-headed myosin V, where backwards loads of around 2 pN lead to an detachment rate $t_{d2}^{-1} = 1.5$ s$^{-1}$~\cite{purcell2005force} independent of ambient ATP and ADP concentrations.  We thus take $t_{d2}^{-1}$ as the transition rate from $1^\prime \to 6$.  The difference between the overall detachment rates for the trailing and leading heads, with $t_{d1}^{-1}$ eight-fold larger than $t_{d2}^{-1}$, underlies the gating~\cite{Veigel2002NCB,Veigel2005NCB,purcell2005force} mechanism:  the trailing head is much more likely to detach first.  Since detachment of the leading head occurs with the ADP still bound, the head can directly reattach back to the forward binding site, with one caveat:  the post-powerstroke configuration creates an energy barrier to reattachment due to geometry (since the lever arm has to adopt a strained stance) and so as before we introduce a factor $b$ to scale the rate to the forward site.  Thus the overall reattachment rate from $6 \to 1^\prime$ is $b(t_{fp}^+)^{-1}$.
   
   \item {\it Backward step} ($1^\prime \to 6 \to 1$):  This pathway starts with leading head detachment ($1^\prime \to 6$) like the leading stomp, but the detached head after diffusion finds the backward binding site.  The post-powerstroke conformation with ADP bound is geometrically favorable for binding to this site (because the lever arm does not have to be bent) and hence the rate from $6 \to 1$ is $(t_{fp}^{-})^{-1}$.  Note that this description of backward stepping is approximately valid for load forces near or below stall, but does not include additional features that may become important for larger superstall forces ($F > 3$ pN), such as power stroke reversal~\cite{Sellers2010NSMB}.
   
   \item {\it Termination}:  For the kinetic pathways described above, if the motor is in one of the three states where only one head is bound to actin (2, 3, or 6), then detachment of the second head (with rate $t_{d1}^{-1}$) will lead to dissociation of the entire motor from the actin filament, terminating the run.
   
\end{enumerate}

In the kinetic network described above, mechanical forces enter in two ways:  the first is through the internal strain that leads to the gating mechanism, and the second is through the external load on the junction that affects the diffusive search and hence the first passage times $t_{fp}^\pm$.  We have already taken into account the asymmetry due to internal strain by using experimentally estimated values for $t_{d1}$ and $t_{d2}$, but the force dependence of $t_{fp}^\pm$ has not been specified.  This is precisely what the coarse-grained polymer description allows us to do.  We take advantage of a separation of time scales:  when either head detaches, the equilibration time $t_r$ over which the polymer legs relax to an approximately equilibrium distribution of conformations is fast compared to the first passage times, $t_r \ll t_\text{fp}^\pm$.  Brownian dynamics simulations and analytical arguments~\cite{Hinczewski13PNAS} show that $t_\text{r} \lesssim 5$ $\mu$s, while the fastest first passage times are at least $t_\text{fp}^\pm \sim {\cal O}(0.1\:\text{ms})$.  A summary of all the time scales in the problem is shown in Fig.~\ref{myoV_times}.  Thus to an excellent approximation the detached head fully explores some equilibrium distribution of positions, ${\cal P}(\mathbf{r})$, before reaching either of the binding sites.  In this case the first passage time to reach the forward (+) or backward (-) binding site at position $\mathbf{r}_\pm$ is inversely proportional to ${\cal P}(\mathbf{r}_\pm)$, the probability density of finding the detached head at that position~\cite{Hinczewski13PNAS}:
\begin{equation}\label{e1}
t_{fp}^\pm \approx \frac{1}{4\pi D_{h} a {\cal
    P}(\mathbf{r}_\pm)},
\end{equation}
Here $D_{h}$ is the diffusion coefficient of the detached head (which can be estimated as $D_{h} \approx 5.7 \times 10^{-7}$ cm$^{2}$/s from the crystal structure using HYDROPRO~\cite{Ortega2011BJ}). The capture radius $a \approx 1$ nm is the distance between the head and binding site for which the interactions become strong enough that binding occurs, comparable to the Debye screening length under physiological conditions.  Eq.~\eqref{e1}, which can be derived from standard first passage time analytical approaches like the renewal method~\cite{vankampen}, is extremely useful:  it converts the dynamical problem of finding mean first passage times of a complex diffusion process into the more tractable problem of calculating the equilibrium distribution ${\cal P}(\mathbf{r})$ of the end-point of a two-legged semiflexible polymer system.  The latter can be found analytically by extending an earlier mean-field theory for semiflexible polymers~\cite{thirumalai1998}, incorporating the orientational constraint of the bound leg due to the post-powerstroke conformation of the lever arm with respect to the motor head.  Only a handful of structural parameters enter into the theory, determining ${\cal P}(\mathbf{r})$:  the contour length $L = 35$ nm and persistence length $l_p \approx 310$ nm of the polymer leg; the angle of the orientational constraint with respect to the actin filament, $\theta_c \approx 60^\circ$; and a parameter $\nu_c$ describing the strength of the orientational constraint.  The first three are all known from earlier experimental estimates~\cite{Moore2004,Dunn2007NSMB,Craig09PNAS}.  $\nu_c$ is along with $b$ one of the two free parameters in the entire theoretical description, and can be estimated based on fitting to experimental data.  

The full expression for ${\cal P}(\mathbf{r})$ and its derivation can be found in~\citet{Hinczewski13PNAS}.  The contours in Fig.~\ref{myoV_traj} illustrate ${\cal P}(\mathbf{r})$ for two different values of backwards load force:  A) $F=0$ pN; B) $F = 2$ pN.  Superimposed are sample trajectories of the detached endpoint, corresponding to a forward and backward step respectively, along with snapshots of the polymer conformation at a time shortly after detachment.  The actin filament lies along the $z$-axis, with $z=0$ corresponding to the location of attached head, and the forward/backward binding sites for the detached head at $z_\pm = \pm \Delta$.  At zero force the peak of the distribution is at $z > 0$ due to the post-powerstroke orientational constraint on the bound leg.  Thus the endpoint probabilities are biased toward the forward binding site and ${\cal P}(\mathbf{r}_+) \gg {\cal P}(\mathbf{r}_-)$.  When $F= 2$ pN the situation is reversed:  the load force pulls the junction backwards, counteracting the post-powerstroke constraint, and the distribution is shifted such that ${\cal P}(\mathbf{r}_-) \gg {\cal P}(\mathbf{r}_+)$.  The dependence of ${\cal P}(\mathbf{r})$ on load force translates into corresponding changes in $t_{fp}^\pm$ through Eq.~\eqref{e1}.

Once $t_{fp}^\pm$ as a function of $F$ is known, a variety of physical quantities can be calculated directly from the kinetic network model~\cite{Hinczewski13PNAS}.  For example both forward steps and trailing stomps have the same mean duration:  the average time $t_{Tb}$ from the detachment of the trailing head to its subsequent reattachment at either the forward or backward site.  Similarly both backward steps and leading stomps have a mean duration $t_{Lb}$ associated with how long the leading head takes to reattach.  These two timescales are given by:
\begin{equation}
t_{Tb} = t_{h} + \frac{t_{fp}^+}{1+b\alpha}, \qquad t_{Lb} = \frac{t_{fp}^+}{b+\alpha}
\end{equation}
where $\alpha \equiv t_{fp}^+/t_{fp}^-$.  Note that $t_{Tb}$ is bounded from below by $t_h$, because hydrolysis is a necessary intermediate step after trailing head detachment, whereas it is not involved after leading head detachment.  The probabilities that the motor takes a forward step (${\cal P}_{f}$), trailing stomp (${\cal P}_{Ts}$), leading stomp (${\cal P}_{Ls}$) and backward step (${\cal P}_{b}$) are:
\begin{equation}\label{e3}
\begin{split}
{\cal P}_{f} &= \frac{g}{1+g}\frac{t_{d1}^2}{(1+b\alpha)(t_{d1}+t_{h})(t_{d1} + t_{Tb} - t_{h})},\\
{\cal P}_{Ts} &=b\alpha {\cal P}_{f}, \quad {\cal P}_{Ls} = \frac{1}{1+g}\frac{b t_{d1}}{(b+\alpha)(t_{d1} + t_{Lb})},\\
{\cal P}_{b} &= b^{-1}\alpha {\cal P}_{Ls},
\end{split}
\end{equation}
where $g \equiv t_{d2}/t_{d1} =8$ quantifies the strength of gating.  These are plotted as a function of load force for the substall regime in Fig.~\ref{myoV_net}B, along with the termination probability ${\cal P}_t = 1- {\cal P}_f -{\cal P}_{Ts} - {\cal P}_{Ls}-{\cal P}_b$.  Forward steps predominate at small forces, but are overtaken by trailing stomps as $F$ approaches the stall value $F_\text{stall} \approx 1.9$ pN, defined as when ${\cal P}_f = {\cal P}_b$.  The force dependence of stomp probabilities has not yet been measured, and so constitutes a prediction of the theory, but there is experimental data on the backward-to-forward ratio ${\cal P}_b/{\cal P}_f$~\cite{Kad2008JBC}.  
\begin{figure}
    \centering
    \includegraphics[width=0.5\textwidth]{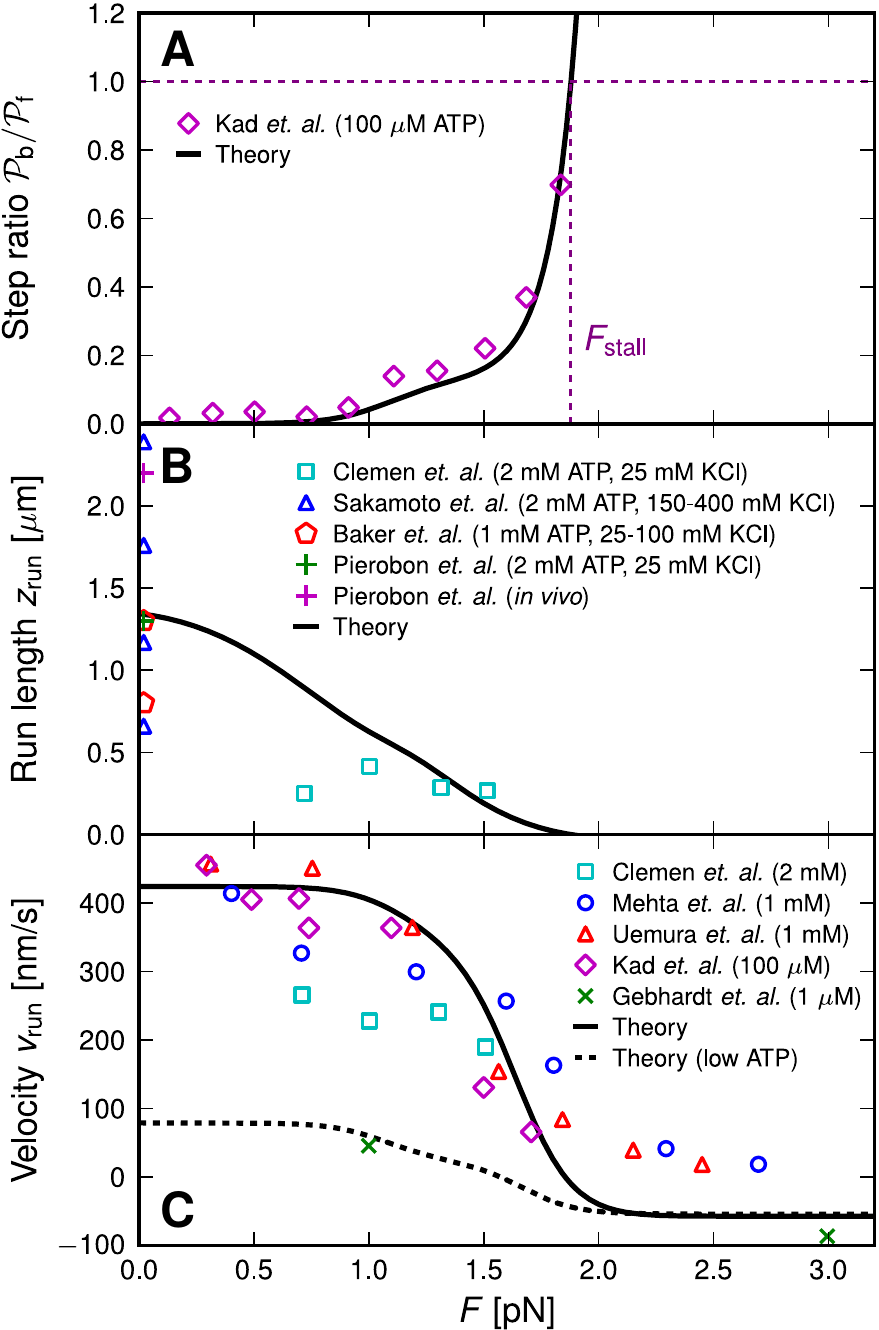}
    \caption{Best-fit theoretical results (solid curves) from the coarse-grained polymer theory for myosin V~\cite{Hinczewski13PNAS} compared to experimental results (symbols) for the following dynamical quantities:  {\bf A.} the ratio of backward to forward steps, ${\cal P}_b/{\cal P}_f$; {\bf B.} the mean run length $z_\text{run}$ on actin before detachment; {\bf C.} mean velocity $v_\text{run}$.  The sources of the experimental data are listed in the legends (Baker {\it et  al.}, 2004; Clemen {\it et  al.}, 2005; Gebhardt {\it et  al.},  2006;  Kad {\it et  al.}, 2008;  Mehta {\it et  al.}, 1999; Pierobon {\it et al.}, 2009; Sakamoto {\it et al.}, 2000; Uemura {\it et al.},  2004) with buffer conditions in parentheses (first value is ATP concentration, second value is KCl concentration).  Where the KCl value is not listed, the concentration is 25 mM.  All experiments / theory are for saturating ATP conditions except (Gebhardt {\it et al.}, 2006) in panel C, where the theory has a modified $t_{d1}^{-1}$ detachment rate to account for low ATP.  Adapted from~\citet{Hinczewski13PNAS}.}
    \label{myoV_exp}
\end{figure}
This is shown in comparison to the theoretical curve in Fig.~\ref{myoV_exp}A, and exhibits excellent agreement.  Other experimentally observable quantities are compared in the remaining panels of Fig.~\ref{myoV_exp}:  B) the mean run length $z_\text{run}$ along the actin before termination, from \cite{Clemen05,Sakamoto00,baker2004myosin,Pierobon09}; C) the mean velocity $v_\text{run} = z_\text{run}/t_\text{run}$, where $t_\text{run}$ is the mean run duration, from \cite{Clemen05,Mehta99Nature,Uemura04,Kad2008JBC,Gebhardt2006PNAS}.  The theoretical expressions for these are:
\begin{equation}\label{e4}
\begin{split}
z_\text{run} &= v_\text{run} t_\text{run}, \quad v_\text{run} \approx \frac{\Delta}{t_\text{d1}}\left(\frac{1}{1+b\alpha} - \frac{\alpha}{g(b+\alpha)}\right),\\
t_\text{run} &\approx \frac{g t_\text{d1}^2}{t_\text{Lb}+ g t_\text{Tb}}.
\end{split}
\end{equation}
The theory curves in Fig.~\ref{myoV_exp} are simultaneous best-fits with only two free parameters, $\nu_c$ and $b$, and overall show that the theory quantitatively captures the main features of the motor dynamics (within experimental uncertainties evident in the scatter of data points collected under different buffer and ATP condtions).  Equally important, the theory provides insights into how motor properties depend on both kinetic parameters (the gating ratio $g$, the reduction in binding rates described by $b$ due to unfavorable lever arm conformations), and structural parameters ($L$, $l_p$, $\theta_c$, and $\nu_c$ that control the distribution during the diffusive search).  This allows clear connections to mutation experiments that perturb the latter, for example by extending or shortening lever arm length to change $L$~\cite{Sakamoto2005Biochem,Oke2010PNAS} (see~\cite{Hathcock19} for a fuller discussion). 

\section{Simulations using Coarse-Grained Models}
Several models that can be simulated readily \cite{Hyeon07PNASa,Hyeon07PNAS,Tehver2010Structure,Zhang12Structure,Nam16PLOSOne,Mugnai17PNAS,Zhang17PNAS,Alhadeff17PNAS,Mukherjee13PNAS,Mukherjee17PNAS,Goldtzvik2018Structure} have been introduced in order address a variety of issues related to stepping kinetics. These include but are not restricted to  the mechanism of stepping of conventional kinesis on microtubule, gating mechanism, and allosteric transitions by which the motor heads communicate with each other and the cytoskeletal filaments. The simulations have to be based on coarse-grained models \cite{Hyeon11NatComm} because of the long time scales and interplay of multiple length scales involved in the motor motility. 

Rather than survey the findings in all these studies, which vary greatly in both details and foci, we describe a particularly illuminating minimal CG mechanochemical model \cite{Craig09PNAS} for myosin V, which beautifully illustrates strain-mediated gating mechanism (required for maintaining processivity) as well as characteristics such as speed and stall force.   
\begin{figure}
    \centering
    \includegraphics[width=0.5\textwidth]{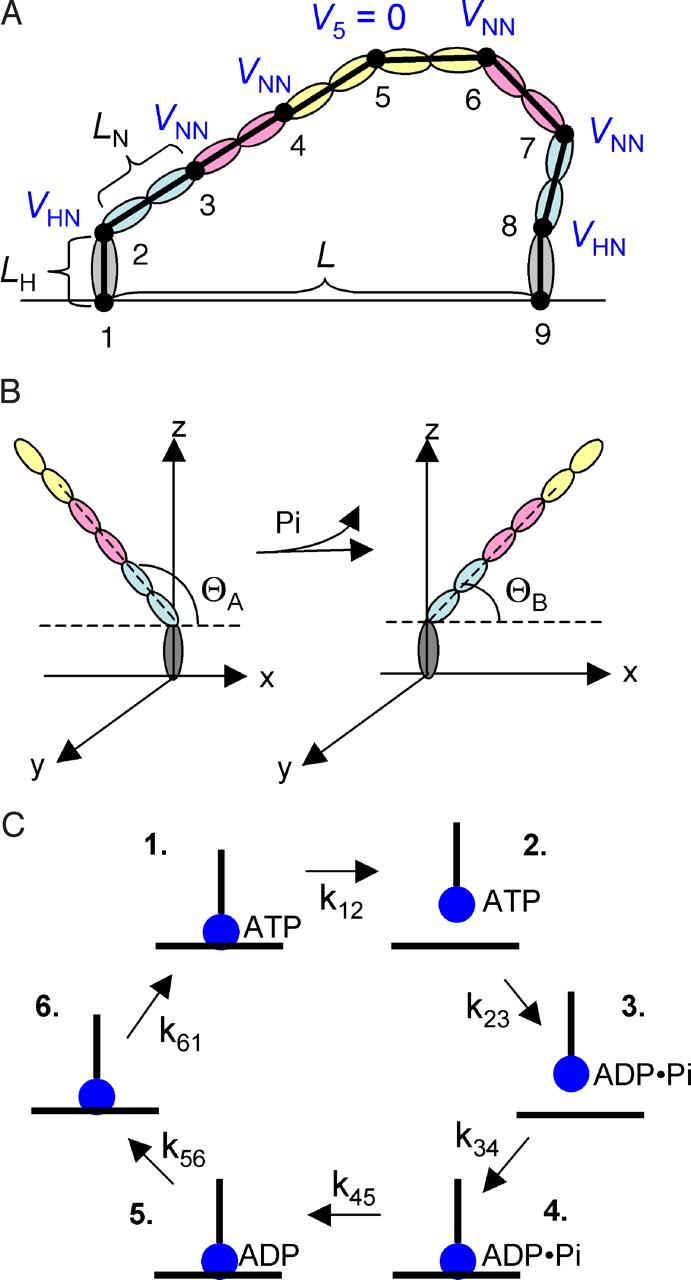}
    \caption{Illustrating a general strategy to construct and simulate a coarse-grained model for myosin V \cite{Craig09PNAS}. The strategy involves making a coarse-grained model based on the structure (see (A) and (B)), which is coupled to the enzyme chemistry given in (C). (A) The lever arm is represented using three rigid segments that are connected to each other. The two lever arms meet at point 5, which rotates freely during the stepping process. (B) The angle between the  lever arm and the head (points 2 and 8) is assumed to change from $\Theta_A$ to $\Theta_B$ upon phosphate release. (C) The reaction cycle with various rates indicated in the figure is coupled to the mechanical model. Brownian dynamics simulations were used to calculate the observable quantities. The figure was reproduced from~\citet{Craig09PNAS}.}
    \label{Craig-Linke}
\end{figure}
The CG model (see Fig. \ref{Craig-Linke}) , which takes the architecture of myosin V in account, was constructed using the following assumptions. First,  the level arm is treated as a semiflexible polymer by representing the six IQ motifs by three interacting moieties. In this sense the model is similar to the subsequent analytical polymer model~\cite{Hinczewski13PNAS}, described above. Second, the junction between the head and the adjacent IQ motifs (points 2 and 8 in Fig. \ref{Craig-Linke}) was treated as a semiflexible joint with equilibrium angles that depend on the nucleotide state of the motor. Such an assumption is justified by comparison to EM images. The forward rotation of the lever arm, which changes the angle from $\Theta_A$ to $\Theta_B$, is taken to be dependent on phosphate release, a crucial step in the reaction cycle of myosin V. Third, the joint between the two lever arms is assumed to fully flexible, which would imply that  the tethered head diffuses freely about this joint (see also \cite{Hinczewski13PNAS}). It should be noted that recent experiments suggest that this may not be the case. It has been pointed out that the angle between the lever arms is constrained \cite{Andrecka15eLife}, which has to be taken into account in describing the diffusive search \cite{Hathcock19}. Fourth, the filamentous actin is treated as a passive one dimensional track with binding sites that are space $\Delta =36$ nm apart, as was also assumed in the  analytical theory \cite{Hinczewski13PNAS}. More importantly, if the tethered head, which has undergone hydrolysis,  diffuses close to a binding site it interacts with the binding site with an attractive electrostatic interaction in order to complete a step. 

A potential energy function based on the mechanical model, which can be simulated using Brownian dynamics, is coupled to the catalytic cycle in one of the motor heads. In order to produce realistic dynamics various rates in the cycle were taken from  experiments (see Table 1 in \cite{Craig09PNAS}). The simulations were successful in reproducing the run length distribution and the value of the stall force ($f_S \approx 2-3$ pN). One of the advantages of the CG simulations is that the stiffness of the head-neck and neck-neck joints, encoded by the terms $V_{HN}$ and $V_{NN}$ could be changed to assess the effect on the motor properties. For example, they discovered that the value of $f_S$ depends on the stiffness of the lever arm. It has to be stiff but not overly so in order to reproduce the experimental data, which was later confirmed theoretically \cite{Hinczewski13PNAS}. 

The strategies used in the models described in this section and the previous one are the following. First, the domains that execute mechanical movements are modeled using available structural data. Second, the mechanical model is coupled to the catalytic cycle, which allows one to predict the dependence of measurable quantities as a function of control parameters such as ATP concentration and external load. The level of coarse-graining in the first step is largely guided by intuition. In \citet{Hinczewski13PNAS} the use of polymer representation afforded analytic solution whereas by discretizing the level arm using discrete connected links Craig and Linke \cite{Craig09PNAS} had to resort to numerical simulations. It is this general strategy that is likely to be successful in tackling the nuances of dynein stepping and perhaps motor functions {\it in vivo}.

\section{Cost-precision Trade-off and Efficiency of Molecular Motors}

\subsubsection{Cost-precision trade-off and its physical bound of molecular motors}
\begin{figure}
\centering
\includegraphics[width=0.5\textwidth]{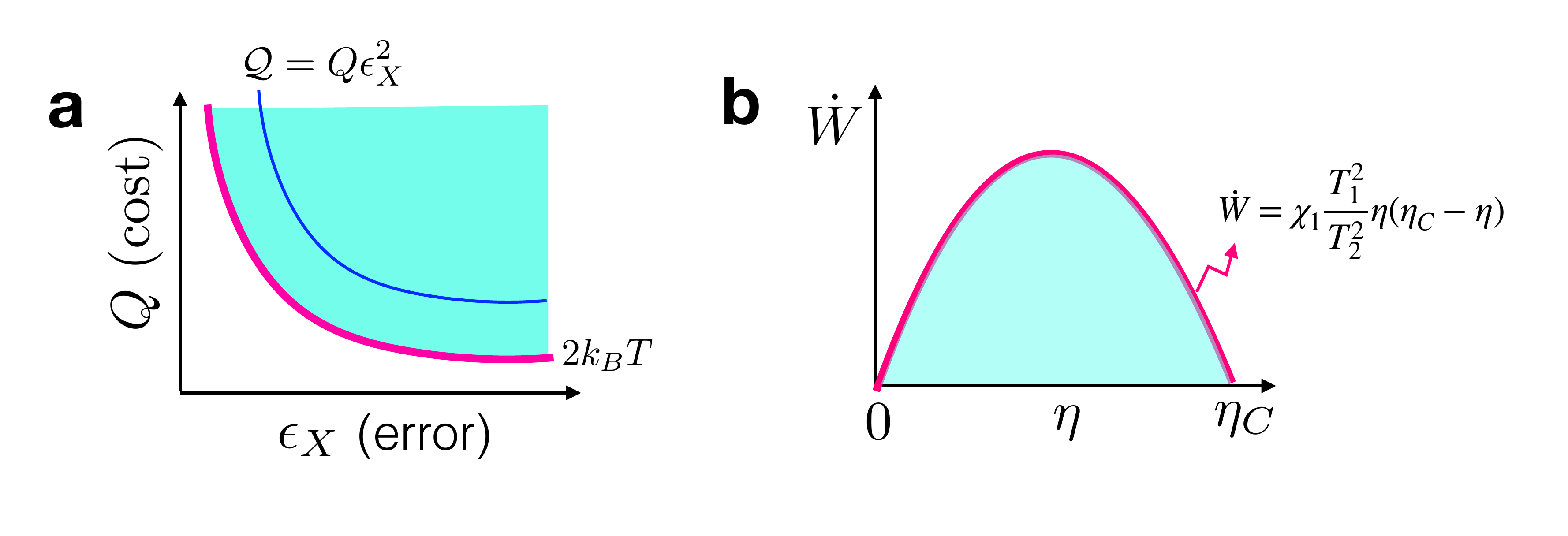}
\caption{{\bf a}. Cost-error tradeoff relation and its physical bound, $\mathcal{Q}=Q\epsilon_X^2>2k_BT$. Accessible region for $\mathcal{Q}$ is in cyan. 
{\bf b}. Power-efficiency tradeoff. Accessible region for power ($\dot{W}$) is demarcated in cyan.  
}
\label{tradeoff}
\end{figure}
From the perspective of thermodynamics, 
biological systems are clearly in non-equilibrium, which means energy is constantly injected and dissipated as heat. 
Because they are subject to incessant thermal and non-thermal fluctuations, 
cellular processes are inherently stochastic and error-prone. 
Thus, a plethora of energy-consuming machineries have evolved to fix any error that may be deleterious to biological functions.
In the presence of large fluctuations inherent to cellular processes, harnessing energy into precise motion and suppressing the uncertainty are critical for accuracy in cellular computation. 
Trade-off relations between the energetic cost and information processing 
have been a recurring theme for many decades in biology \cite{Hopfield74PNAS,ehrenberg1980BJ,bennett1982IJTP,AlbertsBook,mehta2012PNAS,Lan2012NaturePhysics,banerjee2017PNAS}. 
Recently, a concise and fundamental relationship relating cost-precision trade-off and its physical bound, which is called the thermodynamic uncertainty relation (TUR), was first  conjectured by Barato and Seifert \cite{barato2015PRL} and extensively studied in the statistical physics community. 

Barato and Seifert \cite{barato2015PRL} formulated the TUR, 
such that a product ($\mathcal{Q}$) between the heat dissipation ($Q(t)$) and the square of relative error associated with a time-intergrated output observable $X(t)$ of the process, $\epsilon^2_X(t)=\langle\delta X^2\rangle/\langle X\rangle^2$, is independent of measurement time.  
Based on numerical results that extensively sampled the rate constants $k_{ij}$ defining the diverse kinetic networks  
and the linear response theory. They further conjectured that $\mathcal{Q}$ cannot be smaller than $2k_BT$ for any chemical kinetic network described by Markov jump processes, which is succinctly written as,   
\begin{align}
	\mathcal{Q}=Q(t)\times \epsilon_X^2(t)\geq 2k_BT. 
	\label{eqn:inequality1}  
\end{align}
The trade-off parameter $\mathcal{Q}$ quantifies the energetic cost for a given error and is bounded below by  $2k_BT$.  
The time-integrated output observable $X(t)$ can be selected such that it can best represent the dynamic process of interest. For enzyme reactions that catalyze substrate to product, $X(t)$ could be the product concentration, $c(t)$. 
For molecular motors moving along one-dimensional track, displacement (travel distance) $x(t)$ is a natural output observable to represent their dynamic processes. 

For  motors, the relative error associated with the motor displacement decreases with time as  $\sqrt{(\delta x(t))^2}/\langle x(t)\rangle\propto 1/\sqrt{t}$. 
Thus, if one were to decide the displacement of a motor precisely, a longer time trace should be generated, which demands more  free energy injection (ATP hydrolysis) and heat dissipation.   
The greater the heat dissipated from the process, the smaller is the error. 
For molecular motors with output observable $x(t)$, 
Eq.\ref{eqn:inequality1} can be written in terms of three quantities (heat dissipation rate $\dot{Q}$, diffusivity $D$, velocity $V$) that depend on control parameters such ATP concentration and external load. 
\begin{align}
\mathcal{Q}=Q(t)\times \frac{\langle \delta x(t)^2\rangle}{\langle x(t)\rangle^2}=\dot{Q}\frac{2D}{V^2}\geq 2k_BT. 
\label{eqn:inequality2}
\end{align}
Notice that $\mathcal{Q}$ depends on a specific type of motor as well as the conditions of $[\text{ATP}]$ and $f$.  

Since the Barato and Seifert's original conjecture, there has been impressive progress in the field. 
TUR has been reinterpreted as the inequality relation between generalized current and total entropy production rate ($\sigma^{tot}=dS/dt$), which can be written as 
\begin{align}
\sigma^{tot}\frac{\text{Var}(j)}{\langle j\rangle^2}\geq 2. 
\label{eqn:entropybound}
\end{align}  
General and elegant proofs for TUR have not only been given for the case of Markov jump processes on kinetic networks by employing the large deviation theory \cite{Gingrich2016PRL}, but also can be deduced from the equality relation for the Fano factor of entropy production for over-damped Langevin processes \cite{Pigolotti2017PRL}. 

More recently, Dechant and Sasa \cite{dechant2018PRE} have generalized Eq.\ref{eqn:entropybound} to underdamped processes in the following form 
\begin{align}
\sigma^{tot}\geq B\langle j\rangle^2. 
\label{eqn:dechant}
\end{align}
where $B(>0)$  is a model dependent parameter, which can be reduced to $2/\text{Var}(j)$ for over-damped Langevin systems or Markov jump processes on networks.  
In fact, right hand side of the inequality is always greater than zero, namely, it recovers the second law of thermodynamics $\sigma^{tot}\geq 0$. Therefore, one interpretation of 
Eq.\ref{eqn:dechant} is that it provides tighter bound to the entropy production in terms of the square of generalized current.  
Although $B$ is not specific but model dependent, Dechant \emph{et al.}\cite{dechant2018PRE} employed the above relation to derive a power-efficiency trade-off for engine operating between two heat sources with temperature $T_1>T_2$. 
\begin{align}
\dot{W}\leq\chi_1\frac{T_1^2}{T_2^2}\eta(\eta_C-\eta). 
\end{align}
where $\eta=\dot{W}/\dot{Q_1}$ and $\eta_C=1-T_2/T_1$ are the thermodynamic efficiency and Carnot efficiency.  
The inequality gives the upper bound to the power generated using the two heat sources. 
But, when $\eta$ approaches to $\eta_C$, $\dot{W}$ approaches to 0 as well if the model-dependent parameter $\chi_1$ is finite.  

For pedagogical purpose, it is worthwhile to consider simple examples that can demonstrate the significance of the physical bound of TUR.  

(i) In the context of the foregoing one-state hopping model to which dynamics of molecular motors can be mapped, the rate of heat dissipation from the process is bounded as, 
\begin{align}
\frac{\dot{Q}}{k_BT}=(u-w)\log{\frac{u}{w}}\geq \frac{2(u-w)^2}{(u+w)}
\end{align}
where the inequality was discussed in Ref. \cite{shiraishi2016PRL}. 
Given a single step displacement $d_0$, 
the velocity and diffusivity of the particle moving along the reaction coordinate is $V=d_0(u-w)$ and $D=(d_0^2/2)(u+w)$, respectively. 
Therefore, inserting the expressions of $\dot{Q}$, $V$, and $D$ to Eq.\ref{eqn:inequality2} it is easy to see that $\mathcal{Q}\geq 2k_BT$. 
The lower bound of $\mathcal{Q}$ in this model is attained when $u=w$, which corresponds to the detailed balance (equilibrium) condition. 
Provided that $u=u^*[S]$ and $v=v^*[P]$, namely, the forward and backward rate constants vary with the substrate ($S$) and product ($P$) concentrations, the detailed balance condition is attained when $[S]/[P]=[S]_{eq}/[P]_{eq}$ with $u^*/v^*=[P]_{eq}/[S]_{eq}$.   

(ii) Hyeon \emph{et al.} \cite{Hyeon2017PRE} studied the over-damped Lagevin motion on tilted washboard potential, which obeys 
\begin{align}
\gamma \dot{x}(t)=f-U'(x)+\xi(t)
\end{align}
with $U(x+L)=U(x)$ and $\langle \xi(t)\rangle=0$, $\langle\xi(t)\xi(t')\rangle=2\gamma k_BT\delta(t-t')$, where the driving of quasi-particle on the potential is controlled by the non-conservative force $f$. 
They showed that TUR parameter $\mathcal{Q}(f)$ for this problem is a non-monotonic function of $f$, attaining its physical bound $2k_BT$ at both $f\ll |U'(x)|$ (near equilibrium) and $f\gg |U'(x)|$ (far from equilibrium). 
For $f\ll |U'(x)|$, the quasi-particle undergoes diffusion on a rough surface. 
On the other hand, for $f\gg |U'(x)|$ the particle slides along a smooth gradient, without feeling the effect of confining potential, dissipating energy with a rate $\dot{Q}=\gamma V^2$ against friction.  
Since the diffusivity of particle in this case follows the Stokes-Einstein $D=k_BT/\gamma$, together with the velocity $V$ one can obtain $\mathcal{Q}=2k_BT$. 
$\mathcal{Q}(f)$ reaches its maximum value near the critical point where the potential barrier confining the particle is about to vanish. 

(iii) As far as the specific models are concerned as in (i) and (ii), the minimal uncertainty condition ($\mathcal{Q}_{\text{min}}=2$ $k_BT$) is attained not only under the detailed balance condition but when there is no confining potential.   
It has been suggested that $\mathcal{Q}_{\text{min}}$ is attained when the heat dissipated from the process is normally distributed as $P(Q)\sim e^{-(Q-\langle Q\rangle)^2/(2\langle\delta Q^2\rangle)}$ and that TUR measures the deviation of heat distribution from Gaussianity \cite{Hyeon2017PRE}. 
\\

\subsubsection{Transport Efficiency}
There are a number of ways to assess the ``efficiency'' of engines or machines \cite{Brown:2017:PNAS}. 
Historically, the efficiency of heat engines has been discussed in terms of the thermodynamic efficiency, the aim of which is to maximize the amount of work extracted from two heat sources with different temperatures \cite{CallenBook}. 
For non-equilibrium machines driven by chemical forces that are constantly regulated without shortage in the live cell, the power production could be a more pertinent quantity to maximize. 
Meanwhile, for transport motors in the cell, 
the TUR parameter $\mathcal{Q}$ can be used to assess the efficiency of suppressing the uncertainty in dynamical process by means of energy consumption, and thus is quite pertinent for evaluating the transport efficiency of a motor (or motors) \cite{dechant2018JStatMech}. 
The connection between $\mathcal{Q}$ and the transport efficiency is clear. 
If a motor transports cargos at a high speed ($V\sim \langle x(t)\rangle/t$) with small fluctuations ($D\sim \langle\delta x(t)^2\rangle/t$) (which leads to punctual delivery to a target site) but consuming only a small amount of energy ($\dot{Q}$), such a motor would be considered efficient for cargo transport.    
A motor efficient in the cargo transport would be characterized by a small $\mathcal{Q}$ with its minimal bound $2k_BT$, or as originally suggested by Dechant and Sasa \cite{dechant2018JStatMech}, one can consider using the definition $\eta_T=2k_BT/\mathcal{Q}$ which is bounded between 0 and 1. 

$\mathcal{Q}$ can be used to assess the ``transport efficiency'' of biological nanomachines and to study how it changes with varying conditions of $f$ and [ATP]. 
To evaluate $\mathcal{Q}$, measurement should be first carried out for $\dot{Q}$, $D$, and $V$ (see Eq.\ref{eqn:inequality2}). 
While $V$ and $D$ are straightforward to calculate ($V=\lim_{t\rightarrow\infty}dx(t)/dt$ and $D=\lim_{t\rightarrow\infty}(1/2) d(\delta x(t))^2/dt$), experimental measurement of $\dot{\mathcal{Q}}$ may be nontrivial. 
Although there are some reports on direct measurements of heat dissipation rate at single cell level \cite{song2019CurrBiol,rodenfels2019DevelCell}, direct measurement of heat dissipation at single molecule level is not yet known. 
Nevertheless, $\dot{Q}$ be estimated by considering a \emph{physically suitable minimal} kinetic network model. 
For a given cyclic kinetic network defined by multiple chemical states ($i=1,2,\ldots N$) and transition rates ($\{k_{ij}\}$) connecting them, 
there are straightforward methods \cite{Koza1999JPA,Lebowitz:1999,Hwang2017JPCL} to associate the measured $V$ and $D$ with $\{k_{ij}\}$. 
As long as all the rate constants $\{k_{ij}\}$ defining the kinetic network are known, it is then straightforward to calculate $\dot{Q}$ value as discussed in details in the early part of this review. 

\begin{figure}
\centering
\includegraphics[width=0.5\textwidth]{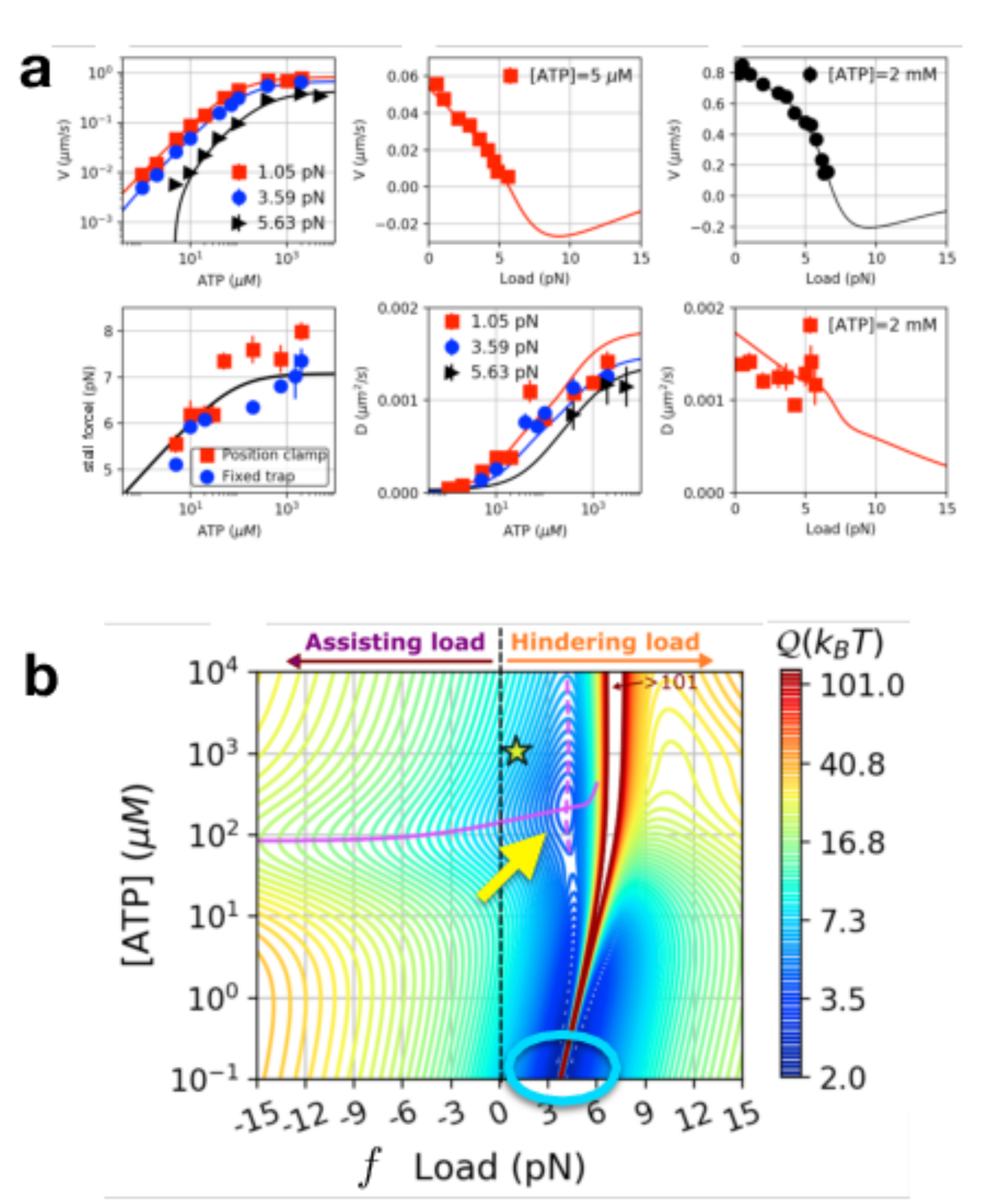}
\caption{{\bf a}. Analysis of motility data to determine the parameters for rate constants ($\{k_{ij}\}$ in Eq.\ref{eqn:kij_f}) for double cycle network model. 
{\bf b}. Diagram of $\mathcal{Q}$ as a function of $\cATP$ and $f$.}
\label{Q_diagram}
\end{figure}

\begin{figure}
\centering
\includegraphics[width=0.5\textwidth]{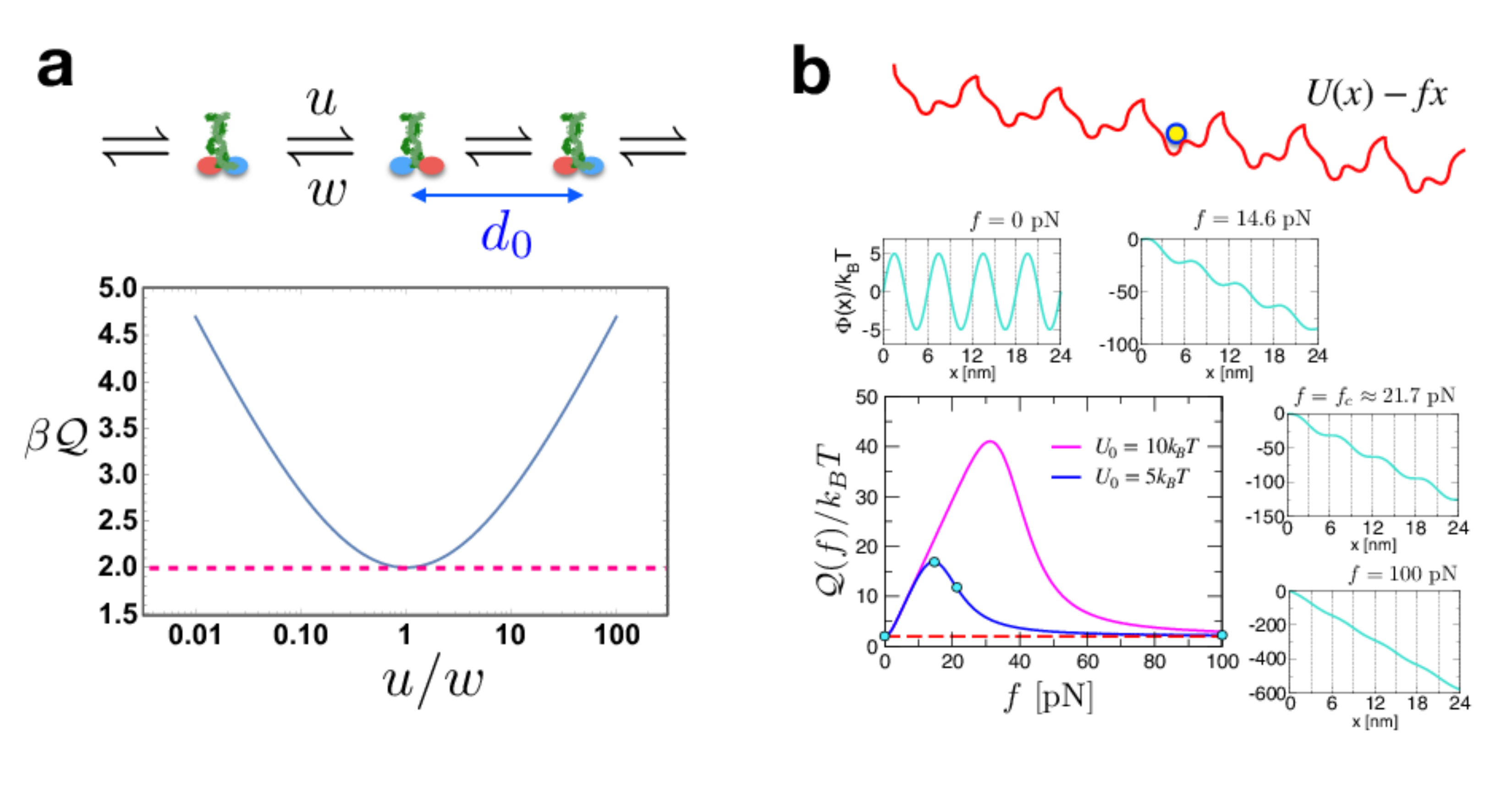}
\caption{Two case studies for TUR. {\bf a}. One-state hopping model with forward and backward rate constants $u$ and $w$. 
{\bf b}. Brownian motion in tilted wash-board potential. $\mathcal{Q}(f)$ was evaluated as a function $f$ using a specific periodic function, $U(x)=U_0\sin{2\pi x/L}$ with $L=6$ nm for $U_0=5$ and $10$ $k_BT$.}
\label{TUR_examples}
\end{figure}

For conventional kinesin, whose chemomechanical properties were extensively studied by several groups, motility data for varying $\cATP$ and load conditions are available in the literature. 
The double-cycle network model, depicted in Fig.\ref{double_cycle}b, with the form of rate constant for each edge can be used to analyze those data, which allows one to determine the dependence of $\{k_{ij}\}$ on $\cATP$ and $f$ and finally to build a diagram of $\mathcal{Q}(f,[\text{ATP}])$ as shown in Fig.\ref{Q_diagram}. 
The TUR diagram, $\mathcal{Q}([ATP],f)$ (Fig.\ref{Q_diagram}), exhibit several features that are worthwhile to explore:  

(i) In terms of the load direction, assisting ($f<0$) and hindering ($f>0$), $\mathcal{Q}([ATP],f)$ is asymmetric. 
This result differs from that of the one-state hopping example for TUR calculated in Fig.\ref{TUR_examples}a. 
The fundamental difference arises from the asymmetric effect of the force on the kinetic rate, particularly on $k_{25}$ and $k_{52}$ with $\theta\neq 1/2$. 
 
(ii) $\mathcal{Q}$ is locally minimized first at a condition of small load ($f\gtrsim 0$) and low $\cATP$ (indicated by cyan ellipse), and second at $f\approx 3$ pN and $\cATP\approx 300$ $\mu M$ (indicated by the yellow arrow). 
The first minimum is closer to 2 $k_BT$ bound; yet, this minimum was attained near the detailed balance condition where $\cATP$ concentration is small and balanced with $\cADP$ and $[\text{P}_i]$. 
It does not have much of biological relevance given that molecular motor works out of equilibrium. In fact, the second local minimum is of particular interest given that the condition is not so far from the cellular condtion $\cATP\approx 1$ mM and $f\approx 1$ pN, indicated by yellow arrow (Here, note that $f\approx 1$ pN is a rough estimate of cellular environment replete with obstacles such as cytoskeletal filaments and road blocks). 
The second local minimum is the very point at which the transport efficiency defined in terms of $\mathcal{Q}$ is optimized. 

(iii) The high $\mathcal{Q}$ region ($\mathcal{Q}>100$ $k_BT$) at $f\approx 3-7$ pN is due to the stall condition. As explained in detail in rationalizing the double-cycle network model, 
energy should be still consumed and heat should be dissipated ($\dot{Q}>0$) at stall condition ($V\approx 0$). This particular condition renders $\mathcal{Q}$ divergent at stall conditions.

\begin{figure}
\centering
\includegraphics[width=0.5\textwidth]{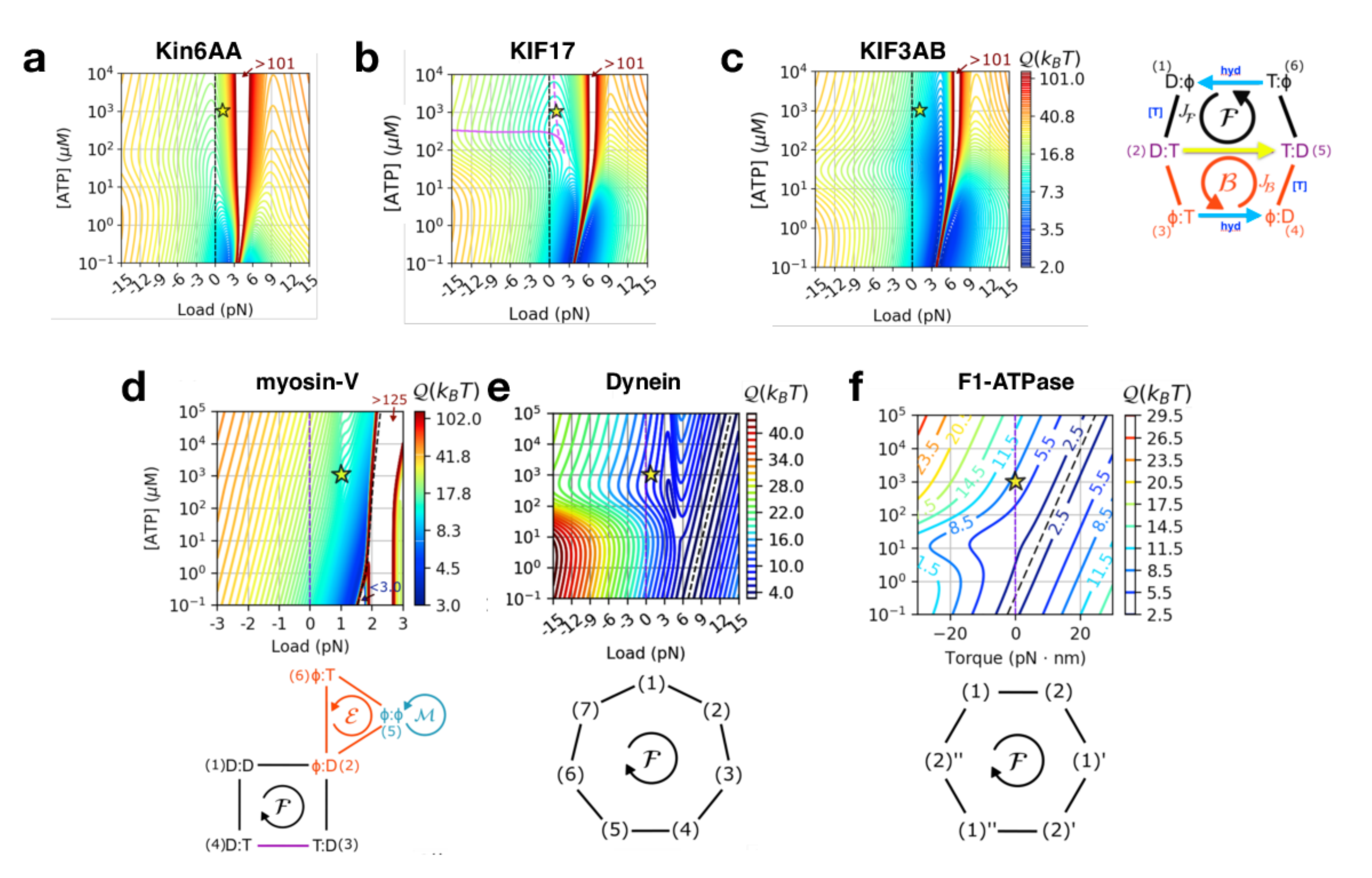}
\caption{$\mathcal{Q}(f,[\text{ATP}])$ for {\bf a}. Kin6AA (kinesin-1 mutant); {\bf b}. KIF17;  {\bf c}. KIF3AB; {\bf d}. myosin V; {\bf e}. Dynein; {\bf f}. F$_1$ATPase. 
The diagrams were calculated by directly analyzing the motility data available in \cite{clancy2011nsmb} for {\bf a}, \cite{milic2017PNAS} for {\bf b}, {\bf c}, and using the same double-cycle network model as that of kinesin-1.  
For other motors in {\bf d}, {\bf e}, and {\bf f}, the kinetic network and rate constants were used as in the Ref. \cite{Bierbaum2011BJ} for myosin V, Ref.~\cite{sarlah2014BJ} for dynesin, and Ref.\cite{gerritsma2010BRL} for F$_1$-ATPase.  
}
\label{other_motors}
\end{figure}

Finally, the structure of $\mathcal{Q}(f,[\text{ATP}])$
is sensitive to the design of motor structure as well as the motor type. 
First, the diagram of $\mathcal{Q}(f,[\text{ATP}])$ for a kinesin construct, Kin6AA, whose necklinker is engineered longer than that of wild-type kinesin-1 via insertion of six amino-acids (AEQKLT) \cite{clancy2011nsmb} exhibits great deviation from that of the WT (Fig.\ref{other_motors}a). 
In all, the values of $\mathcal{Q}$ were increased, the stall forces were reduced, and the local minimum observed in the WT is missing.  
KIF17 and KIF3AB \cite{milic2017PNAS} show qualitatively similar structure of $\mathcal{Q}(f,[\text{ATP}])$; however there are different in terms of quantitative details from that of WT. 
Next, myosin V, dynein, and F$_1$-ATPase were analyzed to calculate $\mathcal{Q}(f,[\text{ATP}])$. 
Bierbaum and Lipowsky \cite{Bierbaum2011BJ} employed a tri-cyclic network model to describe the chemomechanics of myosin V in which cycles for forward steps, energy-consuming futile steps, and force-induced mechanical slippage steps were considered. The resulting $\mathcal{Q}(f,[\text{ATP}])$ (Fig.\ref{other_motors}d) shows ATP-insensitive stall condition where $ \mathcal{Q}$ is divergent with no local minimum as in WT kinesin-1. 
For dynein (Fig.\ref{other_motors}e), the uni-cyclic chemomechanical network model adopted for construction of $\mathcal{Q}(f,[\text{ATP}])$ is in principle not satisfactory, since it would be able to account for the physically correct behavior of dynein dynamics at stall and superstall conditions. In particular, $\mathcal{Q}(f,[\text{ATP}])$ diagram shows minimization to $2k_BT$ at the stall. Nevertheless, a suboptimal local minimum is found at $f\approx 3$ pN and $\cATP\approx 300$ $\mu M$.  
Finally, $\mathcal{Q}(\tau,[\text{ATP}])$ (torque $\tau$ instead of load $f$) for F$_1$ATPase is shown in Fig.\ref{other_motors}f calculated based on uni-cyclic kinetic network model \cite{gerritsma2010BRL}. 
For F$_1$-ATPase, which shows near-reversibility and hence characterized 
   with high efficiency \cite{Toyabe2010PRL}, the use of uni-cyclic network model would be reasonable although this conclusion should be reached with care (see \cite{sumi2019NanoLett}). 
   
\begin{figure}
\centering
\includegraphics[width=0.5\textwidth]{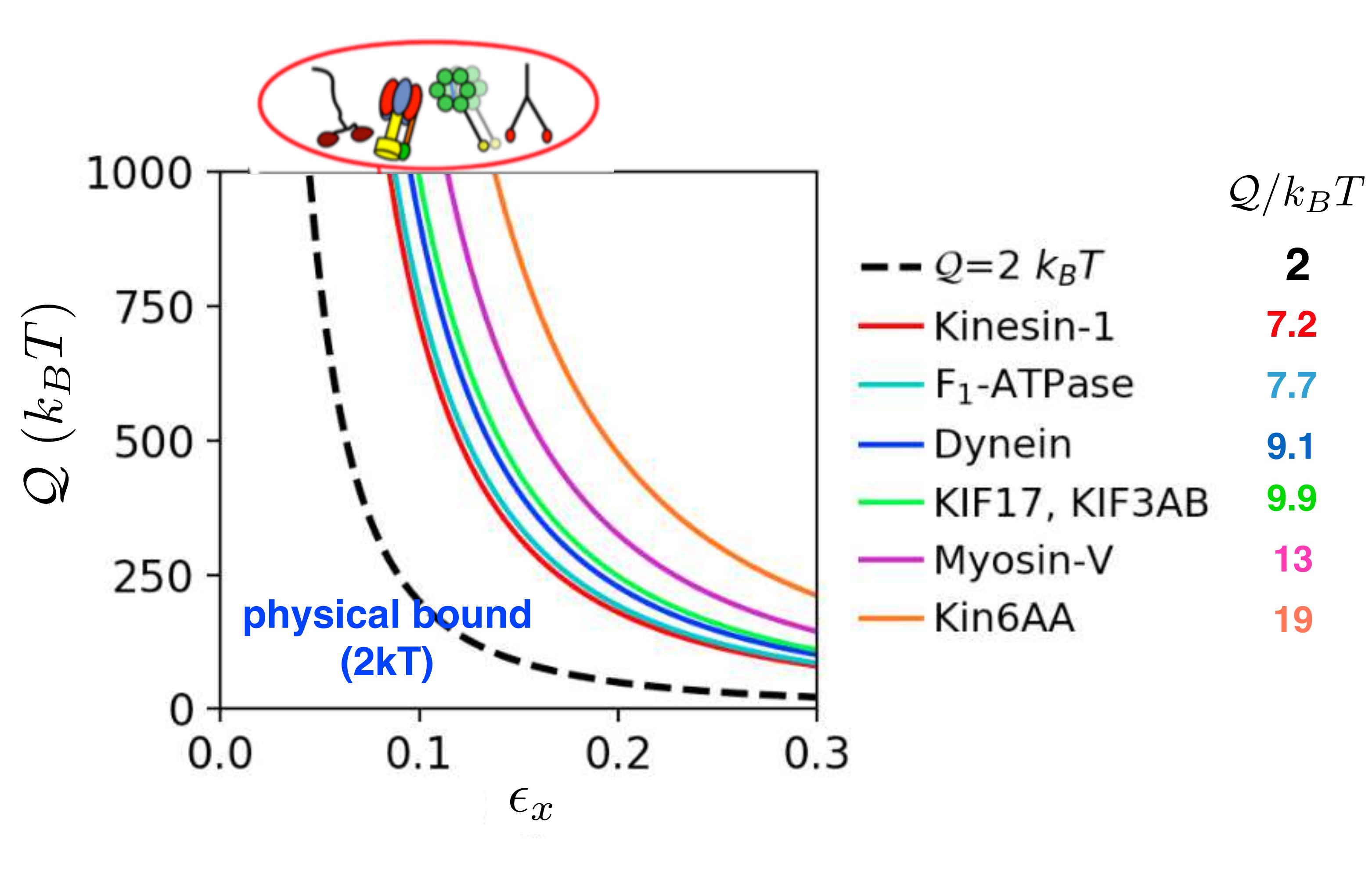}
\caption{$\mathcal{Q}$ values for various transport motors calculated at $\cATP=1$ mM and $f=1$ pN (except for F$_1$-ATPase calculated at $\cATP=1$ mM and $\tau=0$ pN$\cdot$nm)
}
\label{motors_compare}
\end{figure}

Some of the biological motors, kinesin family and dynein, studied here are found to be semi-optimized in terms of $\mathcal{Q}$ under the cellular condition, which alludes to the role of evolutionary pressure that has shaped the present forms of molecular motors in the cell.  
In addition, 
the efficiency quantified in terms of $\mathcal{Q}$ for various molecular machines presented here are ranged between 7 and 20 $k_BT$ (Fig.\ref{motors_compare}). 
Given that all these machines function out-of-equilibrium, it is of great interest to discover that the value of $\mathcal{Q}$ calculated at the working cellular condition is not significantly far from its physical bound 2 $k_BT$. This may arise from the fact that the molecular machine analyzed here are tightly coupled machine, meaning that ATP hydrolysis is almost always transduced to a machanical step even though energy-wasting futile steps still remain as a possibility. 

As long as the underlying mechanism, which offers a clear model for chemomechanical kinetic network, is known, TUR can be studied for any time trace generated from cyclic process at NESS. 
Other energy-intensive processes, for example, error-correction processes, circadian cycle, and chaperonin action, would exhibit high $\mathcal{Q}$.
\\

\section{Molecular chaperones} 
Chaperones are another class of cellular molecular machines that expend free energy change associated with ATP binding and catalysis to facilitate the folding of certain  proteins and RNA that cannot fold spontaneously under cellular conditions. Because they serve a key function in maintaining protein homeostasis they are deemed as essential for the survival of the organisms. 
Among a large class of molecular chaperones, the function of the heat-shock GroEL-GroES machinery found in {\it E.Coli}, referred to as cheperonin, is now quantitatively understood thanks to experimental and theoretical advances. The {\it in vivo} function of  GroES-GroEL machinery in {\it E. Coli.} is to rescue substrate proteins that are otherwise destined for aggregation. Just like molecular motors, GroEL undergoes a catalytic cycle involving a  series of large scale structural changes in response to ATP binding, hydrolysis, and release of ADP and phosphate. 
GroEL/GroES machine anneals the population of misfolded proteins driving them to the folded state by repeatedly going through rounds of the catalytic cycle, which has been referred to as the Iterative Annealing Mechanism (IAM) \cite{Todd96PNAS,Tehver08JMB}. 

\begin{figure}
	\centering
	\includegraphics[width=0.5\textwidth]{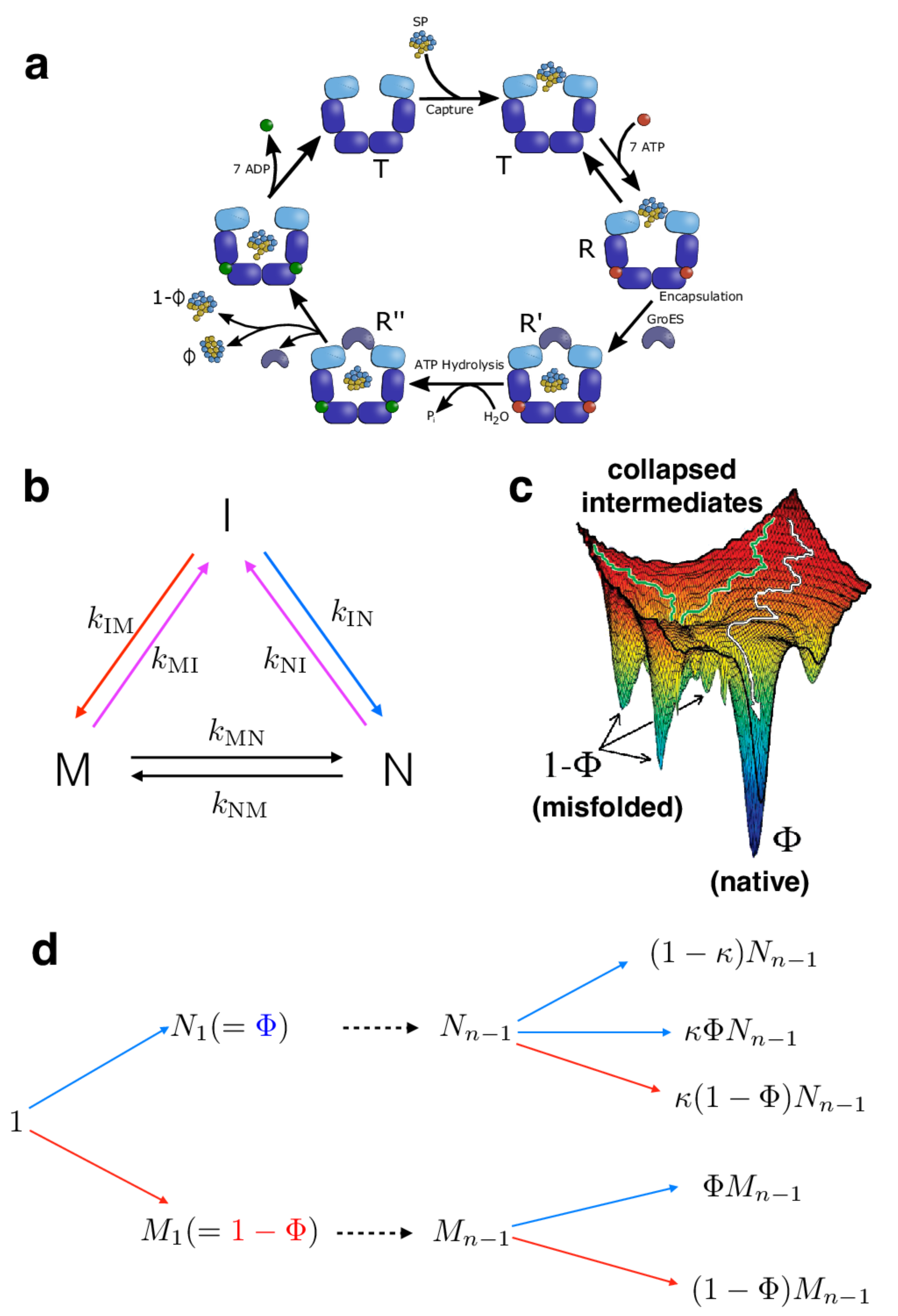}
	\caption{Chaperone-assisted folding of substrate.  
	{\bf a}. Schematic of the iterative annealing mechanism (IAM) illustrated for the hemicycle of GroEL. The $T \rightleftharpoons R$ transition starts when ATP and the substrate protein (SP) bind. 
	GroES binding and ATP hydrolysis engenders the $R^\prime \rightarrow R^{\prime\prime}$ transition with a fraction of the SP partitioning to the folded structure. Subsequently ADP and Pi and the SP (folded or not) are released and the $R^{\prime\prime} \rightarrow T$ transition completes the cycle. In the presence of the SP the machine turns over in about a second and ADP release is accelerated by about a hundred fold. The rapid turnover comports well with the predictions of IAM \cite{Todd96PNAS}.
	{\bf b}. A minimal kinetic network model for chaperone-assisted folding of a substrate molecule. 
Transitions between three manifolds of collapsed intermediate ($I$), misfolded states ($M$), and native state ($N$) are represented in terms of rate constants. 
Folding, misfolding, and chaperone assisted unfolding pathways are depicted in blue, red, and magenta, respectively. 
{\bf c}. Schematic of folding landscape of substrate molecules. 
Upon spontaneous folding, the ensemble of intermediate states collapsed from unfolded ensemble reach the native and misfolded basins of attractions with the proportion of $\Phi$ and $1-\Phi$, respectively. 
{\bf d}. Schematic of the generalized iterative annealing mechanism of chaperone-assisted substrate folding, from which the recursion relation for the native yield from $n$-th annealing process was derived. 
$N_i$ and $M_i$ denote the proportion of native and misfolded states from the $i$-th annealing process. The blue and red arrows represent the pathways leading to the native and misfolded states, respectively. 
		}
	\label{chaperone}
\end{figure}

A minimal kinetic network model of chaperonin-assisted protein folding, 
can be constructed, based on considerable experimental evidence,  by assuming that a protein exists either in an intermediate ($I$), misfolded ($M$), or folded (native) ($N$) state as illustrated in Fig.\ref{chaperone}. 
The unfolded state, a transient state right after the protein is synthesized, collapses to the $I$ state, and it further undergoes a spontaneous folding process via the kinetic pathways with rates denoted by $k_{IM}$ and $k_{IN}$.  Only a fraction ($\Phi$) of the entire population folds correctly to the $N$ state and the remaining fraction ($1-\Phi$) is misfolded to  the $M$ state. The process of the initial population of molecules with $\Phi$ ($1-\Phi$) reaching the folded (misfolded) state is termed the Kinetic Partitioning Mechanism (KPM). The KPM, which was first theoretically proposed~\cite{Guo95BP}, has been used to explain the folding of  of proteins~\cite{Kiefhaber95PNAS} as well as RNA~\cite{Pan97JMB,Thirumalai05Biochemistry}. 
Typically, substrate proteins (SPs or ribozymes) that require assistance from chaperonin action are characterized by extremely small values of  $\Phi$ ($\ll1$), which implies that the majority of population without chaperones are trapped in the misfolded states, which could potentially aggregate unless rescued by the chaperones.  

The function of the GroEL machine is quantitatively explained by the Iterative Annealing Mechanism (IAM) according to which the chaperone recognizes and acts on misfolded substrate preferentially. Typically, the SPs have exposed hydrophobic residues, which make them prone to aggregation unless recognized by the GroEL-GoES machine. When the SPs bind to GroEL they become disordered as a result of domain movements in GroEL, which imparts a mechanical force ($\approx$ 10pN) that is sufficiently large to unfold (at least partially) the SPs \cite{Thirumalai01ARBB}. As the catalytic cycle proceeds the SPs are encapsulated for a brief period (roughly about two seconds) during which a small fraction folds rapidly during the time they are in the cavity.  It is worth remarking that if they fold (the probability being $\Phi$), they do so while being encapsulated in the cavity of GroEL.  When the catalytic cycle is complete, the SP is ejected from the cavity regardless of whether it is folded or not. If the SP is misfolded then it once again recognized by GroEL and the cycle is iterated until sufficient yield of the folded state is obtained. A key requirement of GroEL-SP interaction is that hydrophobic residues of the SPs must  be exposed, which does not typically occur in folded states. It this ability of chaperone not to recognize native proteins that enables the GroEL-GroES machinery to drive the misfolded states to  the native state over repeated iterations of the catalytic cycle.  More specifically, when the annealing process, corresponding to one catalytic cycle, is iterated $n$ times, the total amount of population that reaches the native state (or native yield) grows as, 
\begin{align}
N_n=1-(1-\Phi)^n
\label{eqn:IAM}
\end{align}
with $n\geq 1$. As $n\rightarrow\infty$, $N_n\rightarrow 1$. 

While it is known that GroEL only recognizes misfolded SPs, 
the IAM concept has to be generalized to RNA  chaperones, which act on the both $N$ and $M$ states, but more favorably on $M$ \cite{Bhaskaran07Nature,chakrabarti2017PNAS}. 
If the proportion of $N$ identified by chaperones in comparison with the $M$ state is defined as $\kappa$ ($0\leq \kappa \leq 1$), the total yield of native state after $n$ rounds of folding (annealing) in the presence of chaperone can be calculated using the following mathematical formulation:   

(i) Let $N_n$ and $M_n$ be the yields of native and misfolded state, respectively, after the $n$-th round of the annealing process. Note that the total amount of substrate proteins is conserved at all times, which implies that $N_n+M_n=1$ for all $n$. 
In the first round of annealing, a fraction $N_1(=\Phi)$ folds to the $N$ state and $M_1(=1-\Phi)$ partitions to the $M$ state;   

(ii) In the $n$-th round of annealing, chaperones recognize $N_{n-1}$ and $M_{n-1}$ differentially by the factor $\kappa$. 
Whereas $(1-\kappa) N_{n-1}$ is left unrecognized by chaperones, $\kappa N_{n-1}$ is unfolded and $\Phi$ of them refold to yield $\kappa\Phi N_{n-1}$ native states and $\kappa (1-\Phi)N_{n-1}$ misfolded states.  
On the other hand, the entire population of $M_{n-1}$ is unfolded and $\Phi$ of them refold to yield $\Phi M_{n-1}$ native states and $(1-\Phi)M_{n-1}$ misfolded states. 
Therefore, after the $(n-1)$-th round of chaperone action, the native yield $N_n$ is determined as 
$N_{n}=1-M_n=1-(1-\Phi)M_{n-1}-\kappa(1-\Phi)N_{n-1}$; 

(iii) From the resulting recursion relation of $N_n=(1-\kappa)(1-\Phi)N_{n-1}+\Phi$ with $N_1=\Phi$, 
we obtain the following expression,   
\begin{align}
N_n=\Phi\frac{1-(1-\kappa)^n(1-\Phi)^n}{\kappa+(1-\kappa)\Phi},
\end{align}
native yield after $n$ iterations.  After a sufficient number of iterations ($n\rightarrow \infty$), the system reaches steady state, $N_{\infty}\rightarrow \Phi/(\kappa+(1-\kappa)\Phi)$. 
 In the case of GroEL only the misfolded state is recognized by chaperones ($\kappa=0$).
 Therefore, the GroEL-GroES machinery drives the entire population of substrate proteins to the native state $N_{\infty}\rightarrow 1$ (Eq.\ref{eqn:IAM}). 

As described for molecular motors, the thermodynamic aspects of chaperonin-assisted folding can be succinctly captured by mapping the folding process of substrate molecule onto a kinetic network model. 
A uni-cyclic reversible kinetic network model consisting of the above-mentioned three states ($I$, $M$, and $N$) suffices to capture the non-equilibrium nature of chaperone-assisted protein folding. The relevant master Equation describing the network is,
\begin{equation}
\partial_t{\bf P}(t)=\mathcal{W}{\bf P}(t)
\end{equation}
where ${\bf P}(t)=(P_I(t),P_N(t),P_M(t))^T$ and, 
\begin{align}
{\mathcal{W}}= \begin{pmatrix}
  -(k_{IN}+k_{IM}) & k_{NI} & k_{MI} \\
  k_{IN} & -(k_{NI}+k_{NM}) & k_{MN} \\
  k_{IM} & k_{NM} & -(k_{MI}+k_{MN} 
 \end{pmatrix}\nonumber
\end{align}
The probability $P_i(t)$ of each state $i$ evolves with time as, 
\begin{align}
{\bf P}(t)={\bf P}^{ss}+c_1\vec{u}_1e^{-\lambda_1 t}+c_2\vec{u}_2e^{-\lambda_2 t}
\end{align}
with $\lambda_2>\lambda_1>0$, and it reaches the steady state value $P_i^{ss}$ at $t\rightarrow \infty$. 
The steady state population $P_i^{ss}$ and steady state current along the reaction cycle ($J=k_{ij}P_i^{ss}-k_{ji}P_j^{ss}$) can be expressed solely using the rate constants, as already discussed in depth in the foregoing section describing the general aspects of the various cycles. 
Because the function of the chaperones is to promote the formation of the folded state our primary interest is the steady state yield of the native state, which is given by, 
\begin{equation}
P^{ss}_N=\dfrac{k_{MI}([C],[T])k_{IN}+k_{MN}(k_{IM}+k_{IN})}
{\Sigma([C],[T])},
\end{equation}
where 
\begin{equation}
\begin{aligned}
&\Sigma([C],[T])=\\
&k_{MI}([C],[T])k_{NI}([C],[T])\\
&+k_{NI}([C],[T])(k_{MN}+k_{IM})+k_{MI}([C],[T])(k_{IN}+k_{NM})\\
&+(k_{IM}+k_{IN})(k_{NM}+k_{MN}).
\end{aligned}
\end{equation}
The partition factor $\Phi$ can be expressed in terms of rate constants as $\Phi=k_{\text{IN}}/(k_{\text{IN}}+k_{\text{IM}})$. 
Because of chaperone action, $k_{\text{IN}}$ and $k_{\text{MI}}$ can be significant (in particular $k_{\text{IN}}\gg k_{\text{NI}}$), whereas $k_{\text{NM}}$ and $k_{\text{MN}}$ are negligible. 
Under this condition, the native yield simplifies to:
\begin{align}
P_N^{ss}&\simeq \frac{k_{\text{MI}}k_{\text{IN}}}{k_{\text{MI}}k_{\text{NI}}+k_{\text{NI}}k_{\text{IM}}+k_{\text{MI}}k_{\text{IN}}}\nonumber\\
&\simeq \frac{k_{\text{IN}}}{\dfrac{k_{\text{NI}}}{k_{\text{MI}}}k_{\text{IM}}+k_{\text{IN}}}=\frac{\Phi}{\kappa+(1-\kappa)\Phi}=N_{\infty}
\end{align} 

In the absence of either of chaperone or ATP which redistributes the population of proteins into NESS, 
$[C]$, $[T]$-dependent rate constants vanish ($k_{ij}([C],[T])=0$).  
In this case, the steady state population of $N$ state becomes
\begin{align}
P_N^{ss}([C]=0\text{ or }[T]=0)&=\frac{1}{1+k_{NM}/k_{MN}}\nonumber\\
&=\frac{1}{1+e^{-\Delta G_{NM}/k_BT}}\nonumber\\
&=P_N^{eq}
\label{eqn:equil}
\end{align}
Replacing the two $[C]$, $[T]$-dependent rate constants ($k_{IN}$ and $k_{IM}$) to zero is tantamount to blocking the chaperone action and placing  the $M$ and $N$ states in isolation.  
As long as they are isolated for long enough time greater than $k_{NM}^{-1}$ and $k_{MN}^{-1}$, the system would finally reach the equilibrium native yield ($P_N^{eq}$), as dictated in Eq.\ref{eqn:equil} assuming that aggregation reaction can be neglected. 
However, equilibrating $M$ and $N$ states via the transition paths of $M\rightleftharpoons N$ is impractical in the light of the time scale of cellular processes because $k_{NM}^{-1}$ and $k_{MN}^{-1}$ would far exceed the biologically meaningful time scale.  These arguments suggest that chaperones drive the substrates out of equilibrium. In the process, they optimize the yield of the folded SPs or RNAs per unit time, which we discuss further below.    
\\

\section{Universal characteristics of Helicases}
\begin{figure}
    \centering
    \includegraphics[width=0.5\textwidth]{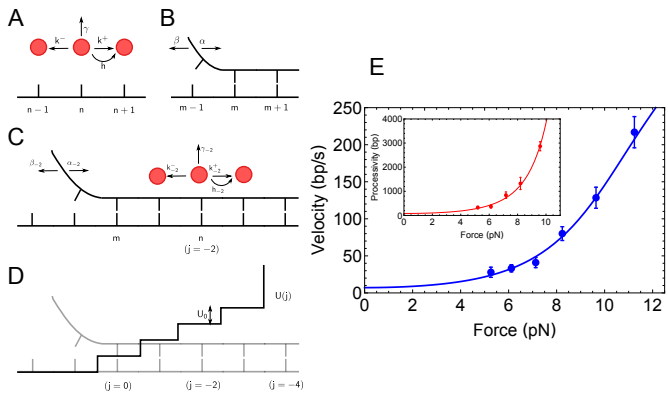}
    \caption{Helicase model.
    Panels (A-D): Schematic of a model showing the unwinding of nucleic acids by helicases. (A) Translocation process of a helicase, with $k^+$, $k^-$, and $\gamma$ being the rates of forward, backward, detachment, respectively; $n$ is the position on the nucleic acid.  
    (B) Opening ($\alpha$) and closing ($\beta$) rates of a base pair. 
    (C) Interactions between helicase and nuceic acids modify the stepping kinetics of  the helicase and and the opening and closing rates of the base pair. The base pair location is $m$ and $j = m - n$.  
    (D) A model for the interaction with $U_0$ being the strength. For a passive helicase $U_0$ is zero.
    Panel (E): Simultaneous fits of the force-dependence of the velocity  and processivity (given in the inset) of T7 helicase as a function of an external load. The blue and red lines are the results using the theory described elsewhere \cite{Chakrabarti19BJ} and the data points in circles are taken from experiments \cite{Johnson07Cell}.}
    \label{Fig:Helicase}
\end{figure}
Helicases, which are molecular motors found in all organisms, unwind double stranded (ds) DNA and dsRNA when they encounter a junction between single strand (ss)-ds nucleic acids~\cite{lohman1992,lohman_mechanisms_1996,DelagoutteQRevBiophys02,DelagoutteQRevBiophys03}. Separation of dsDNA strands is required for DNA replication as well as DNA repair. Malfunction of helicases causes genomic instability and are also implicated in cancer.  In addition, certain RNA chaperones are also deemed to have helicase activity, which means they are able to unwind helices in RNA in order to facilitate its folding \cite{Mohr02Cell,Russell2013RNAbiology}. We refer the reader to a number of articles that describe a variety of cellular functions associated with helicases \cite{bianco_2000, lohman_non2008,pyle_translocation_2008, rocak_dead-box_2004, Bustamante11Cell,DongJBiolChem95,Velankar99Cell,MariansStructure00,PangEMBO02,VenkatesanJBiolChem82}. Helicases,  classified into six super families based on their sequences \cite{GorbalenyaCOSB93,IyerJStructBiol04}, are described as active or passive~\cite{lohman1992}. Active helicases destabilize the base pairs of the dsDNA, perhaps by exerting a force, thus separating  the two strands. If the helicase is passive then it binds to the ssDNA whenever thermal fluctuations transiently open the base pairs. As  the strands of the dsDNA are separated the helicase translocates along the ssDNA. This strand separation and translocation are intimately related.  Several ensemble experiments have provided glimpses into the stepping mechanism   \cite{wong_allosteric_1992, ali_kinetic_1997,  levin_helicase_2002, lucius_general_2003, jeong_dna-unwinding_2004} of helicases.   In addition, these experiments have also been insightful in deciphering how they interact with single strand--double strand (ss-ds) junctions, and how often they dissociate from their track. The most detailed picture of the functions of a number of helicases have come from single molecule laser optical tweezers
(LOT) and Magnetic Tweezers (MT) experiments. These experiments provided the kinetics of stepping and nucleic acid unwinding \cite{PatelARB00,PerkinsBiophysJ04,DessingesPNAS04,LionnetNucAcidsRes06}.  Such measurements are instrumental in not only formulating theories and simulations but also in refining them as additional high precision experiments become available.

In a remarkable and influential paper, Betterton and Julicher (BJ) \cite{Betterton03PRL,Betterton05PRE,Betterton05JPhysCondMatt} presented a theory, which describes quantitatively the coupling between translocation and strand separation. The framework used in this theory, which is another illustration of the SKM,  has been most instrumental in understanding the differences between active and passive helicases.  The BJ model assumes that the helicase moves forward (backward) at a rate $k^+$ ($k^-$) when it is very far from the ss-ds junction. This aspect of the motor movement is similar to Fig. 5 with $\gamma$=0.  When the motor is very far from the ss-ds junction  the helicase merely translocates along the ss nucleic acids.  Similarly, in isolation a nucleic acid base opens at a rate $\alpha$ and closes at a rate $\beta$. Depending on whether it is a AT or a GC base pair the rates are different but the inequality $\beta \gg \alpha$ holds in both cases. Since isolated base pair opening rates are due to thermal fluctuations they satisfy $\frac{\alpha}{\beta} = e^{-\Delta G_{bp}/k_BT}$ where $\Delta G_{bp}$ is stability of the base pair. However, interactions with base pairs modify these rates.  Let $n$ be the position of the helicase on the track and $m$ be the location of the ss-ds junction (Fig.~\ref{Fig:Helicase}). 
Upon approaching the ds-ss junction the helicase interacts with NA, which was modeled using a variety of potentials - all based on some combination square well like potentials.  Passive helicases, characterized by $U_0 \rightarrow \infty$, opportunistically step when the base is open. For active helicases, which forcibly rupture the base pair interactions,  $U_0$ is finite but depends on $j = m - n$.   The rates $k^+$ and $k^-$ and $\alpha$ and $\beta$ are modified when the helicase interacts with the NA \cite{Betterton05JPhysCondMatt}. To describe the action of the helicase one has to keep track of its position on the track as well the ss-ds location. Helicase-NA interactions modify all the relevant rates making them position ($j$) dependent. Using the model, with detachment rate set to zero,  the velocity of the helicase assuming that it steps one base pair at a time, is  given by,
\begin{equation}
V=\frac{1}{2}\sum \limits_{j}\left( k_{j}^{+}+\alpha _{j}-k_{j}^{-}-\beta
_{j}\right) P_{j},
\label{HelicaseV}
\end{equation}
where $P_{j}$ is the probability of observing the helicase and junction that are separated separated by $j$, $\alpha_j$ is the rate at which the junction opens when the helicase and junction are at separation j. The corresponding rates for base pair closing, forward, and backward stepping rates of the helicase are $\beta_j$, $k_j^+$, and $k_j^-$, respectively. 
 
Although experiments (see for example \cite{Manosas10NAR}) have been analyzed using the theory sketched above  the BJ model has to be generalized in order to make precise comparison with experiments. (1) To account for the average helicase processivity ($\langle \delta \rangle m$), which was not addressed in the original formulation, the consequences of detachment of the motor have to be considered. As pointed out earlier, almost all of the chemical kinetics models ignore detachment, which of course is unrealistic because the run-length in motors or number of base pairs that are disrupted if finite. (2) A theory that allows for arbitrary step size ($s$) is needed instead of assuming that $s$ is unity. Helicases, such as PcrA and NS3 interact with and possibly destabilize  several base pairs that are down stream of the ss-ds junction \cite{Velankar99Cell,Cheng07PNAS}. In other words, step size $s$ exceeds unity. (3) Experiments also apply external force and measure the changes in the processivity and velocity as a function of force, $f$.  A viable theory should produce tractable expressions for both the mean velocity and $\langle \delta m \rangle$ as a function of of quantities. (4) Finally, how the sequence of the NA affects $\langle \delta m \rangle$ and $V$ needs to be considered in order to draw general conclusions. 

An analytically solvable model that accounts for the first three effects stated above has been proposed recently \cite{Chakrabarti19BJ}, which was preceded by a less general theory ($s$ was set to unity) \cite{Pincus15BJ} that investigate sequence effects ob the mean velocity and $\langle \delta \rangle m$. These studies produced a number of unexpected predictions for the helicase velocity and processivity as a function of  external force and DNA sequence. (i) It was predicted that, regardless of the underlying architecture and unwinding kinetics of the helicase or the precise DNA sequence, processivity exhibits a universal increase with applied external force. This finding, which has subsequently has been validated experimentally \cite{Li16NAR,Bagchi18NAR}, was used to suggest that helicases may have evolved to maximize processivity rather than velocity. (ii) The theory, which quantitatively accounts for the experiments for force-dependent $V$ and $\langle \delta m \rangle$ for T7 replisome (Fig. 6 from BJ 2019), shows that $s=2$ base pairs. (iii) Normally, in analyzing experimental data back NTP-dependent stepping rates are neglected. This is justified while the helicase translocates along ss NA. For instance, in T7 helicase the ratio of the forward to back stepping rate is $\sim$ 270, a value that is not that different from kinesin at zero resistive force. The probability of back-stepping is $\approx$ 0.3\%. However, when T7 unwinds dsDNA the probability of back-stepping increases to 26\%. This estimate is not that dissimilar to observations in XPD helicases belonging to a different (SF2) family, where it was shown that at 1mM ATP concentration the back-stepping probability is $\sim$ 10\%.   (iii) Interestingly, many helicases do not function {\it in vitro} unless an external force is applied. Under {\it in vivo} conditions partner proteins that bind to single strands and impart a force at the ss-ds junction to destabilize the base pairs \cite{Pincus15BJ} are needed.  That this is the case has been shown for UvrD helicase \cite{Comstock15Science}, which behaves as a processive motor only when 2pN force is applied. It was noted previously that even though these associated proteins may or may not increase the unwinding velocity of a helicase they should universally increase the processivity of the helicase. 

The theories for helicase stepping are not complete because they do not resolve many of the challenging problems. First, there is no quantitative explanation for the very broad velocity distribution \cite{Johnson07Cell} measured during unwinding of the weakly active ring-shaped T7 helicase. It is challenging to calculate $P(v)$ for the recent model \cite{Chakrabarti19BJ}, which is the minimal that accounts for the $F$-dependent mean velocity and $\delta m$ accurately. Second, there is very little understanding of the structural basis of the universal increase in $\langle \delta m \rangle$ as a function of $F$ and the underlying dramatic variations in sequence and architectures of the motor. Perhaps, carefully designed simulations might shed light on this issue \cite{Yu06BJ,Ma18eLife}. Third, a recurring theme in many aspects of motors is that the dynamics is heterogeneous, exhibiting characteristics reminiscent of glasses. It was discovered that there are substantial molecule-to-molecule variations in the unwinding speed of  {\it E. Coli.} RecBCD helicase even if all the enzymes are prepared under the same condition. To account for this observation it has been suggested that the functional landscape is likely partitioned into a number of metastable states and the initial preparation quenches the enzyme into a specific substate \cite{Kirkpatrick15RMP} The helicase ergodically explores all the conformations within a single metastable state but the transitions to other states could only be achieved by resetting the ATP concentration \cite{Liu13Nature}. The emergence and relevance of glass-like heterogeneous behavior, under ambient conditions, is not understood theoretically,  and remains a challenge not only in the context of helicases but also in other biological systems as well \cite{Solomatin10Nature,Altschuler10Cell,Hyeon12NatChem}.

\section{Discussion}
The combination of excellent experiments and few theoretical approaches touched on here have greatly advanced our understanding of how biological machines work. However, we still do not have a complete understanding of how these machines work even in {\it in vitro} conditions. This should not be a surprise because some believe that the link between allosteric communication in hemoglobin  (Hb) and oxygen transport is not fully understood despite over fifty years of intensive study. It is unclear if at present there is an analogue of Hb, considered the hydrogen molecule for allostery, in molecular machines. The investigation of  these machines, one at a time {\it in vitro},  seems to raise unsolved problems in each individual case. We outline a few of the challenging problems. Clearly, the list is far from being exhaustive.
\medskip

\subsection{Sometimes Details Matter}  Here, we have only described coarse-grained theoretical methods, which are impervious to the molecular details. There are several examples (we mention two) in which spectacular changes occur by a single or few amino acid substitutions. 
(i) About twenty years ago, Endow and Higuchi \cite{Endow00Nature} made a single amino acid substitution in the neck linker (NL) region of a Ncd, a motor that is related to kinesin. 
The wild type Ncd walks towards the minus end of the microtubule in contrast to conventional kinesin. 
Upon substituting an asparigine (a polar amino acid residue) to lysine (positively charged) in the NL, an element that is responsible for the motor to walk predominantly on a single protofilament of the MT, it was found that Ncd moves in both the plus and minus direction on the MT.   
(ii) Myosin VI, unlike myosin V, walks towards the filamentous actin  minus end and is the only known member  belonging to the myosin super family with this property. 
In humans there are reports of three mutations that cause deafness. 
A point missense mutation (replacement of aspartic acid by tyrosine (D179Y) in the so-called U50 domain of the motor) leads to deafness in mouse \cite{Hertzano08PlosGenet}. 
It has been suggested \cite{Pylypenko15PNAS}, using a variety of experimental methods,  that D179Y mutant leads to a premature release of the phosphate P$_i$, a product of ATP hydrolysis, from the detached head, thus preventing it from rapidly binding to F-actin. 
In the meanwhile, ATP does bind to the leading  head, which detaches the motor from actin, thus preventing processive motion.   
Besides these examples there are many others, such as the link between mutations in $\beta$-cardiac myosin and hypertrophic cardiomyopathy. 
None of these observations are amenable to theoretical treatments exposed here, in which molecular details are ignored. 
Detailed simulations, if possible, could provide biophysical insights but linking such studies to functions is a daunting task. 
These anecdotal examples should remind us that in the search of principles of generality in biology one should not forget that molecular details matter and could dramatically influence functions. 
\medskip

\subsection{Efficiency and optimality} A naive assessment  of efficiency may be made by estimating the theoretical stall force, based on the available free energy due to ATP hydrolysis, and comparing it to the measured stall force \cite{kolomeisky07arpc}.
Consider F$_1$-ATPase for example. This is part of the F$_0$F$_1$-ATP synthase responsible for synthesizing ATP. 
This rotary motor undergoes precise 120$^\circ$ rotations in the absence of an external applied torque ($\tau_{tor}$) at low ATP concentrations. 
By controlling $\tau_{tor}$ using the electro rotation method and the chemical potential by choosing appropriate ATP, ADP, and P$_i$ concentrations, it is possible to measure the probability ($p_s$) of rotation  in the synthetic direction (ADP and P$_i$ are consumed to generate ATP and  the probability  $p_h$ in the reverse hydrolytic direction could be measured (see for example \cite{Toyabe11PNAS}). 
From the linear dependence of $k_B T ln \frac{p_s}{p_h}$ on it was found that the output energy at stall is roughly equal to the chemical potential. 
This implies that F$_1$-ATPase operates at near 100\% efficiency.

For myosin motors, which take roughly a $d$=36 nm step, the maximum force that can be exerted is $\fmax \approx \frac{\Delta G_{ATP}}{d}$, which is approximately 2.5 pN assuming that $\Delta G_{ATP} \approx 22 k_BT$. 
The measured stall force ($\fstall$) is roughly in this ball park, which suggests that myosin motors operate efficiently ($\eta = \frac{\fstall}{\fmax}$ is very high). A similar argument for kinesin yields ($d$=8.1nm) $\fmax \approx$ 12pN whereas measurements report values close to 8 pN. Thus, there is about a 30-35\% decrease in $\eta$ for kinesin motors. 
A precise computation of efficiency should be undertaken by considering the network for a given motor that captures many aspects of motor motility. 
The arguments given above hold roughly if the motility can be described in a periodic one dimensional tilted potential with two equivalent sites with a transition state, which is close to the initial site. 
Treatments \cite{Schmiedl08EPL,Golubeva12EPL,Seifert11PRL,Wagoner19PNAS} using more elaborate models indicate that the the motor efficiency would be much less and would depend on the details of the network dynamics. 

A question that is related to efficiency is optimal performance. In the context of the biological machines discussed here, performance should be measured by velocity of movement, processivity, and for molecular chaperones the time-dependent production of folded state. As mentioned above, studies in the last decade have addressed how biological machines might optimize speed by considering models that are used to analyze force-velocity curves in motors. One of the lessons is that speed might be optimized if the motor takes many sub-steps instead of a single step \cite{Wagoner16JPCB}. These studies have not considered processivity (run length) as a function of ATP and external force. For helicases, it appears that maximization of velocity is not as relevant as optimization of processivity. In addition, for GroEL it appears that the rate of production of the folded state per unit time is maximized even at the consumption of lavish amount of energy, which would render this machine highly inefficient. Whether questions pertaining to optimal performce, given available free energy, must consider simultaneously many functional requirements  remain an open problem. It is possible that optimality could depend on specific function carried out by a class of machines. 
\medskip 

\subsection{ Specificity versus Promiscuity}
 The {\it E. coli} chaperonin has evolved to be a promiscuous machine in that it facilitates the folding of a variety of misfolded substrate proteins (even those that are not in the {\it E. coli} proteome) that are unrelated by sequence, size, or the structure of the folded state.  Using directed evolution methods a mutated GroEL/GroES, referred to as GroEL$_{3-1}$ was constructed \cite{Wang02Cell}.   The altered GroEL$_{3-1}$ contained a single mutation in GroES (tyrosine was replaced by histidine) and two mutations (valine was substituted for alanine and glycine for aspartic acid) in  GroEL. It was found that GroEL$_{3-1}$ had enhanced ATPase activity compared to the wild type. More importantly, it was found that GroEL$_{3-1}$,  with a highly polar environment in the cavity compared to the wild-type, dramatically increased the folding of green fluorescent protein (GFP). However, the enhanced specificity of folding GFP with ease came at the expense of substantial reduction in the capacity of GroEL$_{3-1}$ to facilitate the folding of several other proteins. Thus, in this instance nature has solved the tension between specificity and promiscuity by evolving an all purpose {\it E. coli} chaperonin that can process the folding of a large class of proteins, albeit not as efficiently. In eukaryotes there has been a great expansion in the number of chaperone classes possibly to enable the larger and more complex proteome. This example suggests that evolutionary constraints might have to a part of addressing issues related to optimality. This might imply that the simple network used to explain the out of equilibrium performance of GroEL/GroES has to expanded to describe optimality in the functions of chaperone networks in eukaryotes. 
\medskip

\subsection{Biological complexity} 
We circle back to Fig.~\ref{Fig:KinDynMyo}, which is a schematic illustration of  transport of melanosomes  (vesicles containing the light absorbing pigment melanin found in amphibians). The vesicles are either dispersed throughout the cytosol or aggregate near the cell center. The transportation of melanosomes is clearly complex and is controlled by the interplay of multiple motors involving kinesin-2, a plus end directed MT motor, and dynein, that walks towards the minus end of the MT. In addition, actin bound myosin V is also involved in the transport. It is suspected that a low number of motors (about 1-2 kinesin-2 motors and roughly 1-3 dyneins) move melanosomes during aggregation \cite{Levi06BJ}. In contrast, pigment dispersion is mediated by kinesin-2 as well as assistance by myosin V. Because both these motors compete for the same attachment site (p150Glue on dynactin, see Fig.~\ref{Fig:KinDynMyo}) it follows \cite{Levi06BJ} that myosin V is released during aggregation and vesicle transport is dominated by dynein. The need to change movement of melanosomes, powered by multiple motors, is thus determined by function, which in this case is related to their dispersion or aggregation.   

Although individual motors predominantly move unidirectionally on cytosketal filamentous tracks there are reports that motors could change directions as well. A very impressive {\it in vitro} illustration of bidirectional motility of the complex of dynein with dynactin (a complex that is attached to the cargo and  activates dynein (Fig.~\ref{Fig:KinDynMyo}) was reported sometime ago \cite{Ross06NCB}.  It was found that the complex moves processively in both directions on MT with ATP-dependent velocities that are not significantly different in either direction \cite{Ross06NCB}. Although the authors provide a qualitative picture of the mechanism of bidirectional transport a theory for such unexpected behavior is lacking. It is not even clear, at least to us, the level of coarse-graining needed to construct such a theory, which is clearly needed to unveil the complexity of vesicle transport. 

\section{A Final remark}
The eventual goals of understanding biology through the lenses of physics are to create theoretical tools that are capable of describing biological functions under crowded and noisy cellular conditions, and in the process discover general physical principles that control life processes. It is likely that as the scale at which living systems are examined increases it may be possible to describe cellular processes using functional modules \cite{Hartwell99Nature}, which is a coarse grained view of biology. However, it would be hard to anticipate the functions of such modules from their components, which are the molecules of life. Furthermore, interactions between modules could lead to new functions not encoded in isolated modules, as illustrated by many examples described in our perspective. After all ``more is different" \cite{Anderson72Science}. Because ``Nature is an excellent  tinkerer, not an engineer" (a quote attributed to Francois Jacob), and tinkering in biology involves stochastically altering existing modules to evolve new functions  without time constraints or any ultimate design as a goal \cite{Jacob77Science}, it is likely that concepts in many fields of science would have to be used to develop an integrated view of biology.

\section*{Acknowledgments}
We are grateful to:
Matthew Caporizzo,
Shaon Chakrabarty,
Yale Goldman,
Yonathan Goldtzvik,
Sabeeha Hasnain,
William O. Hancock,
David Hathcock,
Chris Jarzynski,
Anatoly Kolomeisky,
George H. Lorimer,
Micheal Ostap,
George Stan,
Ryota Takaki,
Riina Tehver,
Huong Vu,
Ahmet Yildiz, and
Zhechun Zhang
for several useful discussions and collaborations involving the topics covered in this review. DT is especially grateful to Prof. Michael E. Fisher for innumerable inspiring discussions for over twenty years not only on molecular motors but also on the potential role of physics in biology.  This work was supported in part by a grant from National Science Foundation (CHE 19-00093) and the Collie-Welch Chair (F-0019) administered through the Welch Foundation.


%

\end{document}